\documentclass{aa}

\usepackage{longtable}
\usepackage{graphicx}
\usepackage{txfonts}
\usepackage{natbib}
\usepackage[dvipsnames,usenames,table]{xcolor}
\usepackage[normalem]{ulem}

\usepackage{longfigure}

\usepackage{txfonts}
\begin{document} 

\title{Meandering periods and asymmetries in light curves of Miras: Observational evidence for low mass-loss rates\thanks{Table~\ref{Tab:All data} is only available in electronic form at the CDS via anonymous ftp to cdsarc.u-strasbg.fr (130.79.128.5) or via http://cdsweb.u-strasbg.fr/cgi-bin/qcat?J/A+A/.}}

\author{
P. Merchan-Benitez\inst{\ref{inst_extrema1}}
\and
S. Uttenthaler\inst{\ref{inst_iap}}
\and
M. Jurado-Vargas\inst{\ref{inst_extrema2}}
}

\institute{
Faculty of Science, University of Extremadura, 06011 Badajoz, Spain\label{inst_extrema1}; \email{pedromer@hotmail.com}
\and
Institute of Applied Physics, TU Wien, Wiedner Hauptstra\ss e 8-10, 1040 Vienna, Austria\label{inst_iap};\\ \email{stefan.uttenthaler@gmail.com}
\and
Department of Physics, Faculty of Science, University of Extremadura, 06011 Badajoz, Spain\label{inst_extrema2}; \email{mjv@unex.es}
}

\date{Received November 30, 2022; accepted January 30, 2023}


\abstract
{Some Miras -- long-period variables in late evolutionary stages -- have meandering pulsation periods and light curve asymmetries, the causes of which are still unclear.}
{We aim to understand better the origin of meandering periods and light curve asymmetries by investigating a sample of solar-neighbourhood Miras. We characterised this group of stars and related their variability characteristics to other stellar parameters.}
{We analysed observations from several databases to obtain light curves with maximum time span and temporal coverage for a sample of 548 Miras. We determined their pulsation period evolution over a time span of many decades, searched for changes in the periods, and determined the amplitude of the period change. We also analysed the Fourier spectra with respect to possible secondary frequency maxima. The sample was divided into two groups with respect to the presence of light curve asymmetries ("bumps"). IR colours and indicators of the third dredge-up were collected to study the sample stars' mass-loss and deep mixing properties.}
{Our analysis revealed one new star, T~Lyn, with a continuously changing period. The group of Miras with meandering period changes is exclusively made up of M-type stars. The Fourier spectra of the meandering period Miras have no prominent additional peaks, suggesting that additional pulsation modes are not the cause of the meandering periods. We confirm that light curve bumps are more common among S and C Miras and show, for the first time, that Miras with such bumps have 
lower mass-loss rates than those with regular, symmetric light curves. Also Miras with meandering period changes have relatively little mass loss.}
{We conclude that Miras with strongly changing periods (including meandering periods) or asymmetries in their light curves have relatively low dust mass-loss rates. Meandering period changes and light curve asymmetries could be connected to He-shell flashes and third dredge-up episodes.}


\titlerunning{Miras with meandering periods}
\authorrunning{Merchan-Benitez et al.}
\maketitle

\section{Introduction}\label{sec:Intro}

Mira stars are long-period variables in the Asymptotic Giant Branch (AGB) phase of stellar evolution. They are characterised by large amplitudes of variation ($\Delta V>2\fm5$) and pulsation periods $P\sim100-1000$\,d. Most Miras have periods that are stable over time spans of decades or even centuries. Nevertheless, some stars experience significant changes in the pulsation period. A division of Miras with changing pulsation periods into three different groups was made by \citet{Zijlstra2002a}:

\begin{itemize}
\item \textbf{Continuous Period Changes (CPC)}, with a continuous and prolonged increase or decrease of the period over time and no evidence of epochs with stable periods. Only $\sim 1\%-2\%$ of Mira-type variables show CPCs. Highlights are the period changes observed in R~Aql, R~Hya, and W~Dra, of $\Delta P/P \sim 15\%$ or more \citep{Wood1981}.
\item \textbf{Sudden Period Changes (SPC)}, with a sudden and rapid period change after a relatively long phase of stable pulsation period, with a total variation similar to the CPCs but reached in a much shorter time. For example, the period changes observed in T~UMi \citep{Gal1995}, R~Cen \citep{Hawkins2001}, and RU~Vul \citep{Uttenthaler2016a} are noteworthy. Also this type of period variation is observed in only $\sim 1\%-2\%$ of Mira-type stars.
\item \textbf{Meandering Period Changes (MPC)}, with periods going up and down by up to $\sim10\%$ of the average period on time scales of a few decades. The rate of change is comparable to the previous rates, but the total period change is in general slightly smaller than in the other two classes. For example, the variations observed in S~Ori amount to $\sim 9\%$ of its average period \citep{Merchan-Benitez2002}. The fraction of observed Mira stars that undergo MPCs is higher than the two other groups, around $\sim10\%$. This fraction is of the order of 15\% if only stars with periods $P>400$\,d are considered. 
\end{itemize}

Several hypotheses have been put forward to explain these period changes. Most importantly, violent ignitions of the He-burning shells called thermal pulses (TPs) are expected to have an important impact on the stellar structure and, thus, on the pulsation period of the outer envelope. When a TP sets in, the star significantly shrinks, and its pulsation period shortens. Since this is predicted to happen within a few decades, it is thought that the onset of a TP may explain the SPC class of Miras with strongly {\it decreasing} periods. From theoretical considerations \citep{Wood1981}, it is expected that $\sim1\%$ of Miras undergo this phase in the TP cycle at any one time. At later phases of the TP cycle, the changes in stellar radius and, thus, pulsation period become slower, so the CPC class of Miras may also be explained as a result of a TP.

From an observational point of view, stars undergoing TPs can be distinguished based on nucleosynthesis products such as radioactive technetium (Tc) and $^{12}$C mixed to the stellar surface in deep-mixing events called the third dredge-up (3DUP). A 3DUP event may occur several hundred years after the TP ignition \citep{Herwig2005}, and Tc can be detected in optical stellar spectra \citep{Little1987,Uttenthaler2011}. However, the absence of Tc does not necessarily mean the absence of TPs because they may have yet to be powerful enough to drive 3DUP in their aftermath.

\citet{Templeton2005} analysed decades-long light curves of Mira-type variables, finding that only $\sim 10\%$ showed significant period variations on time scales of decades. Among them, only eight ($\sim 1.6\%$) showed highly significant monotonic period changes, a fraction consistent with that expected for stars in the early post-TP phase, where the models predict the largest period change. The rest of the stars with significant but non-monotonic period changes belong to the MPC group in the \citet{Zijlstra2002a} classification.

The physical mechanism to account for these MPCs is unclear, but various theories have been put forward in recent decades. \citet{Templeton2005} speculate that MPCs might be related to TPs, too, although the timescales of these variations (several decades) are much shorter than those predicted for global changes induced directly by a TP. \citet{Ostlie1986} worked with several AGB model stars in the mass range $0.8 - 2.0 M_{\sun}$ and obtained Kelvin-Helmholtz cooling timescales $\tau_{KH}$ between 6 and 200 years. Thus, the timescales observed in the MPCs are similar to the Kelvin-Helmholtz cooling timescale of solar-like stars' envelopes. Therefore, these shorter period variations may be thermal relaxation oscillations in response to the global changes caused by a TP. Another explanation, in this case non-evolutionary, is that the pulsations of Mira variables are intrinsically nonlinear by nature and exhibit low-dimensional chaotic behavior \citep{Kiss2002}. Amplitude (not period) variations in the light curve of some of these stars could be due to the nonlinear interaction of two or more pulsation modes, similar to those found in RV~Tauri stars \citep{Buchler1996}. Other theories suggest that the pulsations of Mira-type variables may be strong enough to modify the stars' internal structure \citep{Ya'Ari1996}. It could modify the entropic structure of the star over time, causing both a change in pulsation mode and a readjustment of the equilibrium structure.

Also the shapes of the light curves of Mira variables have been studied. Here, the presence of asymmetries such as so-called "bumps" or "humps" are of particular interest. Classification systems of Mira light curves have, for example, been established by \citet{Campbell1925} and \citet{Ludendorff1928}. The latter separated the stars into three main groups ($\alpha$, $\beta$, and $\gamma$) and further divided them into a total of ten subgroups. \citet{Vardya1988} included an asymmetry factor $f$ in the classification, defined as the rise time in proportion to the mean pulsation period, and found that only a fraction of $\sim20\%$ of the Miras shows substantial deviations from a symmetric light curve with $f\leqslant0.4$ or $0.5\leqslant f$. \citet{Lebzelter2011} used the parameter $\chi^2$, defined as the sum of the squared differences between the light curve of a star and a sinusoidal reference curve, and found that $\sim30\%$ of the sample stars had light curves that significantly deviate from a purely sinusoidal shape. Lebzelter also found a connection between atmospheric chemistry and the light curve shape, with a higher fraction of S and C stars with non-sinusoidal variations compared to M stars. However, no simple correlation between the light curve shape and various colour indices was found. On the other hand, \citet{Lockwood1971} found that the stars' spectral types change along the rising and descending branches, but they do not appear to be directly related to the asymmetries and bumps observed. They determined that the stellar temperature increases continuously from minimum to maximum light, whether or not there are bumps. In summary, no clear explanation for light curve asymmetries exists to date. 

Several papers have focused on CPC and SPC stars, but there are not many systematic studies on MPC stars. In this paper, we analyse the periods and light curves of a sizeable sample of solar neighbourhood Miras, focusing on the MPC class to shed more light on the origin of this type of period change. We search multi-decade light curves for period changes, quantify the amplitude of period change, and inspect information on the 3DUP activity and mass loss of the sample stars. The mass-loss properties of Miras with changing pulsation periods, particularly the MPC group, have hardly been studied in the literature, with few exceptions \citep{Zijlstra2002b}.

\section{Sample stars and data reduction}\label{sec:sample-and-data}

\subsection{Sample selection and data collection}\label{sec:selection}

We used the long-term observations collected in four databases to search for period changes: AAVSO (American Association of Variable Star Observers), AFOEV (French Association of Variable Star Observers), ASAS (All Sky Automated Survey), and DASCH (Digital Access to a Sky Century at Harvard). Most data are visual or $V$-band observations of the AAVSO and AFOEV. In these two databases, the visual observations are heterogeneous because they are made by a large number of observers with slightly different ocular responses and observing methods, so they will be subsequently subjected to an averaging process, see below. On the other hand, the large amount of observational data collected allows for obtaining long-term average light curves with good accuracy, reaching, in some cases, more than 120 years of observations. The $V$-band photometric data collected from the ASAS database stem from observations obtained by a CCD photometric tracking program at Las Campanas Observatory, Chile, in the declination range $-90$\degr\ to $+28$\degr\ between 2000 and 2009. The limiting magnitude is about 14\fm5, and about 500 observations per star have been collected. Finally, the historical data in the DASCH database come from an archive of digitised photographic plates of the Harvard College Observatory \citep{Laycock2010}, dating back several decades to over 100 years. However, the temporal coverage of the ASAS and DASCH databases alone is too low to analyse the long-term period evolution. The ASAS database covers intervals of about ten years, while in the DASCH database, we can find very old observations but generally very dispersed in time. Therefore, although they are a great tool to use in combination with the other databases, they alone do not usually allow a detailed study of the period changes.

In addition, we used the relatively short-term, high-cadence ASAS-SN \citep[All-Sky Automated Survey for Supernovae,][]{Shappee2014} Catalog of Variable Stars III \citep{Jayasinghe2019} to analyse light curve shapes and search for asymmetries (Sect.~\ref{sec:LC-shapes}). For the stars with insufficient ASAS-SN observations or not well covered rising branches of the light curve, we turned to $V$-band photometric data from the AAVSO database and, ultimately, to binned light curves as described in Sect.~\ref{sec:binning}.

The starting point was to review in the General Catalog of Variable Stars \citep[GCVS,][]{Samus2017} those variables marked as Mira type, from which we found around 8000 stars. Of these, we discarded in the first instance those with a full $V$-band amplitude of less than 2\fm5 (i.e., those stars not fulfilling the amplitude criterion of Miras), obtaining a first selection of about 1500 Mira-type variables. Many of these stars are very faint, so their light curves in the AAVSO and AFOEV databases are too sparse to analyse the period evolution. Therefore, they were eliminated from our selection by a simple visual inspection of their light curves. Based on these criteria, the final selection contains 548 Mira-type variables, of which 472 are M-type, 43 are S-type, and 33 are C-type. Although the sample size is almost identical to \citet[][547 stars]{Templeton2005}, we have only 465 stars in common. The largest part of the difference may be explained by the fact that our time series are augmented by DASCH data where possible, and our selection criteria do not include stars such as semi-regular variables or symbiotics.

\begin{figure}[!t]
\includegraphics[scale=0.55]{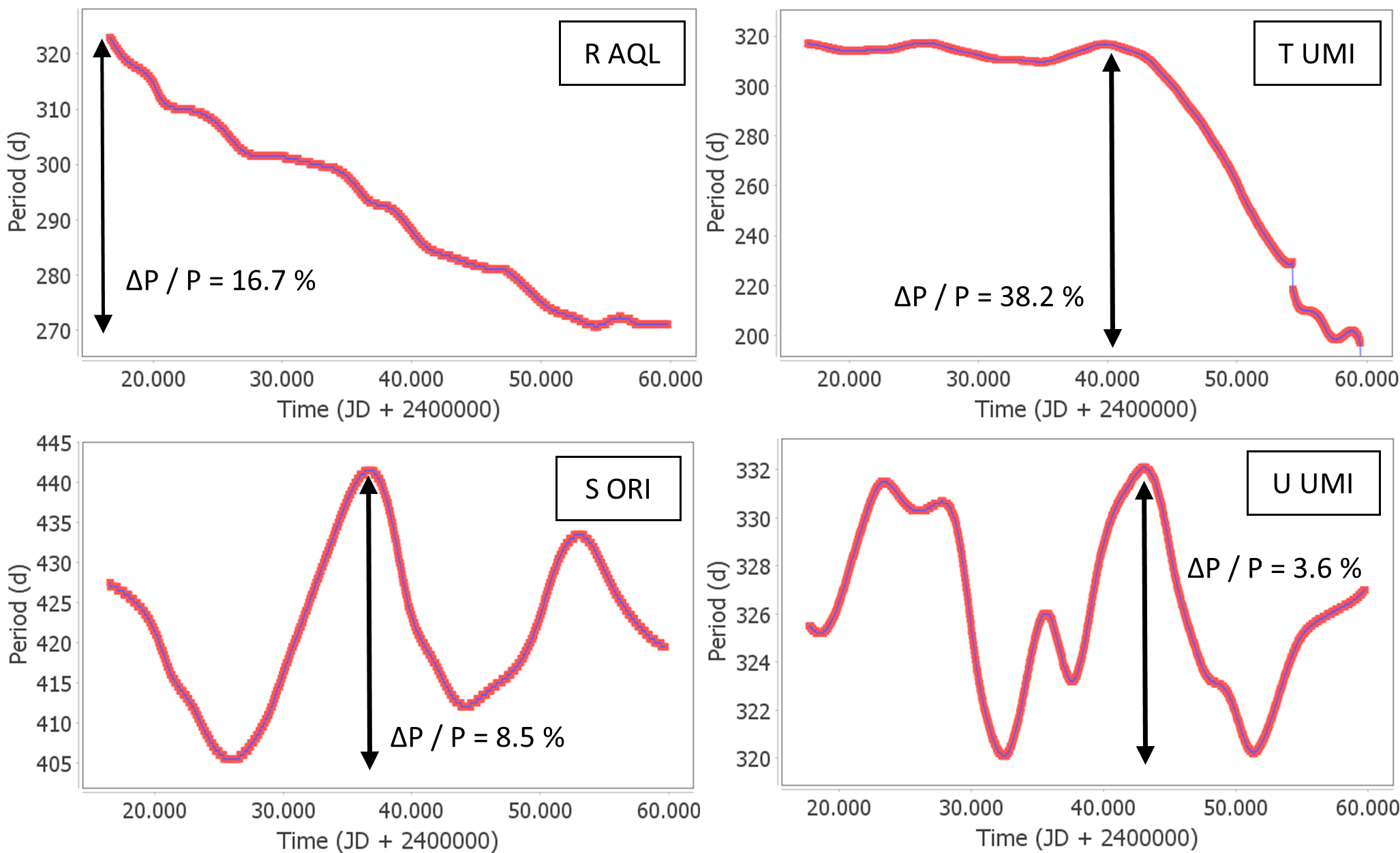}
\caption{Typical examples of period changes found with the VStar program: {\it Top left:} Continuous period change (R~Aql); {\it Top right:} Sudden period change (T~UMi); {\it Bottom left:} Significant meandering period change (S~Ori); {\it Bottom right:} Non-significant meandering period change (U~UMi).}
\label{Fig:VStar}
\end{figure}

In these 548 sample stars, we studied the period changes using the VStar program, which includes the option of performing time-frequency analysis through the Weighted Wavelet Z-Transform \citep[WWZ,][]{Foster1996}. With this tool, we constructed plots of the period versus time and determined the full amplitude of period variations, calculated as the ratio between the observed amplitude of the period variation and the mean period obtained by Fourier analysis of the complete data. We then separated the sample stars with large period changes into the three groups described in the Introduction: stars with continuous, sudden, and meandering period changes. Figure~\ref{Fig:VStar} shows typical examples of these different types of period change in a period vs.\ time plot. This figure includes two examples of MPC: S~Ori as one of the most obvious examples of period change greater than 5\% (which we later call significant MPC, see Sect.~\ref{sec:K-22_vs_DP/P}), and U~UMi with period change less than 5\%. Figure~\ref{Fig:VStar} also illustrates the definition of the full amplitude of period variation calculated for each star.

\subsection{Binning and averaging light curves}\label{sec:binning}

We focus on those Mira-type variables among the 548 sample stars that present either significant meandering period changes above 5\% or clear asymmetries or bumps in their light curves.

\begin{figure}[!t]
\includegraphics[scale=0.58]{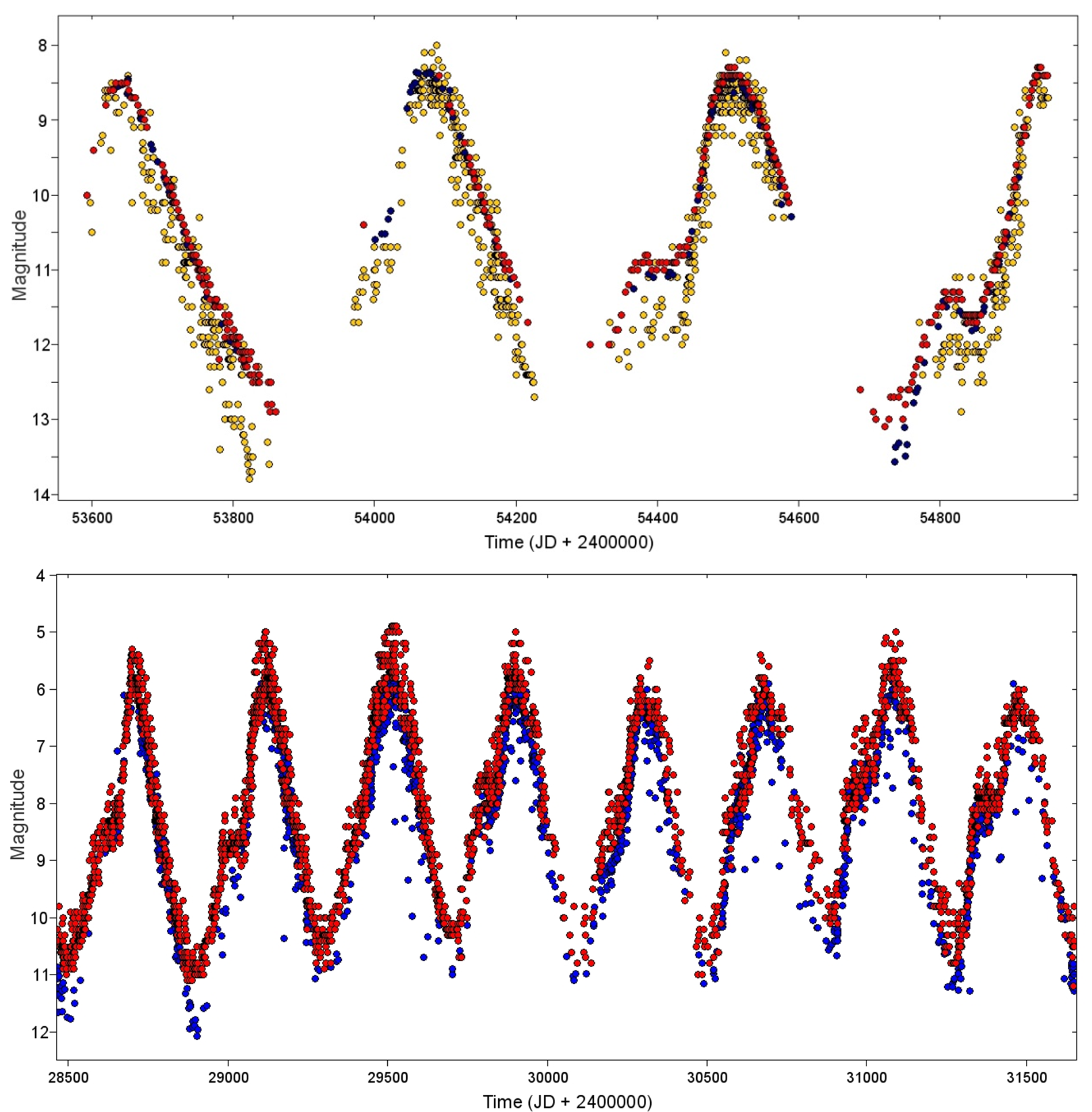}
\caption{Combined individual data from different databases. {\it Top panel:} S~Ori: visual AAVSO (yellow), $V$-band AAVSO (blue), and ASAS (red). {\it Bottom panel:} T~Cep: visual AAVSO (red) and DASCH (blue).}
\label{Fig:comb_data}
\end{figure}

Therefore, for these latter stars, we collected all available data in the four databases, seeking the maximum temporal coverage to analyse their period changes and as much data as possible to analyse in detail the shape of their light curves (e.g., bumps). All these observations were subjected to a validation process during which discrepant data or data with high uncertainty were eliminated. Subsequently, we checked in graphs if they could be combined. As an example, Fig.~\ref{Fig:comb_data} shows sections of the light curves of S~Ori and T~Cep with all the data combined, where we can observe that the minor differences that may exist between the various databases do not exceed the dispersion level of the data. Therefore, data from all sources were merged to calculate the average light curves.

\begin{figure}[!t]
\includegraphics[scale=0.58]{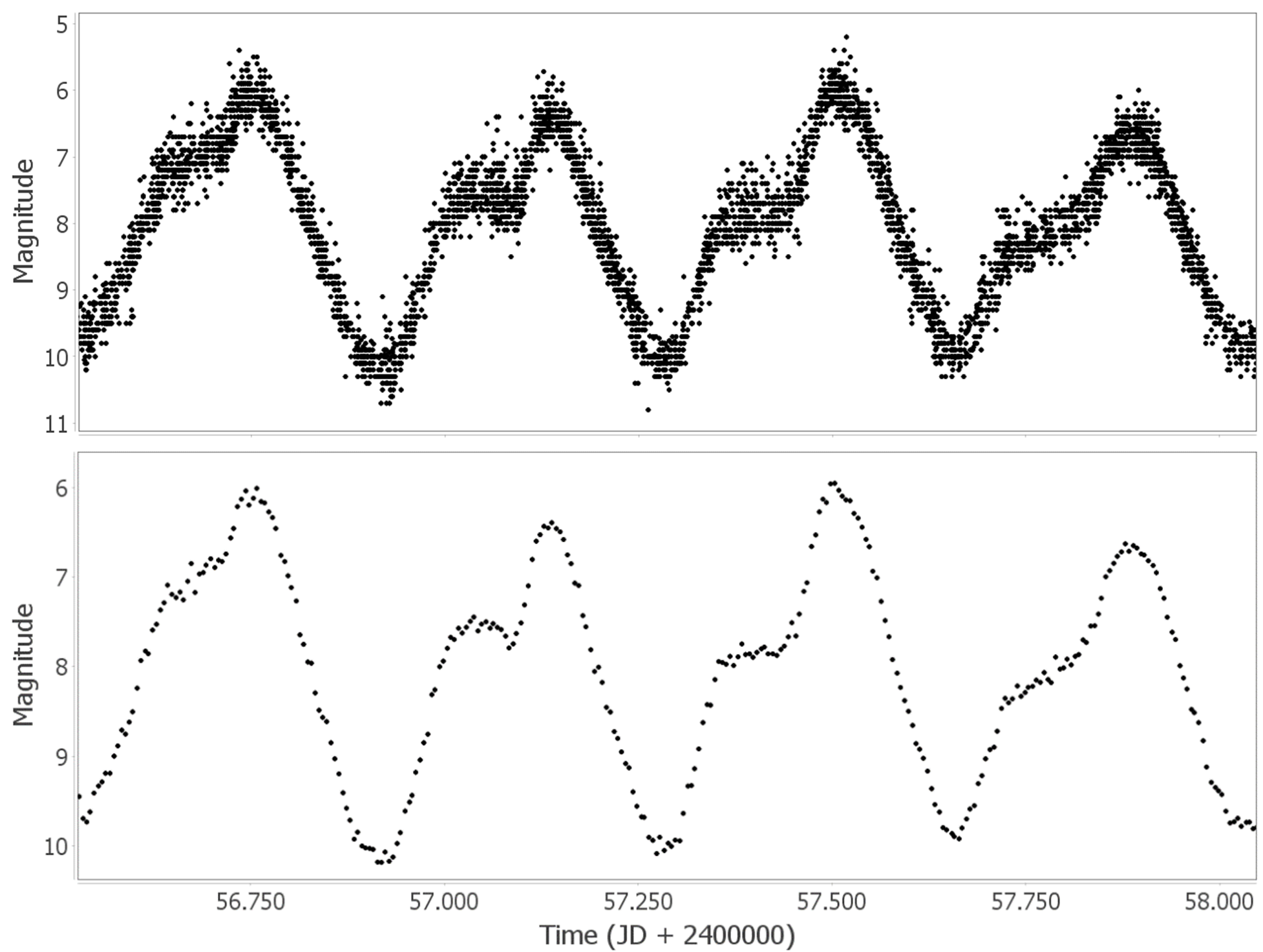}
\caption{Section of the light curve of T~Cep. {\it Top panel:} raw observational data. {\it Bottom panel:} final data after processing and averaging in 5-day bins.}
\label{Fig:data_proc}
\end{figure}

\begin{figure*}[!t]
\includegraphics[scale=1.2]{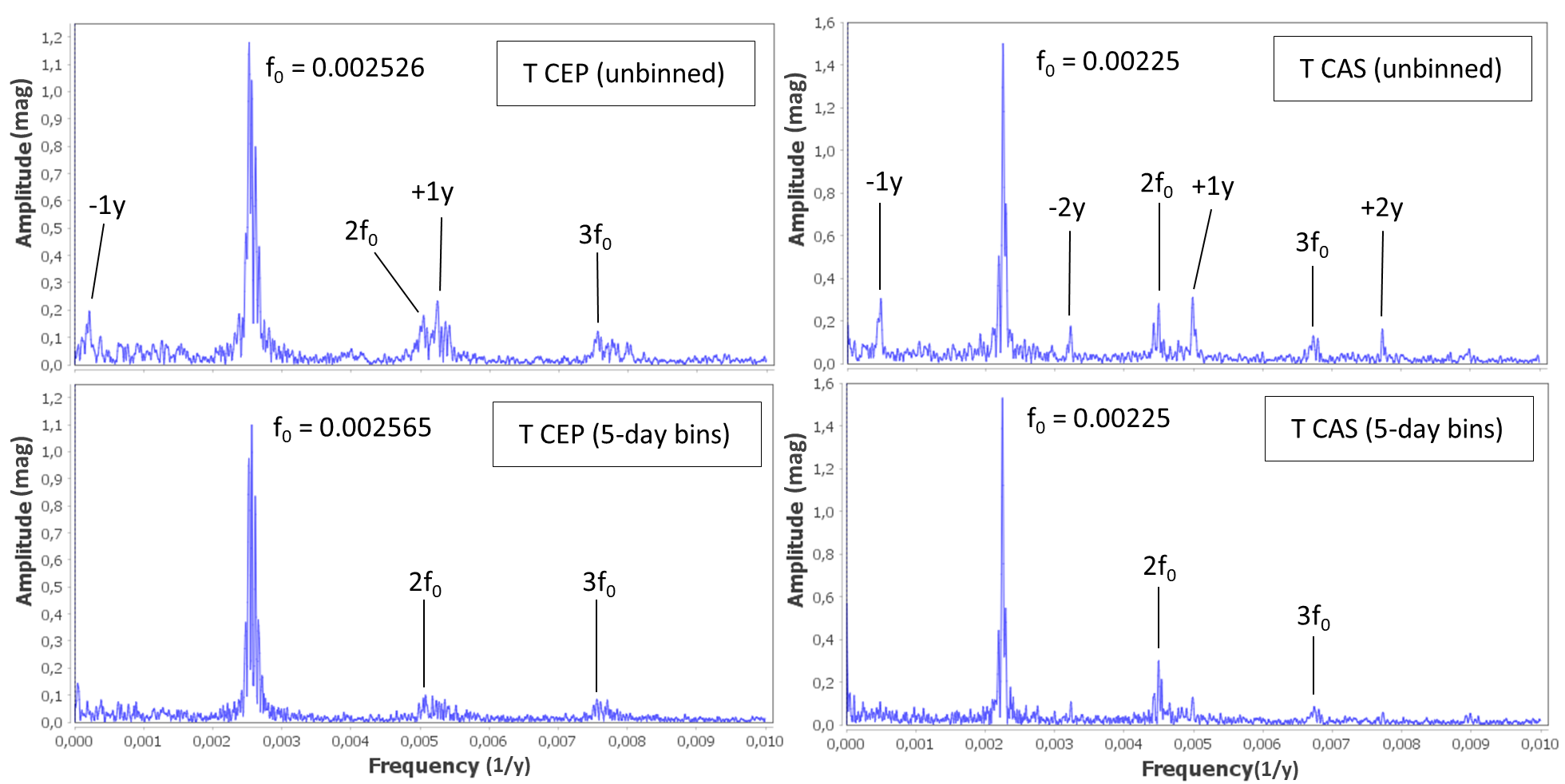}
\caption{Fourier spectra of T~Cep and T~Cas, using unbinned data and 5-day binning, see the figure legend.}
\label{Fig:Kiss_test}
\end{figure*}

For the averaging process of the observations, we tailored a code to calculate the average value of all the individual data with intervals between 3 and 5 days, depending on the star's pulsation period, obtaining between 70 and 80 data points per cycle. Double weight was assigned to the AAVSO, AFOEV, ASAS, and DASCH $V$-band photometric data compared to the AAVSO and AFOEV visual data due to the higher precision of the former. The result can be seen in Fig.~\ref{Fig:data_proc}, which shows, as an example, the final processed light curve of T~Cep. The plots demonstrate that our binning and averaging process results in very smooth light curves, well-suited for further analysis.

\subsection{Frequency test}\label{Frequency test}

To determine the possible impact of the binning process and subsequent averaging of data on the Fourier spectra, as well as to study the possible appearance of artificial peaks, we performed two different tests, cf.\ \citet{Kiss1999}. The first test compared the Fourier spectrum of the raw data with that of the averaged data in 5-day bins. The results of this test are shown in Fig.~\ref{Fig:Kiss_test} for T~Cep and T~Cas. These two stars were chosen because the number of observations differs significantly: 121\,622 and 55\,417, respectively. The dominant frequency obtained is shown in the figure. No additional peaks appear, and the main frequency remains practically unchanged in both position and amplitude. The structure of the main peaks is preserved, and the aliasing peaks diminish. Similar results were obtained by performing this process with the rest of the stars analysed in this study.

\begin{figure}[!t]
\includegraphics[scale=0.57]{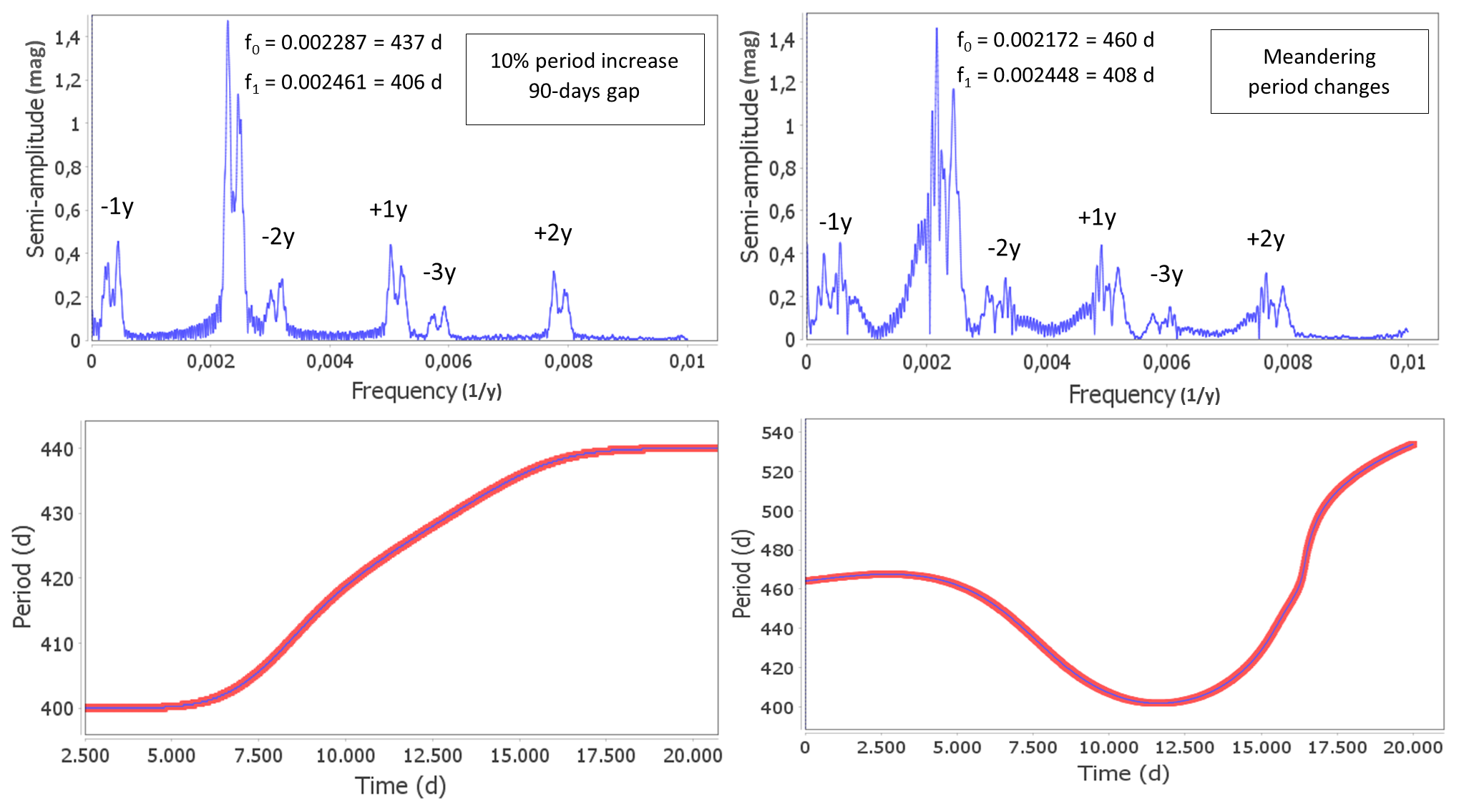}
\caption{Numerical tests with synthetic light curves. {\it Left panels:} Continuous period increase of approximately 10\%, with seasonal gaps of 90 days. {\it Right panels:} Alternating period decrease and increase, simulating MPC type variations.}
\label{Fig:synthetic1}
\end{figure}

The second test was done with synthetic light curves and pursued two aims: i) check the impact of seasonal observation gaps on the appearance of additional peaks in the Fourier spectrum, and ii) check if our analysis program can recover the phases of period increase or decrease similar to those observed in MPC stars. We have generated several synthetic light curves with random observational noise to the magnitude (Gaussian noise with $\sigma=0\fm15$) and the observation date (uniformly distributed shifts of $\pm$2\,d to the five-day bins). The synthetic light curves span a time of about 65 years. With these artificial data, we tried to mimic the observations used for the analyses and the period changes observed in the MPC stars, both in depth (between 5 and 10\%) and in time scale (several decades). These synthetic light curves were generated in two different situations. First, a continuous period increase of 10\% was assumed (between 400 and 440 days over about 25 years), including 20 years of constant period before and after the increase. The second situation assumed a continuous period decrease of 5\% (between 525 and 500 days over 25 years) and stable period phases of about 20 years before and after. In addition, in both situations, we varied the length of the seasonal gaps between approximately 0, 30, 60, and 90 days. The left panel of Fig.~\ref{Fig:synthetic1} shows the case results corresponding to a continuous period increase of approximately 10\% with seasonal gaps of 90 days. It verifies that the result is in good agreement with the input parameters.

We also analysed a synthetic light curve containing phases of rising and falling periods. This light curve's Fourier spectrum shows a broad peak between the frequencies corresponding to the minimum and maximum period but without additional peaks beyond those corresponding to the seasonal aliases. Period vs.\ time plots clearly reveal the evolution of the input period (right panel of Fig.~\ref{Fig:synthetic1}). Therefore, we conclude that our analysis program can reliably recover the actual period changes in the observed stars and allows us to interpret the possible peaks appearing in the Fourier analyses.

\subsection{Additional data}\label{sec:add_data}

Since we aim to study the mass-loss properties of our stars, we need a measure of their mass-loss rates. The best (gas) mass-loss rates are traditionally determined from radio CO rotational lines. However, they are available only for a small fraction of our sample stars. Therefore, we mainly use the $K-[22]$ colour as an indicator of the dust mass-loss rate. Although this IR colour traces only the dust in the stellar outflow, which is a small fraction of the total mass loss, and the dust-to-gas ratio in the outflow is suspected to vary from star to star, \citet[][see their Fig.~5]{McDonald2018} have shown it to be directly related to the total (gas) mass-loss rate above a cut-off of $K-[22]>0\fm55$. Throughout this paper, we will refer to the mass-loss rates of the stars, but keep in mind that the IR colours are only indicators of the \emph{dust} mass-loss rate. We collected photometric data to calculate colour indices such as $K-[22]$ and $[3.4]-[22]$. The $K$-band magnitude was taken from the 2MASS catalogue \citep{Skrutskie2006} or, when available, from the compilation of average magnitudes in \citet{Uttenthaler2019}. The $[3.4]$ and $[22]$ magnitudes were measured by the Wide-field Infrared Survey Explorer ({\it WISE}) space observatory \citep{Wright2010} and are taken from the AllWISE \citep{Cutri2014} or, in case of saturation, from the unWISE catalogue \citep{Schlafly2019}.

Furthermore, information about technetium (Tc) and the carbon isotopic ratio was collected for inspecting the 3DUP activity of our sample stars. Tc data were adopted from \citet{Little1987}, \citet{Uttenthaler2019}, and Uttenthaler et al.\ (in preparation). Literature sources for the $^{12}$C/$^{13}$C comprised of \citet{Lambert1986}, \citet{Ohnaka1996}, \citet{Abia1997}, and \citet{Greaves1997} for the C-type stars; \citet{Dominy1987}, \citet{Schoier2000}, and \citet{Lebzelter2019} for S-type stars; and \citet{Ramstedt2014} and \citet{Hinkle2016} for M-type stars.

Eventually, IR colour indices are available for all stars, Tc observations for 154 stars, and $^{12}$C/$^{13}$C ratios for 47 stars. All data are listed in Table~\ref{Tab:All data} in the Appendix.

\section{Period changes and light curve shapes}\label{sec:Period changes}

\subsection{The stars with the strongest period changes}\label{sec:strong-changers}

We derived the relative period changes $\Delta P/\left<P\right>$ for the 548 sample stars. Here, $\Delta P$ is the full amplitude of the period change, and $\left<P\right>$ is the mean period obtained by our Fourier analysis of the complete time series available. The quantity $\Delta P/\left<P\right>$ should be better suited to identify MPC Miras than a linear fit to the period evolution as used, e.g., by \citet{Templeton2005}. We verified that $\Delta P/\left<P\right>$ is essentially independent of the length of the available time series. However, we cannot exclude that the relative period change might be underestimated for the shortest time series in the sample ($\lesssim20\,000$\,d, equivalent to $\sim55$\,yrs). Applying a limit of variation of $\Delta P/\left<P\right> > 5\%$ \citep[roughly equivalent to the $2-3\,\sigma$ significance level of][]{Templeton2005}, we find a total of 27 Mira variables that surpass this limit; their data are shown in Table~\ref{Tab:period-changes}. Column~4 of this table contains the period change over the analysed time. The choice of the 5\% limit will be justified below in Sect.~\ref{sec:K-22_vs_DP/P}.

\begin{table}
\caption{Mira-type stars with significant period changes. The stars are grouped in MPC, CPC, and SPC classes.}
\label{Tab:period-changes}
\centering
\begin{tabular}[b]{lclrr}
\hline \\ 
GCVS & $\left<P\right>$ & Spectral & $\Delta P/\left<P\right>$ & $K-[22]$\\
     &   [d] & type     &             \% & [mag]   \\\\
\hline \\ 
\textbf{MPC}\\
RU Tau & 588.6 & M3.5e-M6.5      & 9.5 & 1.685\\
S Ori  & 409.2 & M6.5e-M9.5e     & 8.5 & 1.733\\
RS Aql & 417.9 & M5e-M8          & 7.2 & 2.531\\
U CMi  & 411.6 & M4e             & 7.0 & 2.018\\
T Cep  & 390.2 & M5-M9IIIe       & 7.0 & 1.246\\
Z Vel  & 411.4 & M9e             & 7.0 & 2.327\\
T Hya  & 285.7 & M3e-M9:e        & 6.7 & 1.351\\
RU Sco & 370.5 & M4/6e-M7II-IIIe & 6.7 & 2.142\\
T CMi  & 316.2 & M4Se-M8         & 6.5 & 1.867\\
AF Car & 446.5 & M8e             & 6.3 & 1.889\\
S Sex  & 259.0 & M2e-M5e         & 6.0 & 1.704\\
T Ser  & 340.0 & M7e             & 5.9 & 1.567\\
W Lac  & 320.2 & M7e-M8e         & 5.8 & 1.955\\
SU Her & 341.6 & M6e             & 5.7 & 1.963\\
SS Peg & 416.3 & M6e-M7e         & 5.7 & 1.962\\
S Her  & 305.5 & M4.Se-M7.5,Se   & 5.2 & 1.032\\
TY Cyg & 357.0 & M6e-M8e         & 5.1 & 1.868\\
Z Sco  & 345.3 & M5.5e:-M7e      & 5.0 & 1.633\\\\

\textbf{CPC}\\
R Aql & 280.7 & M5e-M9IIIe   & 16.7 & 2.270\\
R Hya & 388.0 & M6e-M9eS     & 15.4 & 0.839\\
W Dra & 279.3 & M3e-M4e      & 14.9 & 3.033\\
Z Tau & 458.5 & S7.5,1e(M7e) & 11.8 & 2.633\\
T Lyn & 409.0 & C5.2e-C7.1e  &  6.4 & 2.148\\\\

\textbf{SPC}\\
T UMi  & 317.3 & M4e-M6e    & 38.2 & 1.212\\
LX Cyg & 477.0 & C(N)	    & 22.6 & 1.602\\ 
R Cen  & 571.5 & M4e-M9.5   & 13.9 & 2.163\\
BH Cru & 528.7 & C(N)       & 9.5  & 1.573\\ 
\hline
\end{tabular}
\tablefoot{Column 1: denomination in GCVS; Column 2: mean period obtained by Fourier analysis of the time series data; 3: spectral type in GCVS; Column 4: relative period change; Column 5: $K-[22]$ colour.}
\end{table}

\begin{figure}[!t]
\includegraphics[scale=0.5]{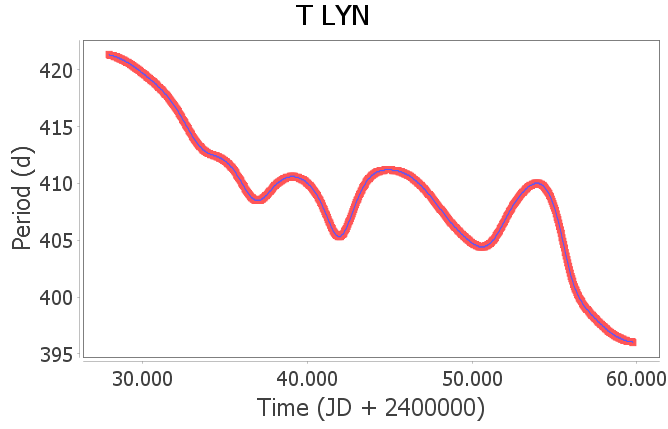}
\caption{Period evolution of the carbon Mira T~Lyn. A strong period decrease has commenced about 15 years ago, continuing the trend seen at the beginning of the available time series.}
\label{Fig:TLyn}
\end{figure}

Five Miras are classified as CPC, namely R~Aql, W~Dra, R~Hya, T~Lyn, and Z~Tau, and four as SPCs, namely R~Cen, BH~Cru, LX~Cyg, and T~UMi. These nine stars account for $\sim$ 1.6\% of the total, which agrees well with the fraction of stars with highly significant period change obtained by \citet{Templeton2005}. The only difference between their study and ours lies in the carbon Mira T~Lyn, for which we found a period change of 6.4\%, which happened relatively continuously. As we can see in Fig.~\ref{Fig:TLyn}, half of the period change in T~Lyn occurred in the last 15 years or so, which probably is why it did not appear among the most significant period changes in the survey of \citet{Templeton2005} almost 20 years ago. We thus propose T~Lyn as a new CPC candidate and recommend monitoring its period evolution.

We also highlight the case of BH~Cru: Prior to 1999, it experienced an {\it increase} of about 25\% relative to the average period given in the GCVS (1970). On the other hand, we detected a period {\it decrease} of 9.5\% since 1999. It has been speculated that the period increase in BH~Cru (and LX~Cyg) is the result of a recent 3DUP event in the aftermath of a TP that increased its atmospheric C/O ratio from $\sim1$ to $>1$ \citep{Whitelock1999,Uttenthaler2016b}. The increased C/O probably provides feedback to the pulsation mechanism because of higher atmospheric opacity. Possibly, the maximum period reached after this event is not long-term stable. BH~Cru is an exciting object to study in-depth, and we recommend monitoring its period evolution.

The group of 18 MPC Miras in Table~\ref{Tab:period-changes} is exclusively made up of M-type stars. Based on the distribution of the spectral types in the sample (472 M-type, 43 S-type, 33 C-type), basic combinatorial considerations yield a probability of $\sim6.6$\% to draw only M-type stars when randomly selecting 18 from the sample. Though it cannot be excluded that this is a chance result, it seems plausible that significant MPCs are much more common among M-type Miras than among the other spectral types. Figure~\ref{Fig:A1} in the Appendix presents the result of the period evolution analysis carried out with the VStar program for these 18 MCP candidate stars. It shows the period vs.\ time and period vs.\ time vs.\ WWZ 2D contour plots for all of them. We can read off from these diagrams that the period meanders with typical timescales of 30 to 75 years.

Interestingly, the distribution of period variations $\Delta P/\left<P\right>$ does not peak around zero, i.e., a completely stable period is not the most common property. Instead, the distribution shows a peak at $\sim2\%$; the median value is 2.4\%. This would also be the precision to which pulsation periods of Miras generally can be determined, also affecting the width of period-magnitude relations of Miras \citep{Soszynski2007}.

\subsection{Fourier analysis of Mira stars with significant MPCs}\label{sec:fourier-spectrum}

Fourier analysis of stellar light curves can reveal important information about how brightness variations occur. It is often a complex process since the Discrete Fourier Transform (DFT) can introduce additional misleading frequencies due to cycle-to-cycle variations, long-term brightness changes, gaps in the light curves, etc., in addition to the non-constant periods of MPC stars. As observed in Sect.~\ref{Frequency test}, the Fourier spectrum shape of MPC stars is dominated by a structure of peaks around the primary frequency ($f_0$) accompanied by its harmonics ($2f_0$, $3f_0$, \dots). Moreover, as the time series are not homogeneous and have gaps, seasonal aliases ($\pm1{\rm y}$, $\pm2{\rm y}$, \dots) appear. In a few cases, the time series are almost uninterrupted, with only one seasonal alias appearing. Finally, due to a physiological effect called the Cerasky effect, the visual observations may additionally have a spurious period of about one year ($f_{\rm y}$) \citep[see for these topics, e.g.,][]{Kiss1999, Templeton2005, Percy2015}.

\begin{figure}[!t]
\includegraphics[scale=0.6]{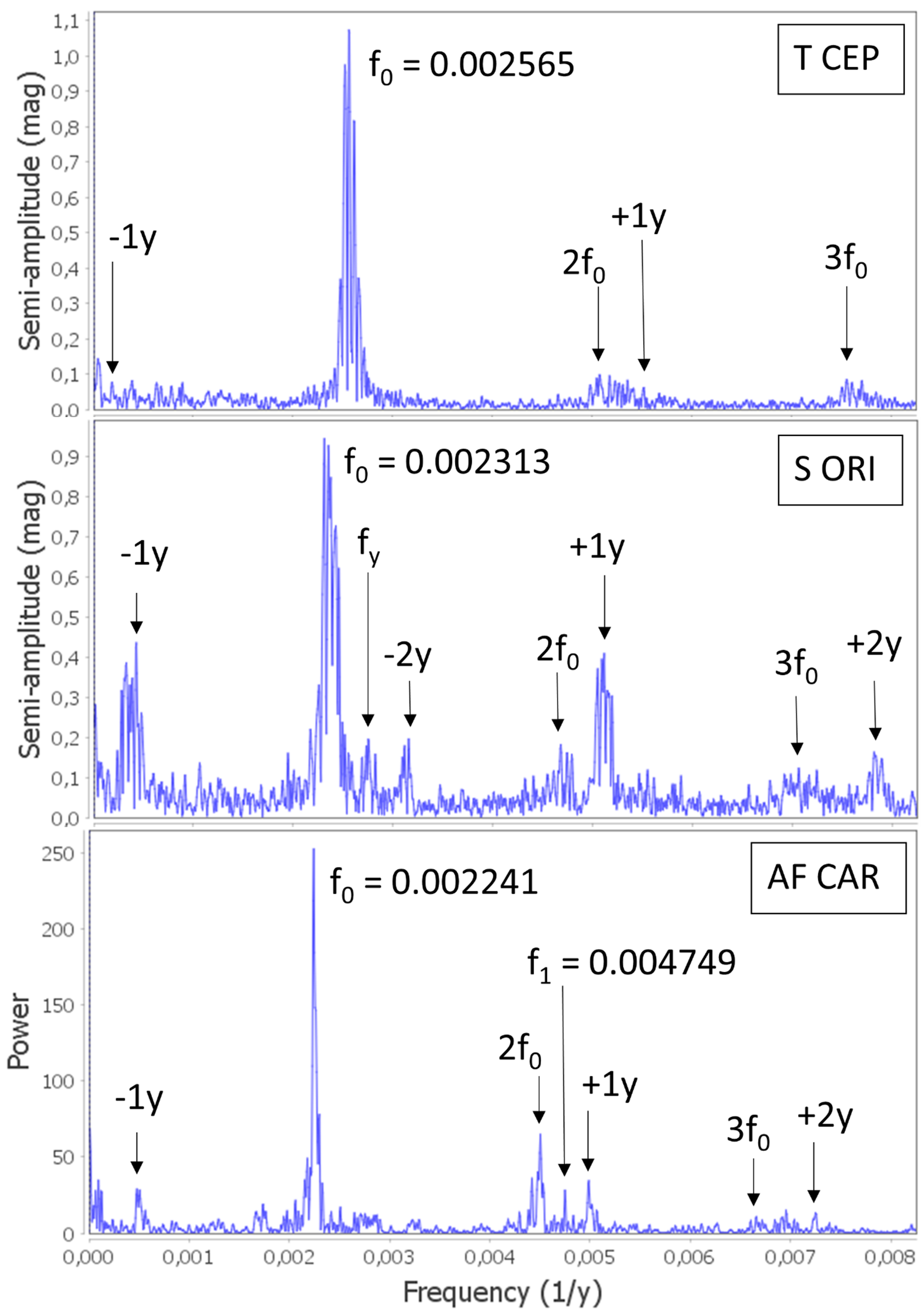}
\caption{Fourier spectra and identification of frequency peaks of three Miras with significant MPCs. Besides harmonics of the main frequency and seasonal aliases, no obvious additional peaks appear in the spectra.}
\label{Fig:FOURIER 1}
\end{figure}

We calculated the Fourier spectra of the 18 MPC stars using the Fourier routine DC-DFT in the AAVSO software package VStar. Then we analysed the structure of the frequency peaks in a semi-amplitude vs.\ frequency diagram. The data were merged, separated into bins, and averaged according to the process described in Sect.~\ref{sec:binning}. In cases where the number of data points is low and gaps are very frequent, the level of background noise makes it difficult to identify peaks, so power vs.\ frequency plots were used to highlight the dominant frequencies. Figure~\ref{Fig:FOURIER 1} shows the Fourier spectra and the identification of obvious peaks for three stars with different characteristics in the collected data. The top panel shows T~Cep, with the highest number of data and an almost uninterrupted time series, so only one seasonal aliasing is minimally noticeable in addition to the primary frequency and its first two harmonics. The middle panel shows the Fourier spectrum of S~Ori, a star with a much smaller number of data and seasonal gaps, which displays a more complex structure of seasonal aliases in addition to its first harmonics. The appearance of the seasonal frequency $f_{\rm y}$ can also be seen. The lower panel of Fig.~\ref{Fig:FOURIER 1} shows the spectrum of AF~Car, the star with the fewest and most scattered data, so we had to resort to the power vs.\ frequency plot with a smoother noise background. Nevertheless, it can be seen that, besides only a faint unidentified peak $f_1$ with a power of $\sim10$\% that of $f_0$, no obvious peaks appear in AF~Car's spectrum beyond those discussed above.

Figure~\ref{Fig:A1} in the Appendix shows the same analysis for the remaining 15 stars, and we can see that none of these Fourier spectra shows prominent additional peaks. Along with AF~Car mentioned above, there are only three other stars (RU~Sco, RS~Aql, and SS~Peg) with some unidentified frequency peaks, marked as $f_1$, but with low amplitude.

\subsection{Light curve shapes}\label{sec:LC-shapes}

We are interested in the light curves of our sample stars because we found that 16 out of the 18 stars with significant MPCs (Table~\ref{Tab:period-changes}) have apparent bumps on the ascending branch of most light cycles. For the remaining two stars, W~Lac and Z~Sco, we cannot state with certainty the presence of bumps because of the relatively low number of observations, which, in addition, are not well distributed in the ascending branch of their cycles. In W~Lac, we found slight bumps in only some of its cycles, while in Z~Sco, slight asymmetries are more or less repetitive in the upper part of its light curve but no clear bumps. The ASAS-SN light curves and phase diagrams were visually inspected, and the stars were qualitatively divided into two large groups:

{\bf Group A (asymmetric):} Stars with apparent asymmetries in their light curves, such as bumps on the rising branch, double maxima, or abrupt changes in the slope of the rising branch. Given the similarity of all these anomalies, they may correspond to a larger or smaller-scale expression of similar phenomena in the stars' atmospheres and represent a relatively homogeneous group. We found a total of 203 Mira-type variable stars with these characteristics. This fraction of $203/548 \approx 37$\% is slightly higher but in fair agreement with the previous studies of \citet[][20\%]{Vardya1988} and \citet[][30\%]{Lebzelter2011}. Group~A would include the stars \citet{Ludendorff1928} classified as $\gamma$-type stars.

{\bf Group S (symmetric):} This group includes the remainder of 336 stars without anomalies in the rising branch of their light curves. This also includes doubtful cases in which the possible observed anomalies occur only in a few cycles or are so weak that we cannot state with certainty their existence. Although this aspect may introduce some uncertainty in our results, the high number of stars in our selection and the use of up to four different sources to inspect the light curves minimises these uncertainties. This group would correspond to the $\alpha$- and $\beta$-type stars of the Ludendorff classification.

\begin{figure*}[!t]
\includegraphics[scale=1.03]{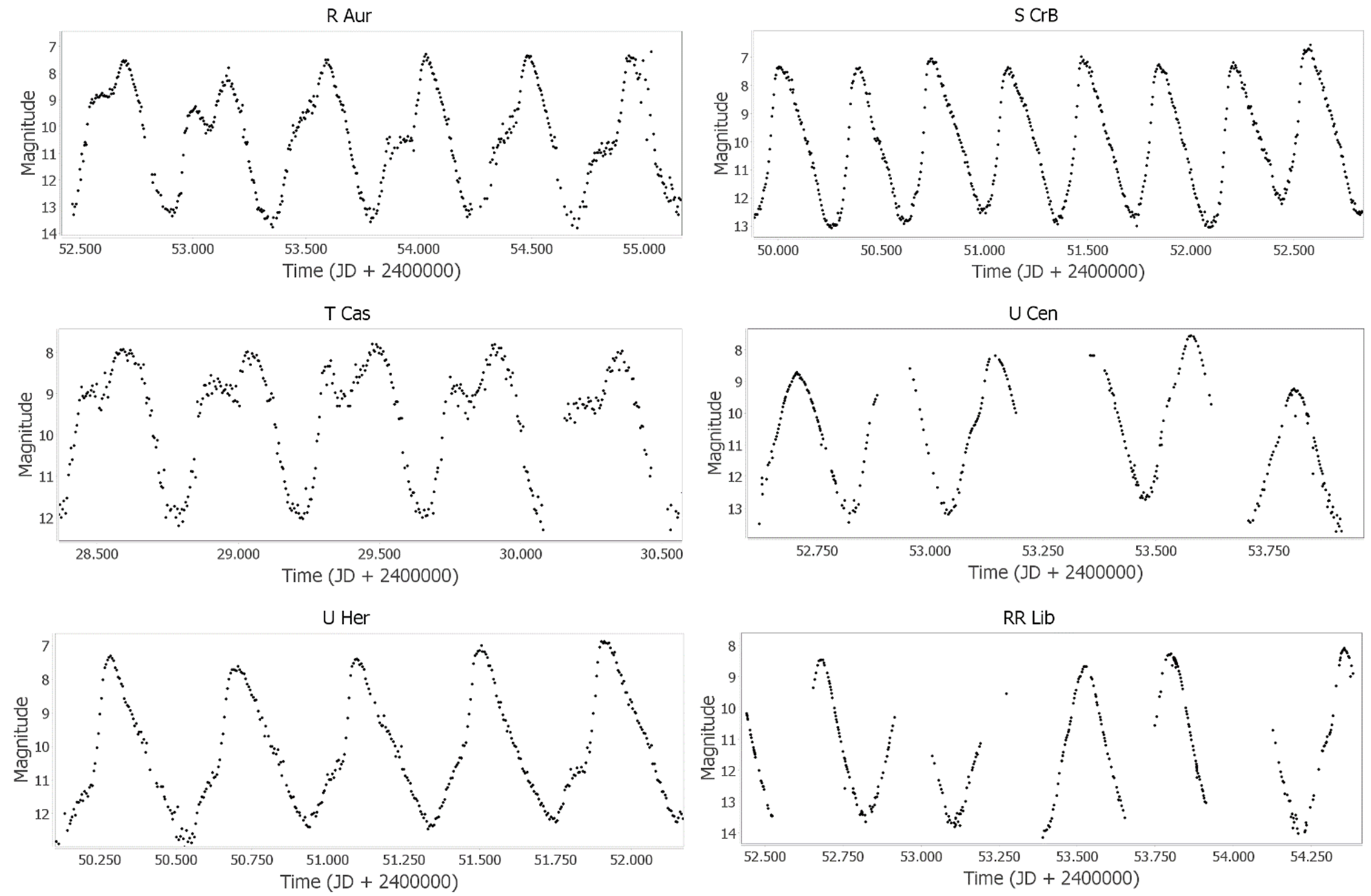}
\centering
\caption{{\it Left column:} Example light curves of Group~A stars. From top to bottom: R~Aur (bumps in the ascending branch), T~Cas (double-peaked maximum), and U~Her (abrupt change in the slope of the ascending branch). {\it Right column:} Example light curves of Group~S stars. From top to bottom: S~CrB (quasi-sinusoidal light curve), U~Cen (slight asymmetries only in some of its cycles), and RR~Lib (gaps in the light curve prevent stating with certainty the presence of asymmetries).}
\label{Fig:MPC-BUMPS}
\end{figure*}

Given this procedure to classify stars in Group~A only if they have evident anomalies, we expect Group~A to be relatively pure. In contrast, Group~S might contain a few stars that would be classified in Group~A had better data been available. In Fig.~\ref{Fig:MPC-BUMPS}, we show example light curves of Miras in Group~A ({\it left columns}) and light curves of Miras in Group~S ({\it right columns}). Figure~\ref{Fig:data_proc} shows a portion of the light curve of T~Cep, another example of a Mira in Group~A. Table~\ref{Tab:All data} in the Appendix collects the classification w.r.t.\ light curve asymmetries.

\subsection{3DUP indicators}\label{sec:3DUP-indicators}

TPs and 3DUP events have been proposed as the basis of the various period change classes. If this is the case, one would expect these stars to show the 3DUP indicator Tc in their spectra and have increased $^{12}{\rm C}/^{13}{\rm C}$ ratios. Table~\ref{Tab:Tc content} shows the data of the stars with significant period changes from Table~\ref{Tab:period-changes} for which information about Tc is available. We found information for only five of the 18 stars with significant MPCs in our sample, in four of which the presence of Tc has been confirmed, or Tc is likely present. Be reminded that the absence of Tc does not necessarily mean the absence of TPs because they may have yet to be powerful enough to drive 3DUP in their aftermath. Nevertheless, it seems likely that many of them, indeed, have undergone 3DUP events. Further observations are required to make a stronger statement.

\begin{table}
\caption{Tc information for stars with significant period changes.}
\label{Tab:Tc content}
\centering
\begin{tabular}[b]{r c | r c | r c}
\hline \\ 
GCVS & Tc & GCVS & Tc & GCVS & Tc\\\\
\hline \\ 
\textbf{MPC} &          & \textbf{CPC} &         & \textbf{SPC}     \\
S Ori        & yes$^1$  & R Aql        & no$^1$  & T UMi  & no$^1$  \\
U CMi        & no$^2$   & R Hya        & yes$^1$ & LX Cyg & yes$^1$ \\
T Cep        & yes$^1$  & W Dra        & no$^1$  & R Cen  & no$^1$  \\
T Hya        & prob$^2$ &              &         & BH Cru & yes$^1$ \\
S Her        & yes$^1$                                              \\
\hline
\end{tabular}
\tablefoot{(1): \citet{Uttenthaler2019}; (2): \citet{Little1987}}
\end{table}

The distribution of the 203 Group~A stars in terms of their spectral types has an interesting pattern. Stars with bumps in their light curves are much more common among S- and C-type than among M-type stars. Specifically, the fraction is $32/40\approx80$\% for the S-type, $24/32\approx75$\% for the C-type, and only $147/467\approx31$\% for the M-type Miras. This is in agreement with the result of \citet{Lebzelter2011} that S and C stars generally show a higher fraction of non-sinusoidal variation than the M-type stars. The higher C/O ratio of more advanced evolutionary stages might favour the presence of asymmetries in the light curves.

\begin{figure}[!t]
\includegraphics[width=\linewidth]{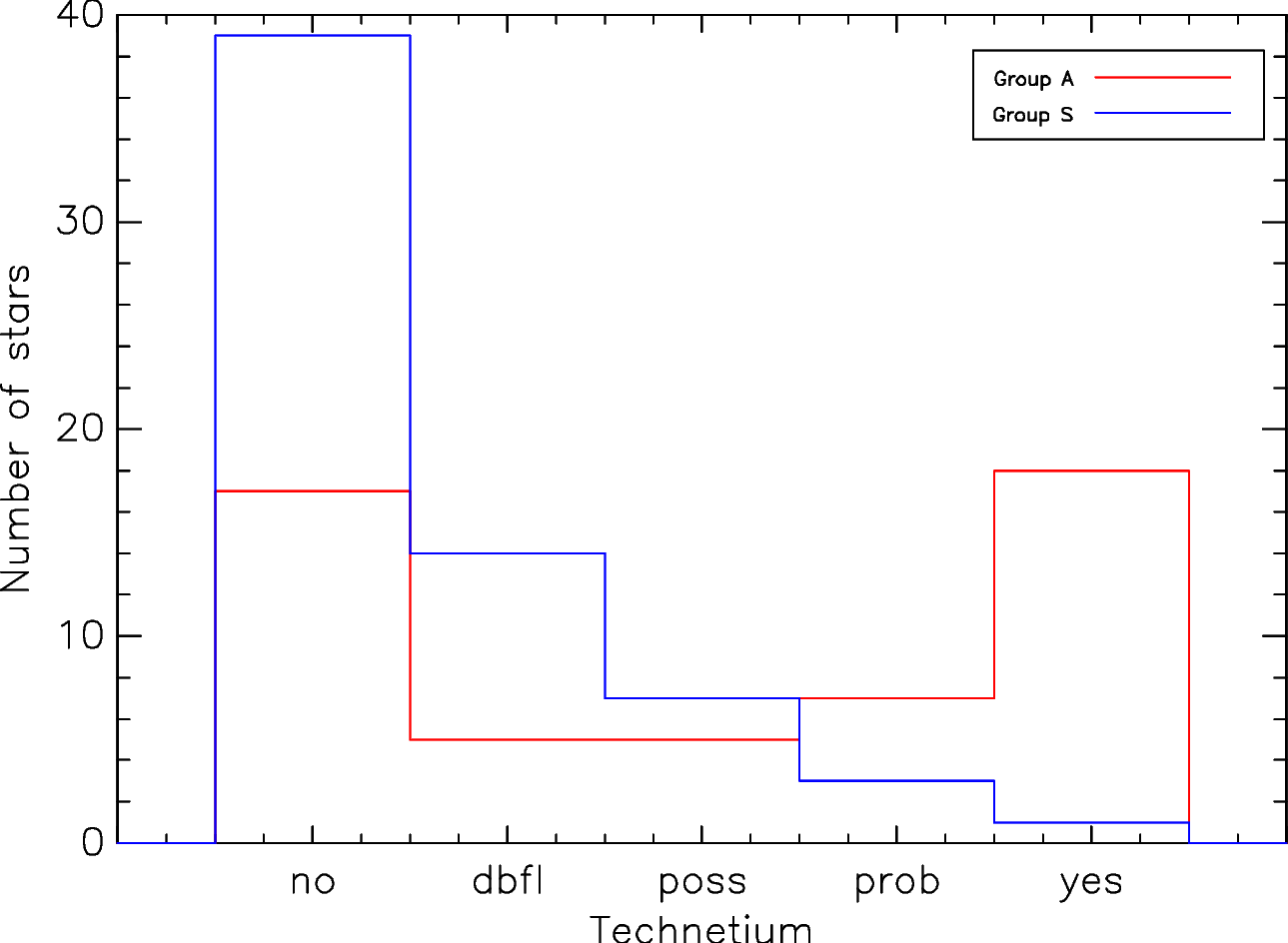}
\caption{Distributions of M-type stars in Groups~A and S with respect to the presence of Tc. The classifications "doubtful" (dbfl), "possibly" (poss), and "probably" (prob) are used only in \citet{Little1987}, whereas "no" and "yes" are used by both \citet{Little1987} and \citet{Uttenthaler2019}. Group~S has a strong tendency to Tc-poor stars.}
\label{Fig:Bumps_and_3DUP}
\end{figure}

This suggests that there could also be a relation between light curve shape and 3DUP activity. All S- and C-type sample stars observed for Tc are confirmed to be Tc-rich, and many also have bumps. The histogram of how the M-type stars in the two groups distribute with respect to the presence of Tc in their spectra is displayed in Fig.~\ref{Fig:Bumps_and_3DUP}. The classifications "doubtful" (dbfl), "possibly" (poss), and "probably" (prob) are used only in \citet{Little1987}, which we retain here. Restricting the selection to the unambiguous "yes" and "no" classifications of \citet{Uttenthaler2019}, we find 17 Tc-poor and 18 Tc-rich M-type Miras in Group~A. On the other hand, there are 37 Tc-poor M-type stars in Group~S, but only one Tc-rich star. Thus, while Group~A contains many Tc-rich stars of all spectral types, Group~S is almost devoid of Tc-rich ones. We thus conclude that, given the prior condition that a Mira of any spectral type is Tc-rich, the probability that it also has bumps in its light curve is of the order of $(18+32+24)/(19+40+32)=0.813$. On the other hand, given the prior condition that an M-type Mira has no bumps, the probability that it is Tc-poor is almost 100\%.

We can also check on the population level if the fractions of Miras with light curve asymmetries and with the 3DUP indicator Tc agree. If these things are connected somehow, the respective fractions should be similar. We better avoid selection biases that could plague especially the Tc observations collected from many literature sources and therefore focus on the M-type Miras. M-type Group~A stars in the complete sample make up a fraction of $147/467=0.315$. On the other hand, M/MS-type, Galactic disc Miras with Tc in the sample of \citet{Uttenthaler2019} make up a fraction of $28/105=0.267$. Given a base fraction of 0.315, the probability of drawing 28 or fewer Tc-rich stars out of 105 M-type stars is 17\%. Thus, we cannot reject with high confidence the notion that the two fractions are identical.

Besides Tc, the carbon isotopic ratio $^{12}$C/$^{13}$C is an indicator of the 3DUP activity of an AGB star. When the first TP takes place on the AGB, the $^{12}$C/$^{13}$C ratio is expected to be of the order $10\lesssim\,^{12}{\rm C}/^{13}{\rm C}\lesssim25$, depending on the initial mass \citep{Hinkle2016}. The lowest values, $10\lesssim\,^{12}{\rm C}/^{13}{\rm C}\lesssim20$, would be found in stars with masses $\lesssim2M_{\sun}$, while the highest values, $20\lesssim\,^{12}{\rm C}/^{13}{\rm C}\lesssim25$, would be found in the most massive stars. For stars undergoing 3DUP episodes, the carbon isotopic ratio will increase as new $^{12}$C synthesised in the He-burning layer is brought to the surface. We may thus identify M-type Miras that underwent 3DUP events by their carbon isotopic ratio elevated to values $^{12}{\rm C}/^{13}{\rm C}\gtrsim25$.

Among the 47 sample stars for which we found $^{12}$C/$^{13}$C isotope ratios in the literature (Sect.~\ref{sec:add_data}), 23 are M-type, 10 are S-type, and 14 are C-type. The $^{12}$C/$^{13}$C is known for two M-type stars with significant MPCs in our sample, S~Ori and T~Cep. The values are reported to be $45\pm10$ and $33\pm10$, in agreement with the fact that both are also Tc-rich. Even the next two stars with non-significant MPCs but with relatively high $\Delta P/\left<P\right>$ values with known isotopic ratios, RU~Her and R~Aur with $\Delta P/\left<P\right>$ of 4.5 and 4.4, have $^{12}$C/$^{13}$C values of 25 and 33 respectively. This agrees with the fact that they are also reported to contain Tc.

The few carbon isotopic ratios available in the literature support the findings from Tc concerning the light curve shapes. The $^{12}{\rm C}/^{13}{\rm C}$ ratio is reported for four stars from Group~S: Two classified for Tc as "doubtful" and "no" have ratios 12 and 16, and two Tc-rich ones have ratios 22 and 26. Also, the carbon isotopic ratios of Group~A stars agree with their Tc classifications: The mean $^{12}{\rm C}/^{13}{\rm C}$ of the nine Tc-poor stars is 13 (range $8-19$), and that of the seven Tc-rich is 28 (range $14-45$). Among the 29 M-type Miras we have in common with \citet{Hinkle2016}, nine have $^{12}{\rm C}/^{13}{\rm C}\geq25$ (taking into account the measurement uncertainty), indicating the possibility of having undergone $^{12}$C enrichment by 3DUP events. Indeed, the literature confirms that most of them also have Tc in their atmosphere, or at least Tc is possibly or probably present. Interestingly, all nine stars are in our Group~A of stars with bumps in the ascending branch of their light curves. The stars' data have been extracted from Table~\ref{Tab:All data} of the Appendix and are shown separately in Table~\ref{Tab:Tc}. This underlines the possibility of a connection between bumps in the light curve and the occurrence of 3DUP events.

\begin{table}
\caption{Data of M-type Miras with $^{12}{\rm C}/^{13}{\rm C}\gtrsim25$.}
\label{Tab:Tc}
\centering
\begin{tabular}[t]{lcccc}
\hline \\ 
GCVS & Bumps & Tc & $^{12}$C/$^{13}$C\\\\
\hline \\ 
R Aur  & yes & yes$^3$  & 33$\pm$13\\
TX Cam & yes & ---      & 21$\pm$6\\
T Cas  & yes & yes$^3$  & 33$\pm$5\\
T Cep  & yes & yes$^1$  & 33$\pm$10\\
U Her  & yes & no$^3$   & 19$\pm$8\\
RU Her & yes & yes$^1$  & 25$\pm$5\\
R Hya  & yes & yes$^1$  & 26$\pm$4\\
S Ori  & yes & yes$^1$  & 45$\pm$13\\
U Ori  & yes & poss$^2$ & 25$\pm$10\\
\hline
\end{tabular}
\tablefoot{Literature sources of Tc content: (1) \citet{Uttenthaler2019}; (2) \citet{Little1987}; (3) Uttenthaler et al., in preparation.}
\end{table}

\section{Mass-loss properties}\label{sec:ml-props}

\subsection{Mass loss as a function of relative period change}\label{sec:K-22_vs_DP/P}

\begin{figure}[!t]
\includegraphics[scale=0.67]{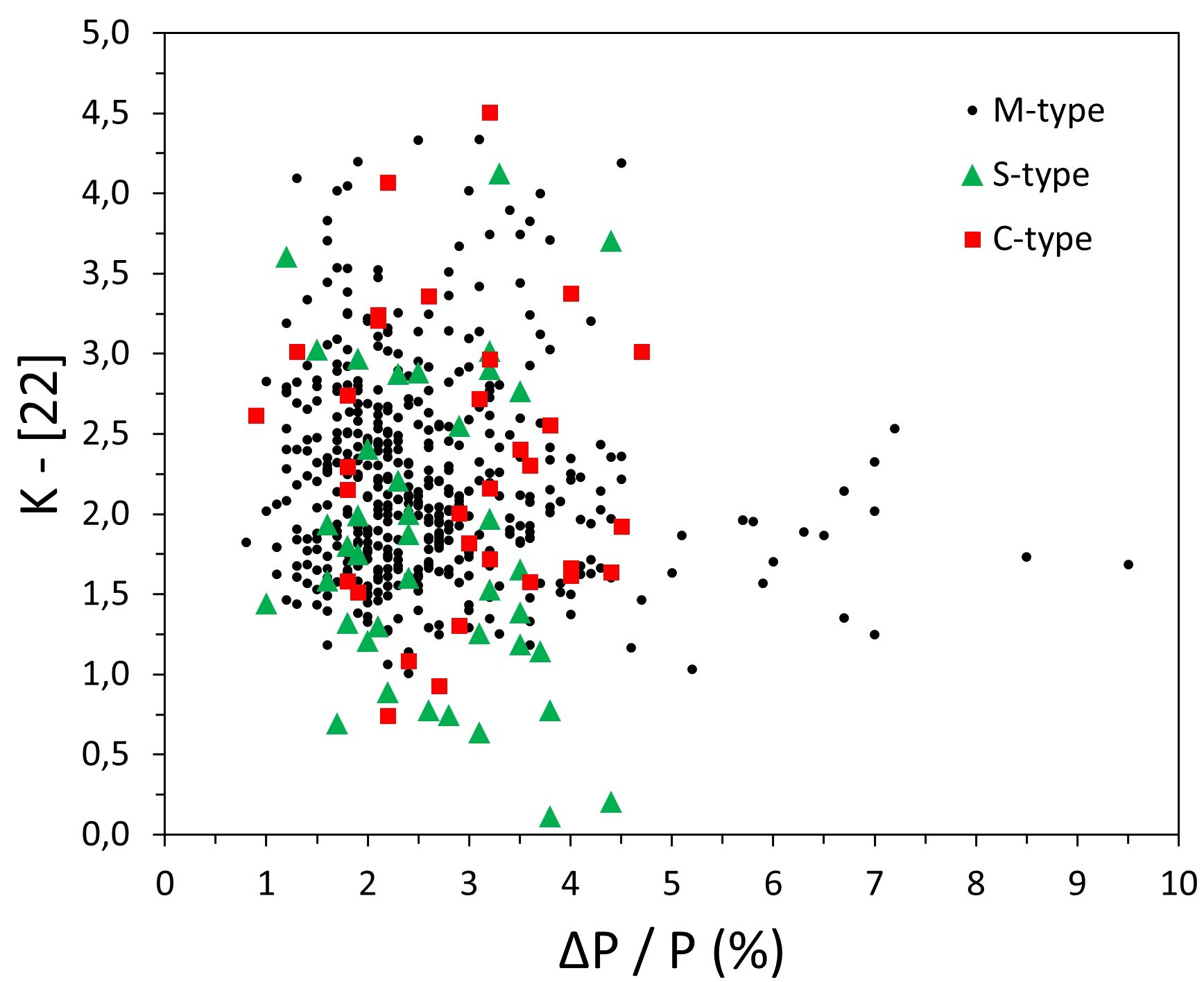}
\caption{$K-[22]$ vs.\ $\Delta P/\left<P\right>$ diagram of our sample stars, excluding the SPC and CPC stars identified in Sect.~\ref{sec:strong-changers}. The chemical spectral types are distinguished by the symbols, see the legend. A few S stars have very low $K-[22]$ colours (dust mass-loss rate), and the "tail" of more strongly varying periods ($>5\%$) is exclusively made up of M stars.}
\label{Fig:MPC - spectral type}
\end{figure}

To connect the period variations presented above to colour indices and mass-loss considerations, it is first insightful to plot the $K-[22]$ dust mass-loss indicator as a function of relative period change $\Delta P/\left<P\right>$. All sample stars except the CPC and SPC Miras are plotted in this diagram in Fig.~\ref{Fig:MPC - spectral type}. Up to a level of $\Delta P/\left<P\right>\sim5\%$, the stars show a considerable spread in $K-[22]$ independent of spectral type, except for the S-type stars, which have several specimens with lower colour index than the rest. Furthermore, we can identify a "tail" of stars with more strongly varying periods. This group has two remarkable properties: (i) its colour indices are confined to the narrow range of $1.0 \lesssim K-[22]\lesssim 2.5$, and (ii) implying a limit of $\Delta P/\left<P\right>\geq5\%$, it is exclusively made up of M-type stars; these are the 18 MPC stars listed in Table~\ref{Tab:period-changes}. This feature led us to define the limit of "significant period change" used throughout this paper. It suggests that the limit of $\Delta P/\left<P\right>\geq5\%$ is suitable to distinguish between random period variations known to occur in Mira stars \citep{Templeton2005} and actual period changes. Stars that change period more than this could form a homogeneous group of MPC stars.

\subsection{Mass loss from stars with significant period change}\label{sec:mass-loss}

\begin{figure}[!t]
\includegraphics[scale=0.72]{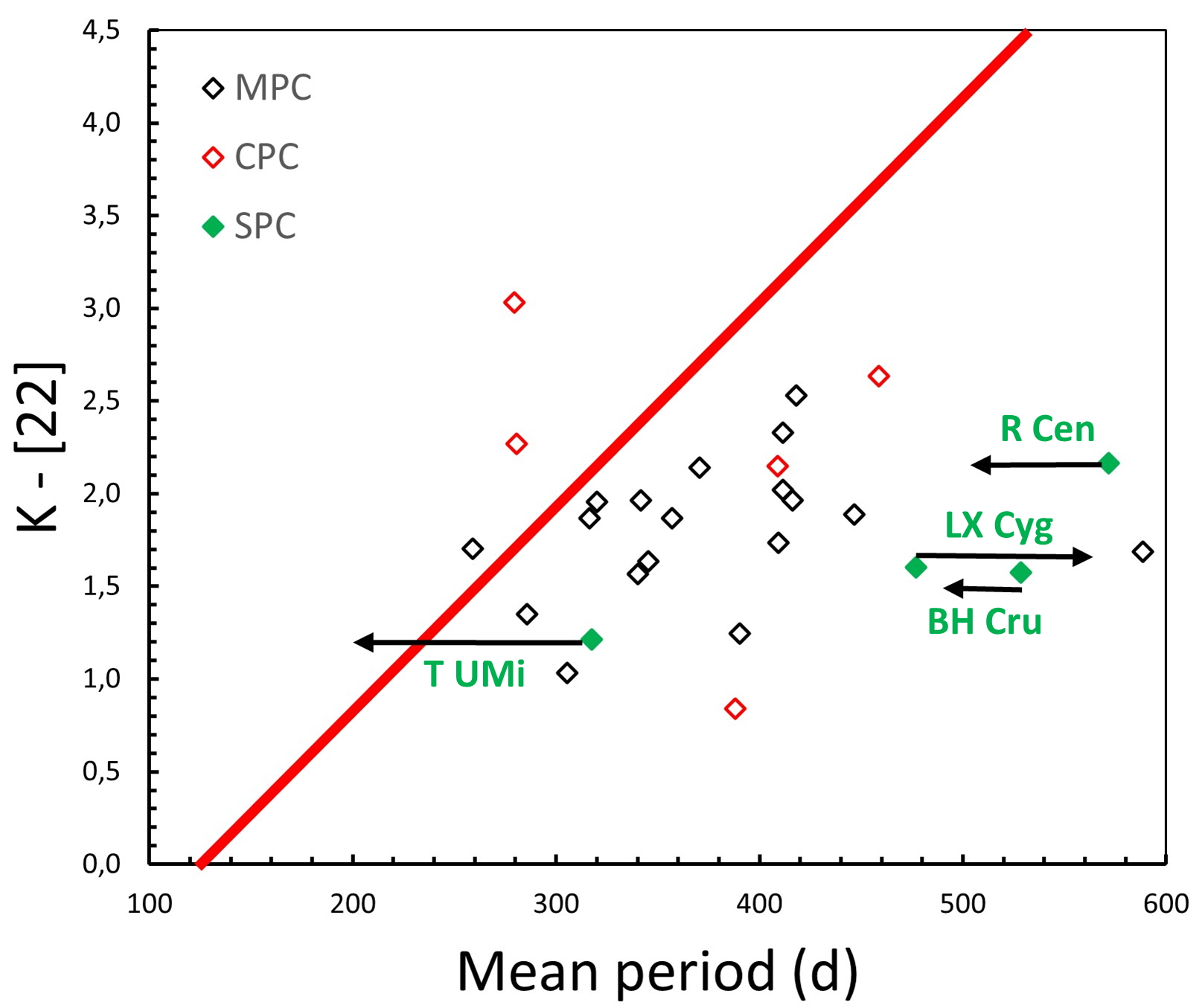}
\caption{$K-[22]$ vs.\ $P$ diagram of the 27 stars in Table~\ref{Tab:period-changes}, see the legend for the identification of the three types of period change. The red solid line is the relation that best separates Tc-poor from Tc-rich Miras as per Equ.~\ref{Eq1}. The arrows attached to the symbols of the SPC stars indicate their recent period evolution.}
\label{Fig:K-22_significant}
\end{figure}

We inspect the $K-[22]$ colour of the 27 stars with significant period changes from Table~\ref{Tab:period-changes} as a function of the mean period in Fig.~\ref{Fig:K-22_significant}. The stars do not appear to have particularly strong dust production since Miras can easily have $K-[22]$ colours above 4\fm0 (Fig.~\ref{Fig:MPC - spectral type}). \citet{Uttenthaler2019} used this colour to study the mass loss from Miras with and without the 3DUP indicator Tc. They demonstrated that Tc-poor and Tc-rich stars occupy two distinct regions in the $K-[22]$ vs.\ $P$ dia\-gram. The optimised linear relation that best separates the two groups takes the form:
\begin{equation}\label{Eq1}
K - [22] = 0.011 \times P - 1.380.
\end{equation}

Tc-poor Miras are generally found above this line, whereas Tc-rich (post-3DUP) Miras are located below. This linear relation is included in Fig.~\ref{Fig:K-22_significant} to guide the eye. The separation's accuracy was found to be 0.87, i.e., the probability that a star is placed on the "correct" side, according to its Tc content, is 87\%. The location of a star in the $K-[22]$ diagram is, therefore, a strong indicator of the star having undergone TP events and subsequent 3DUP.

As already indicated by Fig.~\ref{Fig:MPC - spectral type}, stars with significant MPCs appear to have relatively low $K-[22]$ colours, or low dust mass-loss rates. Indeed, 17 out of 18 (94\%) MPC stars are located below this separating line, where the 3DUP indicator Tc is usually detected. Only one star, S~Sex, is located by a narrow margin above the line. This star has also been proposed as a candidate for having undergone a TP recently because of its substantial period decrease \citep{Merchan-Benitez2000}. The location of the stars in this period-colour diagram supports the notion from Sect.~\ref{sec:3DUP-indicators} that the MPC phenomenon could be related to TP and subsequent 3DUP events.

In Fig.~\ref{Fig:K-22_significant}, we can also observe how the stars classified as SPCs all fall in the Tc-rich zone. This would be expected because a recent TP or 3DUP event has been proposed to cause the sudden period change. SPC stars are characterised by period changes that can be as large as $\sim40$\% within a few decades, so their location in the $K-[22]$ vs.\ $P$ diagram will present some uncertainties depending on the mean period used. For example, looking only at its Mira phase, the period change in T~UMi amounts to 38\%. Note that this star has recently switched from the fundamental Mira pulsation mode to a higher-overtone mode \citep{Molnar2019}. In Fig.~\ref{Fig:K-22_significant}, we indicated the period evolution by arrows. (Note that the colour might have changed, too, but there are insufficient IR data to quantify this.) LX~Cyg and BH~Cru are C-type Miras, so their location in the Tc-rich zone is consistent with expectations. For the latter, the arrow in Fig.~\ref{Fig:K-22_significant} indicates the 9.5\% period decrease since 1999.

On the other hand, the location of R~Cen and T~UMi in the $K-[22]$ vs.\ $P$ diagram is not consistent with the proposed separation of Tc-poor and Tc-rich Miras. Both stars are M-type and Tc-poor but located in the diagram's Tc-rich zone. R~Cen is a candidate for being a fairly massive star undergoing HBB. The $^{22}$Ne neutron source operating in such stars produces only little Tc and 3DUP events are inefficient, so Tc is sufficiently enriched in the atmosphere only until after many TPs \citep{GarciaHernandez2013}. Reconstructing the $K-[22]$ infrared excess from historical observations, \citet{McDonald2020} showed that T~UMi has become significantly less dusty since about 1975. It may well have been located in the Tc-poor region before the sudden period decrease commenced.

In any case, keeping in mind the different spectral types and directions of period evolution, even from these few stars, it is clear that the SPC group appears to be inhomogeneous, and, in fact, the observed period changes may reflect different physical processes going on in these stars (TPs, 3DUP).

Also, the five CPC stars exhibit some diversity: three of them are located in the Tc-rich zone (R~Hya, Z~Tau, T~Lyn) and two in the Tc-poor zone (W~Dra, R~Aql) in Fig.~\ref{Fig:K-22_significant}. The three CPC stars for which their Tc content is known (Table~\ref{Tab:Tc content}) are located in the part of the $K-[22]$ vs.\ $P$ diagram expected for their Tc content. Also, the spectral types are diverse, see Table~\ref{Tab:period-changes}. We remind that the absence of Tc or $^{12}$C enrichment does not necessarily mean that TPs are absent in those stars.

\begin{figure}[!t]
\includegraphics[scale=0.72]{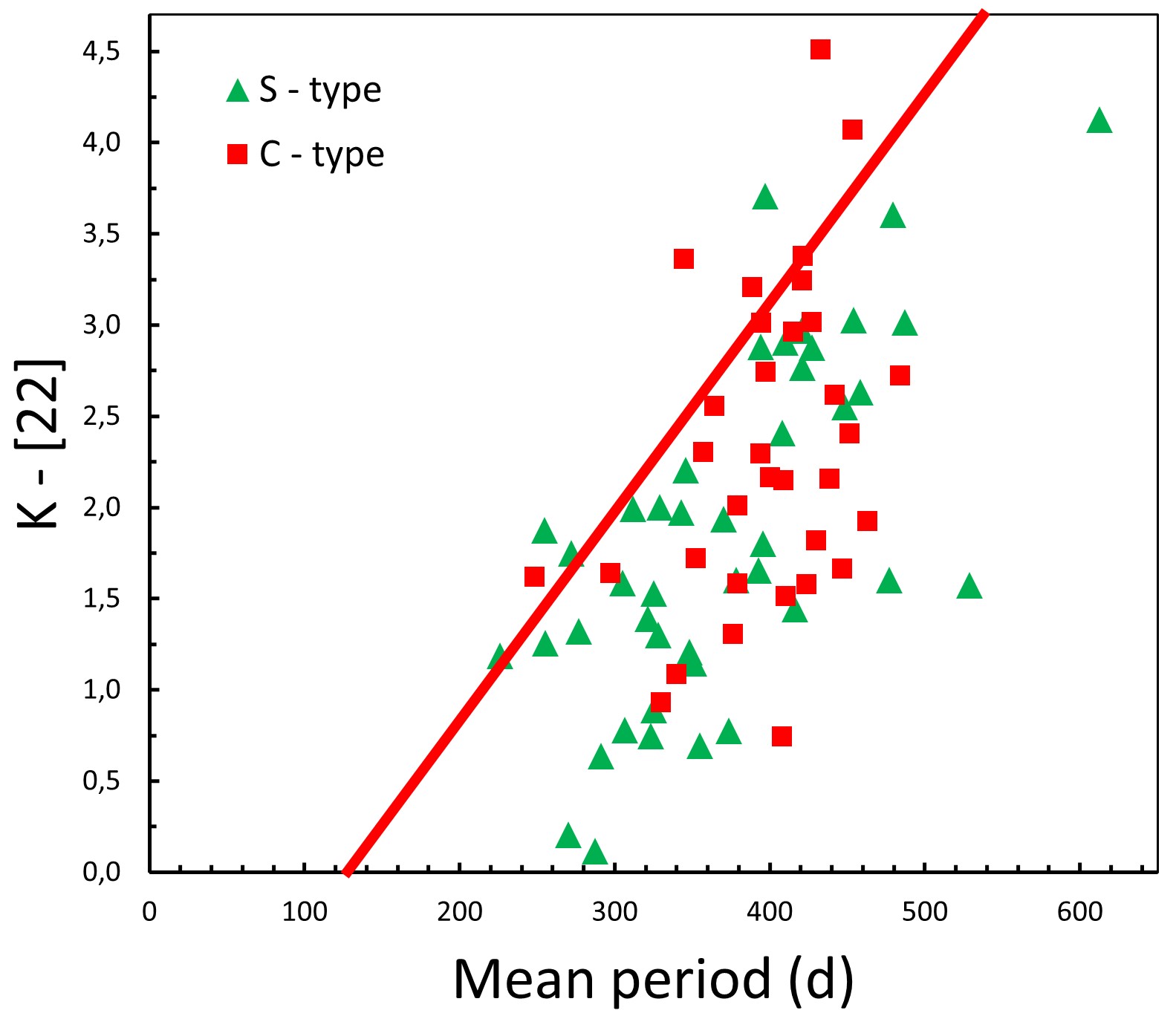}
\caption{$K-[22]$ vs.\ $P$ diagram of the S- and C-type Miras in the sample. The red solid line is the relation that best separates Tc-poor from Tc-rich Miras, according to Equ.~\ref{Eq1}. As expected, most of the stars are located in the area where Tc-rich Miras are found, i.e., below the line.}
\label{Fig:K-22 S-C type}
\end{figure}

The assignment of the Tc-rich area in the $K-[22]$ vs.\ period diagram can be reviewed by inspecting the location of the 43 S-type and 33 C-type sample Miras in such a diagram. These stars are expected to have evolved to their current stage of carbon enrichment by repeated dredge-up of carbon and $s$-process elements (e.g., Tc). Figure~\ref{Fig:K-22 S-C type} shows this diagram. The separation line defined by Equ.~\ref{Eq1} is again included in this plot. It can be seen that 39/43 (91\%) of the S-type and 27/33 (82\%) of the C-type stars are located in the zone expected for Tc-rich stars. Some of the stars above the separating line are there only by a narrow margin. We remind that Mira stars have a non-negligible pulsation amplitude also in the $K$-band and the measurements are often single-epoch observations. Thus, this selection of stars confirms the assignment of the region below Equ.~\ref{Eq1} to Tc-rich, post-3DUP Miras.

\subsection{Mass loss and light curve asymmetries}

\begin{figure}[!t]
\includegraphics[scale=0.68]{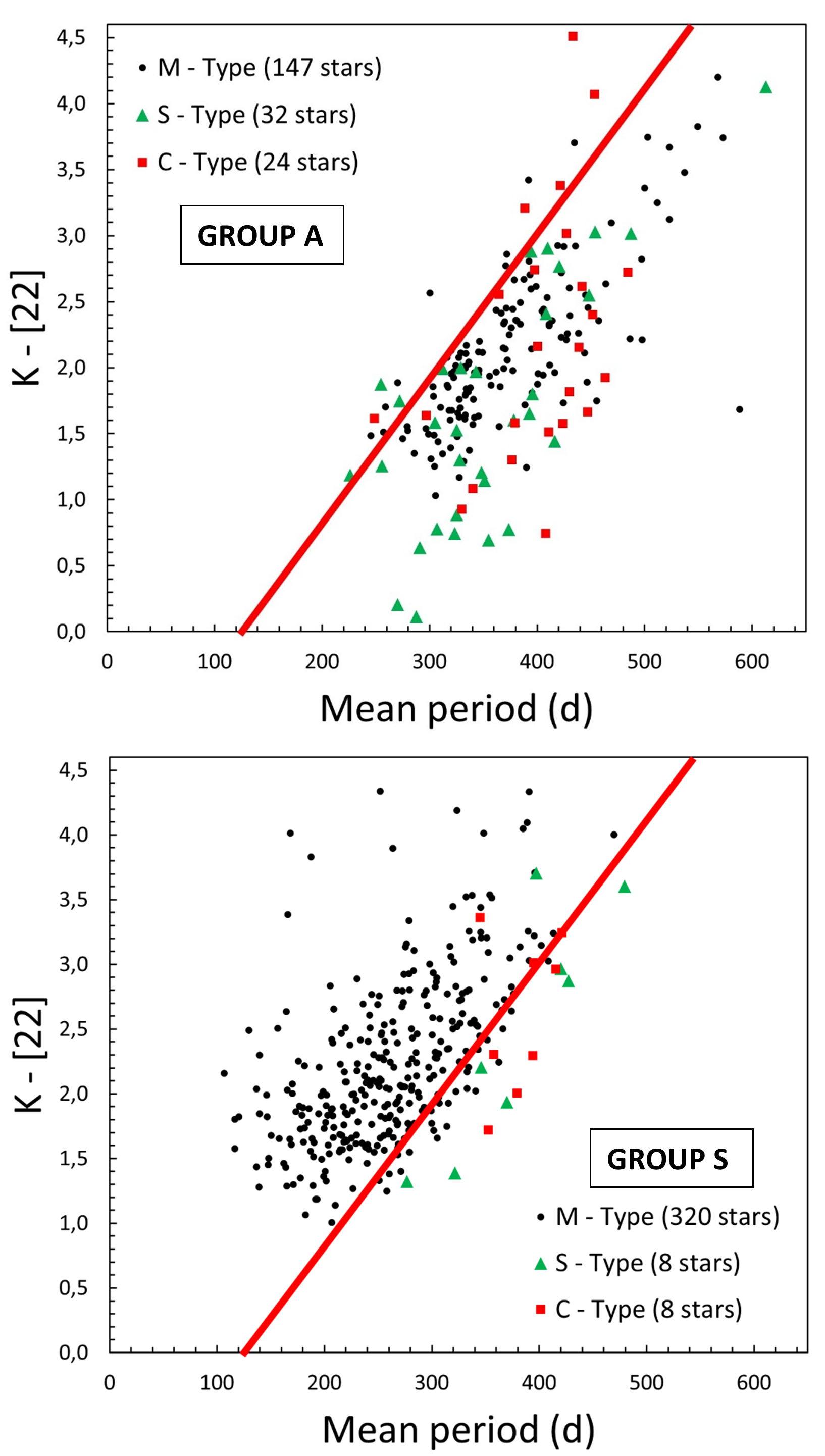}
\caption{$K-[22]$ vs.\ $P$ diagram of Miras. {\it Upper panel:} Group~A, with bumps or asymmetries in the ascending branch of their light curve. {\it Lower panel:} Group~S, with symmetric light curves. The symbols distinguish the three main spectral types, see the legend. The solid red line is the linear relation given by Equ.~\ref{Eq1}. Evidently, Miras without bumps have higher $K-[22]$ colours (dust mass-loss rates) than their siblings with bumps at similar pulsation periods.}
\label{Fig:k22-BUMPS}
\end{figure}

Separate $K-[22]$ vs.\ $P$ diagrams for Groups~A and S are shown in Fig.~\ref{Fig:k22-BUMPS}. Note the different distributions in period: While Group~A has a mean period of 378\,d and starts to be populated at $P\gtrsim240$\,d, Group~S has a mean period of 266\,d and is essentially limited to $P\lesssim400$\,d. From the results about Tc in Sect.~\ref{sec:3DUP-indicators}, it may be expected that many Group~A stars are located below the separation line defined by Equ.~\ref{Eq1}, whereas Group~S stars are above it. Indeed, 92\% of the 203 Group~A stars are located below the separation line. The few stars above the line are relatively close to it. Moreover, this fraction is almost insensitive to the spectral type: $\sim94$\% M-type, $\sim91$\% S-type, and $\sim79$\% of the C-type Miras with bumps are located below the line given by Equ.~\ref{Eq1}.

The lower panel of Fig.~\ref{Fig:k22-BUMPS} shows the $K-[22]$ vs.\ $P$ diagram of the 336 Group~S stars. Note that this group is probably not as pure as Group~A. Nevertheless, we can observe that the stars in Group~S are mainly located above the line ($\sim89$\%) and seem to "avoid" the zone below it. This percentage rises to 94\% if we consider only the M-type stars. Many of the stars below the line are S- or C-type Miras that, in turn, have $K-[22]$ colours higher than their siblings at similar periods in Group~A.

Figure~\ref{Fig:k22-BUMPS} suggests that the same line \citet{Uttenthaler2019} found to separate Tc-rich from Tc-poor Miras also seems to separate stars with and without bumps in their light curves. Surprisingly, stars with Tc in their spectra occupy the same region in the $K-[22]$ vs.\ $P$ diagram as stars with bumps in their light curves, as do stars without Tc and bumps. This suggests an association between 3DUP and the occurrence of bumps in a Mira's light curve. One would not expect such a relation from theoretical considerations or previous observational results.

\begin{figure}[!t]
\includegraphics[scale=0.76]{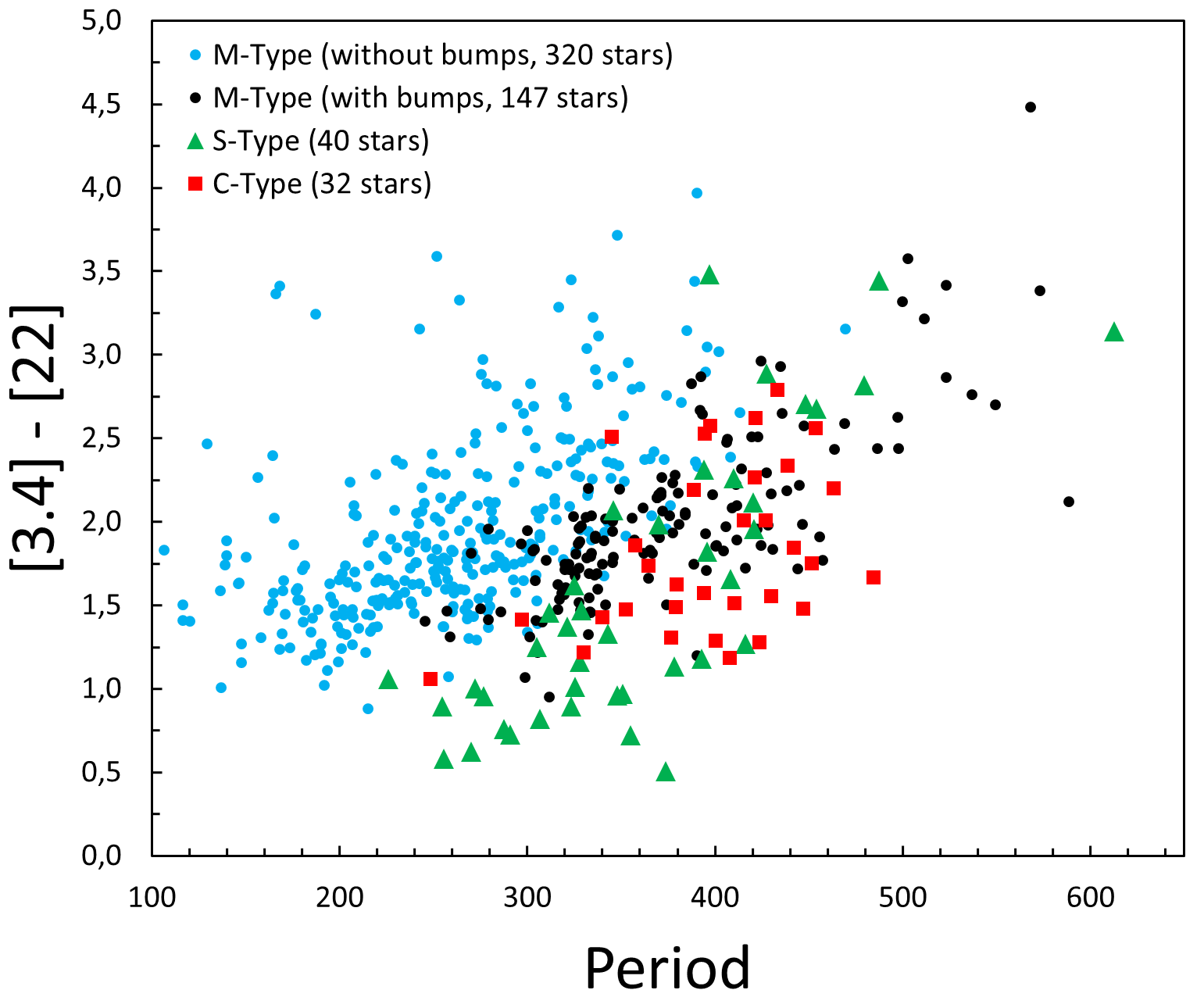}
\caption{$[3.4]-[22]$ vs.\ $P$ diagram of M-type Miras with and without light curve bumps, respectively, as well as S- and C-type Miras.}
\label{Fig:[3.4]-[22] vs Period}
\end{figure}

Instead of the $K$-band magnitude, we can also use the WISE $[3.4]$ band, which has a central wavelength of $\sim3.4\,\mu{\rm m}$. Advantages of the $[3.4]$- over the $K$-band are that it was measured simultaneously with the $[22]$ band and the variability amplitude is lower. Both should reduce the variability scatter in the colour index. The disadvantage is that bright Miras saturate some detector pixels in the $[3.4]$ band, decreasing the accuracy for those stars. For not-too-dusty stars, the $[3.4]$ band is still dominated by photometric emission, and we can establish the $[3.4]-[22]$ colour as an indicator of dust mass-loss rate. Figure~\ref{Fig:[3.4]-[22] vs Period} shows a $[3.4]-[22]$ vs.\ $P$ diagram of our sample stars, distinguished by their symbols into M-type stars with bumps (Group~A), M-type stars without bumps (Group~S), S-type, and C-type stars. Again, this diagram clearly demonstrates the separation of the stars into different regions: M-type stars without bumps have higher $[3.4]-[22]$ colours, i.e., higher dust mass-loss rates, than other stars at similar periods. Note the location of a group of S-type stars at very low $[3.4]-[22]$ colours that barely have dust in their circumstellar envelopes.

\section{Discussion}\label{sec:discussion}

The new evidence from our sample allows for drawing conclusions on the origin of significant MPCs. The Fourier spectra of MPC Miras show no prominent additional peaks beyond those corresponding to the harmonics of the primary frequency. This suggests that hypotheses invoking additional pulsation modes disagree with the evidence collected here, unlike in RV~Tau stars \citep{Buchler1996}.

Following theoretical evolutionary models, Miras with significant MPCs could have a place in the interpulse period of the TP-AGB. TPs repeat quasi-periodically every few 10$^4$ years, depending on stellar mass \citep[e.g.,][]{Schwarzschild1967,Wood1981}. After an initial luminosity drop at the onset of a TP, the star is predicted to be more luminous than before the He-shell flash for a few $10^2$ years. Stars with SPC (e.g., T~UMi) and CPC (e.g., R~Hya) have been suggested to currently pass through this phase \citep[e.g.,][]{Wood1981,Uttenthaler2011}. Subsequently, in a phase of a few 10$^3$ years at most, the He-burning shell is gradually extinguished, causing a decrease in luminosity below the level prior to the TP. In this sense, a possible interpretation for the observations could be that the large structural changes in the star between about 10$^2$ and 10$^3$ years after the TP could be the origin of the significant MPCs, after which the pulsation period will regain stability as the stellar structure is restored. These shorter period variations may be thermal relaxation oscillations in the stellar envelope, with Kelvin-Helmholtz cooling timescales \citep[][$\tau_{KH} \sim 6-200$\,yrs]{Ostlie1986} that broadly coincide with the timescales of period variations found in MPC stars ($\sim30-75$\,yrs).

If significant MPCs have their origin in the aftermath of a TP, their number fraction ($18/548\approx3.3$\%) should reflect the fraction of the TP cycle in which MPCs occur. At a typical Mira mass of $1.8\,M_{\sun}$, state-of-the-art AGB evolutionary models \citep{Marigo2013} predict an interpulse period of $\sim1.2\times10^5$\,yrs. This would mean that the star typically undergoes significant MPCs for $\sim4\,000$\,yrs. Note that this is about half the time it takes the model star to reach the luminosity minimum after the onset of a TP. Indeed, several of the MPC stars in our sample have Tc lines in their spectra, meaning they must have undergone TPs in the past (Sect.~\ref{sec:3DUP-indicators}). However, the puzzling piece of information is that the MPC group is made up exclusively of M-type Miras. Certainly, also S- and C-type Miras undergo TPs; hence the absence of significant MPCs among them is unexpected. Given the distribution of chemical spectral types in our sample, the probability for this selection is 6.6\% -- low, but not low enough to conclude that it is not a chance result. A still larger sample of Miras, including more of type S and C, would be required to test this conclusion.

Since a very high fraction ($16/18$) of the MPC Miras also have bumps on the ascending branch of their light curves, we inspected the light curves of all sample stars for asymmetries and divided them into two Groups A (asymmetric) and S (symmetric). We confirm the conclusion by \citet{Lebzelter2011} that the fraction of light curves with asymmetries is much higher among S- and C-type Miras. Further evidence that the light curve shape could be related to 3DUP events comes from the distribution of Tc in the M-type Miras in Groups~A and S: While in Group~A, they uniformly distribute between having Tc (post-3DUP) and not having Tc, almost all with symmetric light curves are devoid of Tc.

We also find evidence that MPC stars have reduced dust mass-loss rates. In a $K-[22]$ vs.\ $P$ diagram (Fig.~\ref{Fig:K-22_significant}), all MPC stars are located at relatively low values of the $K-[22]$ colour. In addition, through a diagram plotting $K-[22]$ or $[3.4]-[22]$ vs.\ $P$ (see Figs.~\ref{Fig:k22-BUMPS} and \ref{Fig:[3.4]-[22] vs Period}), we demonstrate that stars with bumps have a decreased dust mass-loss rate compared to Miras with symmetric light curves at similar periods. No further relation between Mira light curve shape and other stellar parameters is known to date.

The question is: How do all those things relate? Why do Miras with Tc, meandering period changes, and bumps in their light curves all fall in the same region of the $K-[22]$ vs.\ $P$ diagram? Why do they have reduced $K-[22]$ colours (dust mass-loss rates) compared to Miras at similar periods but without either of these properties?

\citet{Uttenthaler2019} suggested that 3DUP, possibly through the dredge-up of radioactive isotopes, leads to reduced dust formation in the stellar envelope. Ions created upon the radioactive decay of unstable ($s$-process) isotopes could inhibit dust formation. In this way, reduced $K-[22]$, as an indicator of dust mass-loss rate, would also be connected to the occurrence of TPs that drive 3DUP. As discussed above, MPCs could also be related to a recent TP via a readjustment of the envelope structure. The fact that most of the MPC Miras, in addition to having a reduced K-[22], also show Tc in their spectra supports the notion that they are in the TP-AGB phase. However, additional observations will be required to strengthen the case.

Finally, the relation of light curve asymmetries to this aggregate of phenomena needs to be established. It was shown by \citet{Liljegren2016,Liljegren2017}, using numerical experiments with dynamic Mira model atmospheres, that the wind properties of a mass-losing model are susceptible to the luminosity and radius evolution during a pulsation cycle because of the importance of the timing between the dust formation and the radiation pressure. Deviations from a sinusoidal light curve might have a detrimental effect on the mass-loss rate. If dust formation in the stellar envelope is hindered by the effects of 3DUP (i.e., radioactive decay of unstable isotopes), this could affect the light curve shape. Suppose the dust is formed inefficiently in the wake of an outwards propagating shock wave, or it couples less efficiently to the radiation pressure from the central star. Then, significant dust amounts would fall back towards the star instead of being accelerated outwards. The infalling material would hit the outwards moving shock wave of the next pulsation cycle, causing bumps or secondary maxima in the light curve. In this way, bumps in the light curves may be connected to the occurrence of 3DUP, and the stars might end up at a reduced $K-[22]$ colour just where Tc-rich, post-3DUP Miras are located.

\section{Summary and conclusions}\label{sec:conclusio}

We collected long-term light curves of 548 solar-neighbourhood Miras from several databases (AAVSO, AFOEV, ASAS, and DASCH) and analysed them for period changes and variations over timescales of several decades. In total, we identify 27 Miras with relative period changes in excess of 5\%. We confirm all identifications of period-changing stars from earlier studies \citep{Templeton2005} and add one more star with a continuously decreasing pulsation period, T~Lyn. The period of this carbon star has decreased by 6.4\% over the past $\sim80$~years. For BH~Cru, a star that was previously reported to have undergone a sudden period increase by $\sim25$\% before 1999, we find that since then the period decreased again by $\sim9$\%. Eighteen stars exhibit significant meandering period changes (MPCs) with relative period changes $\Delta P/P\geq5.0$\%, all of which are of spectral type M. This MPC group is also characterised by a relatively low mid-IR excess. Furthermore, we find that the median (random) period change of Miras is $\sim2.4$\%, which should also be the accuracy with which periods can be determined from long-term light curves.

Since we noticed that MPC Miras show asymmetries (bumps) in their light curves, we extended our analysis to light curve shapes of the stars. The sample was divided into two groups, one with bumps or other anomalies in their light curves ($203/548 \approx 37$\%), and those seemingly without such asymmetries ($336/548\approx 63$\%). Our sample clearly confirms the suggestion found by \citet{Lebzelter2011} that S- and C-type Miras are strongly over-represented in the group with asymmetric light curves. We searched for further stellar parameters that distinguish these two groups and found that they clearly locate in two different regions of a $K-[22]$ vs.\ $P$ diagram in the sense that, at a given pulsation period, Miras without bumps have higher $K-[22]$ colour (dust mass-loss rate) than those with bumps. No such distinction was reported previously.

Our empirical results are both clear and intriguing. We speculate that the origin of MPCs lies in thermal oscillations of the stellar envelope in the aftermath of a thermal pulse because of the similarity of observed and predicted timescales and the stars' 3DUP activity. However, the fact that MPC Miras in our sample are exclusive of M-type seems puzzling. Furthermore, the co-location in a $K-[22]$ vs.\ $P$ diagram of Miras that can be distinguished by seemingly unconnected characteristics is surprising. At a given pulsation period, Miras with significant period changes or bumps in their light curves have distinctively lower $K-[22]$ colours (dust mass-loss rates) than Miras without. The same is true for Miras with 3DUP activity, as indicated by the presence of Tc in their spectra and increased $^{12}$C abundance, compared to Miras without 3DUP. We speculate that all of this could be related to the occurrence of TPs and 3DUP events, but better and more targeted data are needed to test these relations.

\begin{acknowledgements}
We acknowledge with thanks the variable star observations from the AAVSO International Database contributed by observers worldwide and used in this research. This publication makes use of data products from the Two Micron All Sky Survey, which is a joint project of the University of Massachusetts and the Infrared Processing and Analysis Center/California Institute of Technology, funded by the National Aeronautics and Space Administration and the National Science Foundation. This publication makes use of data products from the Wide-field Infrared Survey Explorer, which is a joint project of the University of California, Los Angeles, and the Jet Propulsion Laboratory/California Institute of Technology, funded by the National Aeronautics and Space Administration. This publication make use of light curves from the ASAS-SN project, specifically from the ASAS-SN Catalog of Variable Stars~III. The DASCH project at Harvard is partially supported by NSF grants AST-0407380, AST-0909073, and AST-1313370.
\end{acknowledgements}

\bibliographystyle{aa}
\bibliography{Meandering-Miras}

\begin{thebibliography}{54}
\expandafter\ifx\csname natexlab\endcsname\relax\def\natexlab#1{#1}\fi

\bibitem[{{Abia} \& {Isern}(1997)}]{Abia1997}
{Abia}, C. \& {Isern}, J. 1997, \mnras, 289, L11

\bibitem[{{Buchler} {et~al.}(1996){Buchler}, {Kollath}, {Serre}, \&
  {Mattei}}]{Buchler1996}
{Buchler}, J.~R., {Kollath}, Z., {Serre}, T., \& {Mattei}, J. 1996, \apj, 462,
  489

\bibitem[{{Campbell}(1925)}]{Campbell1925}
{Campbell}, L. 1925, Harvard Reprint, No. 21

\bibitem[{{Cutri} {et~al.}(2021){Cutri}, {Wright}, {Conrow}, {Fowler},
  {Eisenhardt}, {Grillmair}, {Kirkpatrick}, {Masci}, {McCallon}, {Wheelock},
  {Fajardo-Acosta}, {Yan}, {Benford}, {Harbut}, {Jarrett}, {Lake}, {Leisawitz},
  {Ressler}, {Stanford}, {Tsai}, {Liu}, {Helou}, {Mainzer}, {Gettngs},
  {Gonzalez}, {Hoffman}, {Marsh}, {Padgett}, {Skrutskie}, {Beck}, {Papin}, \&
  {Wittman}}]{Cutri2014}
{Cutri}, R.~M., {Wright}, E.~L., {Conrow}, T., {et~al.} 2021, VizieR Online
  Data Catalog, II/328

\bibitem[{{Dominy} \& {Wallerstein}(1987)}]{Dominy1987}
{Dominy}, J.~F. \& {Wallerstein}, G. 1987, \apj, 317, 810

\bibitem[{{Foster}(1996)}]{Foster1996}
{Foster}, G. 1996, \aj, 112, 1709

\bibitem[{{Gal} \& {Szatmary}(1995)}]{Gal1995}
{Gal}, J. \& {Szatmary}, K. 1995, \aap, 297, 461

\bibitem[{{Garc{\'\i}a-Hern{\'a}ndez}
  {et~al.}(2013){Garc{\'\i}a-Hern{\'a}ndez}, {Zamora}, {Yag{\"u}e},
  {Uttenthaler}, {Karakas}, {Lugaro}, {Ventura}, \&
  {Lambert}}]{GarciaHernandez2013}
{Garc{\'\i}a-Hern{\'a}ndez}, D.~A., {Zamora}, O., {Yag{\"u}e}, A., {et~al.}
  2013, \aap, 555, L3

\bibitem[{{Greaves} \& {Holland}(1997)}]{Greaves1997}
{Greaves}, J.~S. \& {Holland}, W.~S. 1997, \aap, 327, 342

\bibitem[{{Hawkins} {et~al.}(2001){Hawkins}, {Mattei}, \&
  {Foster}}]{Hawkins2001}
{Hawkins}, G., {Mattei}, J.~A., \& {Foster}, G. 2001, \pasp, 113, 501

\bibitem[{{Herwig}(2005)}]{Herwig2005}
{Herwig}, F. 2005, \araa, 43, 435

\bibitem[{{Hinkle} {et~al.}(2016){Hinkle}, {Lebzelter}, \&
  {Straniero}}]{Hinkle2016}
{Hinkle}, K.~H., {Lebzelter}, T., \& {Straniero}, O. 2016, \apj, 825, 38

\bibitem[{{Jayasinghe} {et~al.}(2019){Jayasinghe}, {Stanek}, {Kochanek},
  {Shappee}, {Holoien}, {Thompson}, {Prieto}, {Dong}, {Pawlak}, {Pejcha},
  {Shields}, {Pojmanski}, {Otero}, {Hurst}, {Britt}, \&
  {Will}}]{Jayasinghe2019}
{Jayasinghe}, T., {Stanek}, K.~Z., {Kochanek}, C.~S., {et~al.} 2019, \mnras,
  485, 961

\bibitem[{{Kiss} \& {Szatm{\'a}ry}(2002)}]{Kiss2002}
{Kiss}, L.~L. \& {Szatm{\'a}ry}, K. 2002, \aap, 390, 585

\bibitem[{{Kiss} {et~al.}(1999){Kiss}, {Szatm{\'a}ry}, {Cadmus}, \&
  {Mattei}}]{Kiss1999}
{Kiss}, L.~L., {Szatm{\'a}ry}, K., {Cadmus}, R.~R., J., \& {Mattei}, J.~A.
  1999, \aap, 346, 542

\bibitem[{{Lambert} {et~al.}(1986){Lambert}, {Gustafsson}, {Eriksson}, \&
  {Hinkle}}]{Lambert1986}
{Lambert}, D.~L., {Gustafsson}, B., {Eriksson}, K., \& {Hinkle}, K.~H. 1986,
  \apjs, 62, 373

\bibitem[{{Laycock} {et~al.}(2010){Laycock}, {Tang}, {Grindlay}, {Los},
  {Simcoe}, \& {Mink}}]{Laycock2010}
{Laycock}, S., {Tang}, S., {Grindlay}, J., {et~al.} 2010, \aj, 140, 1062

\bibitem[{{Lebzelter}(2011)}]{Lebzelter2011}
{Lebzelter}, T. 2011, Astronomische Nachrichten, 332, 140

\bibitem[{{Lebzelter} {et~al.}(2019){Lebzelter}, {Hinkle}, {Straniero},
  {Lambert}, {Pilachowski}, \& {Nault}}]{Lebzelter2019}
{Lebzelter}, T., {Hinkle}, K.~H., {Straniero}, O., {et~al.} 2019, \apj, 886,
  117

\bibitem[{{Liljegren} {et~al.}(2017){Liljegren}, {H{\"o}fner}, {Eriksson}, \&
  {Nowotny}}]{Liljegren2017}
{Liljegren}, S., {H{\"o}fner}, S., {Eriksson}, K., \& {Nowotny}, W. 2017, \aap,
  606, A6

\bibitem[{{Liljegren} {et~al.}(2016){Liljegren}, {H{\"o}fner}, {Nowotny}, \&
  {Eriksson}}]{Liljegren2016}
{Liljegren}, S., {H{\"o}fner}, S., {Nowotny}, W., \& {Eriksson}, K. 2016, \aap,
  589, A130

\bibitem[{{Little} {et~al.}(1987){Little}, {Little-Marenin}, \&
  {Bauer}}]{Little1987}
{Little}, S.~J., {Little-Marenin}, I.~R., \& {Bauer}, W.~H. 1987, \aj, 94, 981

\bibitem[{{Lockwood} \& {Wing}(1971)}]{Lockwood1971}
{Lockwood}, G.~W. \& {Wing}, R.~F. 1971, \apj, 169, 63

\bibitem[{{Ludendorff}(1928)}]{Ludendorff1928}
{Ludendorff}, H. 1928, in \textit {Handbuch der Astrophysik} (Verlag Von J.
  Springer, Berlin), Vol. 6, Chap.2, 49

\bibitem[{{Marigo} {et~al.}(2013){Marigo}, {Bressan}, {Nanni}, {Girardi}, \&
  {Pumo}}]{Marigo2013}
{Marigo}, P., {Bressan}, A., {Nanni}, A., {Girardi}, L., \& {Pumo}, M.~L. 2013,
  \mnras, 434, 488

\bibitem[{{McDonald} {et~al.}(2018){McDonald}, {De Beck}, {Zijlstra}, \&
  {Lagadec}}]{McDonald2018}
{McDonald}, I., {De Beck}, E., {Zijlstra}, A.~A., \& {Lagadec}, E. 2018,
  \mnras, 481, 4984

\bibitem[{{McDonald} {et~al.}(2020){McDonald}, {Uttenthaler}, {Zijlstra},
  {Richards}, \& {Lagadec}}]{McDonald2020}
{McDonald}, I., {Uttenthaler}, S., {Zijlstra}, A.~A., {Richards}, A.~M.~S., \&
  {Lagadec}, E. 2020, \mnras, 491, 1174

\bibitem[{{Merch{\'a}n Ben{\'\i}tez} \& {Jurado
  Vargas}(2000)}]{Merchan-Benitez2000}
{Merch{\'a}n Ben{\'\i}tez}, P. \& {Jurado Vargas}, M. 2000, \aap, 353, 264

\bibitem[{{Merch{\'a}n Ben{\'\i}tez} \& {Jurado
  Vargas}(2002)}]{Merchan-Benitez2002}
{Merch{\'a}n Ben{\'\i}tez}, P. \& {Jurado Vargas}, M. 2002, \aap, 386, 244

\bibitem[{{Moln{\'a}r} {et~al.}(2019){Moln{\'a}r}, {Joyce}, \&
  {Kiss}}]{Molnar2019}
{Moln{\'a}r}, L., {Joyce}, M., \& {Kiss}, L.~L. 2019, \apj, 879, 62

\bibitem[{{Ohnaka} \& {Tsuji}(1996)}]{Ohnaka1996}
{Ohnaka}, K. \& {Tsuji}, T. 1996, \aap, 310, 933

\bibitem[{{Ostlie} \& {Cox}(1986)}]{Ostlie1986}
{Ostlie}, D.~A. \& {Cox}, A.~N. 1986, \apj, 311, 864

\bibitem[{{Percy}(2015)}]{Percy2015}
{Percy}, J.~R. 2015, JAAVSO, 43, 223

\bibitem[{{Ramstedt} \& {Olofsson}(2014)}]{Ramstedt2014}
{Ramstedt}, S. \& {Olofsson}, H. 2014, \aap, 566, A145

\bibitem[{{Samus'} {et~al.}(2017){Samus'}, {Kazarovets}, {Durlevich},
  {Kireeva}, \& {Pastukhova}}]{Samus2017}
{Samus'}, N.~N., {Kazarovets}, E.~V., {Durlevich}, O.~V., {Kireeva}, N.~N., \&
  {Pastukhova}, E.~N. 2017, Astronomy Reports, 61, 80

\bibitem[{{Schlafly} {et~al.}(2019){Schlafly}, {Meisner}, \&
  {Green}}]{Schlafly2019}
{Schlafly}, E.~F., {Meisner}, A.~M., \& {Green}, G.~M. 2019, \apjs, 240, 30

\bibitem[{{Sch{\"o}ier} \& {Olofsson}(2000)}]{Schoier2000}
{Sch{\"o}ier}, F.~L. \& {Olofsson}, H. 2000, \aap, 359, 586

\bibitem[{{Schwarzschild} \& {H{\"a}rm}(1967)}]{Schwarzschild1967}
{Schwarzschild}, M. \& {H{\"a}rm}, R. 1967, \apj, 150, 961

\bibitem[{{Shappee} {et~al.}(2014){Shappee}, {Prieto}, {Grupe}, {Kochanek},
  {Stanek}, {De Rosa}, {Mathur}, {Zu}, {Peterson}, {Pogge}, {Komossa}, {Im},
  {Jencson}, {Holoien}, {Basu}, {Beacom}, {Szczygie{\l}}, {Brimacombe},
  {Adams}, {Campillay}, {Choi}, {Contreras}, {Dietrich}, {Dubberley},
  {Elphick}, {Foale}, {Giustini}, {Gonzalez}, {Hawkins}, {Howell}, {Hsiao},
  {Koss}, {Leighly}, {Morrell}, {Mudd}, {Mullins}, {Nugent}, {Parrent},
  {Phillips}, {Pojmanski}, {Rosing}, {Ross}, {Sand}, {Terndrup}, {Valenti},
  {Walker}, \& {Yoon}}]{Shappee2014}
{Shappee}, B.~J., {Prieto}, J.~L., {Grupe}, D., {et~al.} 2014, \apj, 788, 48

\bibitem[{{Skrutskie} {et~al.}(2006){Skrutskie}, {Cutri}, {Stiening},
  {Weinberg}, {Schneider}, {Carpenter}, {Beichman}, {Capps}, {Chester},
  {Elias}, {Huchra}, {Liebert}, {Lonsdale}, {Monet}, {Price}, {Seitzer},
  {Jarrett}, {Kirkpatrick}, {Gizis}, {Howard}, {Evans}, {Fowler}, {Fullmer},
  {Hurt}, {Light}, {Kopan}, {Marsh}, {McCallon}, {Tam}, {Van Dyk}, \&
  {Wheelock}}]{Skrutskie2006}
{Skrutskie}, M.~F., {Cutri}, R.~M., {Stiening}, R., {et~al.} 2006, \aj, 131,
  1163

\bibitem[{{Soszynski} {et~al.}(2007){Soszynski}, {Dziembowski}, {Udalski},
  {Kubiak}, {Szymanski}, {Pietrzynski}, {Wyrzykowski}, {Szewczyk}, \&
  {Ulaczyk}}]{Soszynski2007}
{Soszynski}, I., {Dziembowski}, W.~A., {Udalski}, A., {et~al.} 2007, \actaa,
  57, 201

\bibitem[{{Templeton} {et~al.}(2005){Templeton}, {Mattei}, \&
  {Willson}}]{Templeton2005}
{Templeton}, M.~R., {Mattei}, J.~A., \& {Willson}, L.~A. 2005, \aj, 130, 776

\bibitem[{{Uttenthaler} {et~al.}(2016{\natexlab{a}}){Uttenthaler}, {Greimel},
  \& {Templeton}}]{Uttenthaler2016a}
{Uttenthaler}, S., {Greimel}, R., \& {Templeton}, M. 2016{\natexlab{a}},
  Astronomische Nachrichten, 337, 293

\bibitem[{{Uttenthaler} {et~al.}(2019){Uttenthaler}, {McDonald}, {Bernhard},
  {Cristallo}, \& {Gobrecht}}]{Uttenthaler2019}
{Uttenthaler}, S., {McDonald}, I., {Bernhard}, K., {Cristallo}, S., \&
  {Gobrecht}, D. 2019, \aap, 622, A120

\bibitem[{{Uttenthaler} {et~al.}(2016{\natexlab{b}}){Uttenthaler}, {Meingast},
  {Lebzelter}, {Aringer}, {Joyce}, {Hinkle}, {Guzman-Ramirez}, \&
  {Greimel}}]{Uttenthaler2016b}
{Uttenthaler}, S., {Meingast}, S., {Lebzelter}, T., {et~al.}
  2016{\natexlab{b}}, \aap, 585, A145

\bibitem[{{Uttenthaler} {et~al.}(2011){Uttenthaler}, {van Stiphout}, {Voet},
  {van Winckel}, {van Eck}, {Jorissen}, {Kerschbaum}, {Raskin}, {Prins},
  {Pessemier}, {Waelkens}, {Fr{\'e}mat}, {Hensberge}, {Dumortier}, \&
  {Lehmann}}]{Uttenthaler2011}
{Uttenthaler}, S., {van Stiphout}, K., {Voet}, K., {et~al.} 2011, \aap, 531,
  A88

\bibitem[{{Vanture} {et~al.}(2007){Vanture}, {Smith}, {Lutz}, {Wallerstein},
  {Lambert}, \& {Gonzalez}}]{Vanture2007}
{Vanture}, A.~D., {Smith}, V.~V., {Lutz}, J., {et~al.} 2007, \pasp, 119, 147

\bibitem[{{Vardya}(1988)}]{Vardya1988}
{Vardya}, M.~S. 1988, \aaps, 73, 181

\bibitem[{{Whitelock}(1999)}]{Whitelock1999}
{Whitelock}, P.~A. 1999, \nar, 43, 437

\bibitem[{{Wood} \& {Zarro}(1981)}]{Wood1981}
{Wood}, P.~R. \& {Zarro}, D.~M. 1981, \apj, 247, 247

\bibitem[{{Wright} {et~al.}(2010){Wright}, {Eisenhardt}, {Mainzer}, {Ressler},
  {Cutri}, {Jarrett}, {Kirkpatrick}, {Padgett}, {McMillan}, {Skrutskie},
  {Stanford}, {Cohen}, {Walker}, {Mather}, {Leisawitz}, {Gautier}, {McLean},
  {Benford}, {Lonsdale}, {Blain}, {Mendez}, {Irace}, {Duval}, {Liu}, {Royer},
  {Heinrichsen}, {Howard}, {Shannon}, {Kendall}, {Walsh}, {Larsen}, {Cardon},
  {Schick}, {Schwalm}, {Abid}, {Fabinsky}, {Naes}, \& {Tsai}}]{Wright2010}
{Wright}, E.~L., {Eisenhardt}, P. R.~M., {Mainzer}, A.~K., {et~al.} 2010, \aj,
  140, 1868

\bibitem[{{Ya'Ari} \& {Tuchman}(1996)}]{Ya'Ari1996}
{Ya'Ari}, A. \& {Tuchman}, Y. 1996, \apj, 456, 350

\bibitem[{{Zijlstra} \& {Bedding}(2002)}]{Zijlstra2002a}
{Zijlstra}, A.~A. \& {Bedding}, T.~R. 2002, JAAVSO, 31, 2

\bibitem[{{Zijlstra} {et~al.}(2002){Zijlstra}, {Bedding}, \&
  {Mattei}}]{Zijlstra2002b}
{Zijlstra}, A.~A., {Bedding}, T.~R., \& {Mattei}, J.~A. 2002, \mnras, 334, 498

\end{thebibliography}


\begin{appendix}


\onecolumn
\section{Appendix Figures}

\begin{longfigure}{c}
\caption{}
\label{Fig:A1}\\
\endLFfirsthead
\caption{continued}\\
\endLFhead
\endLFfoot
\endLFlastfoot
\includegraphics[width=17cm, height=22cm]{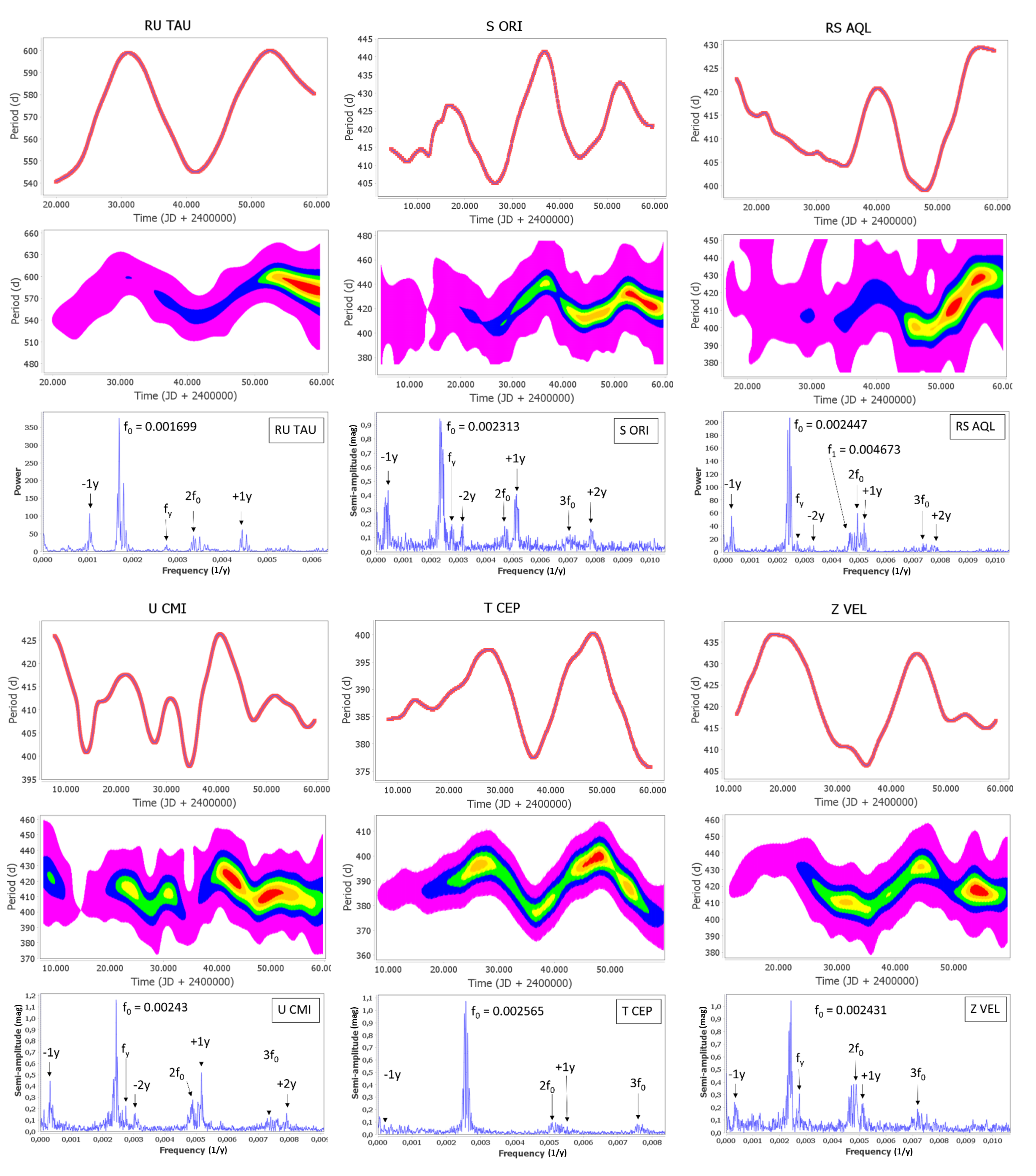}\\
\includegraphics[width=17cm, height=22cm]{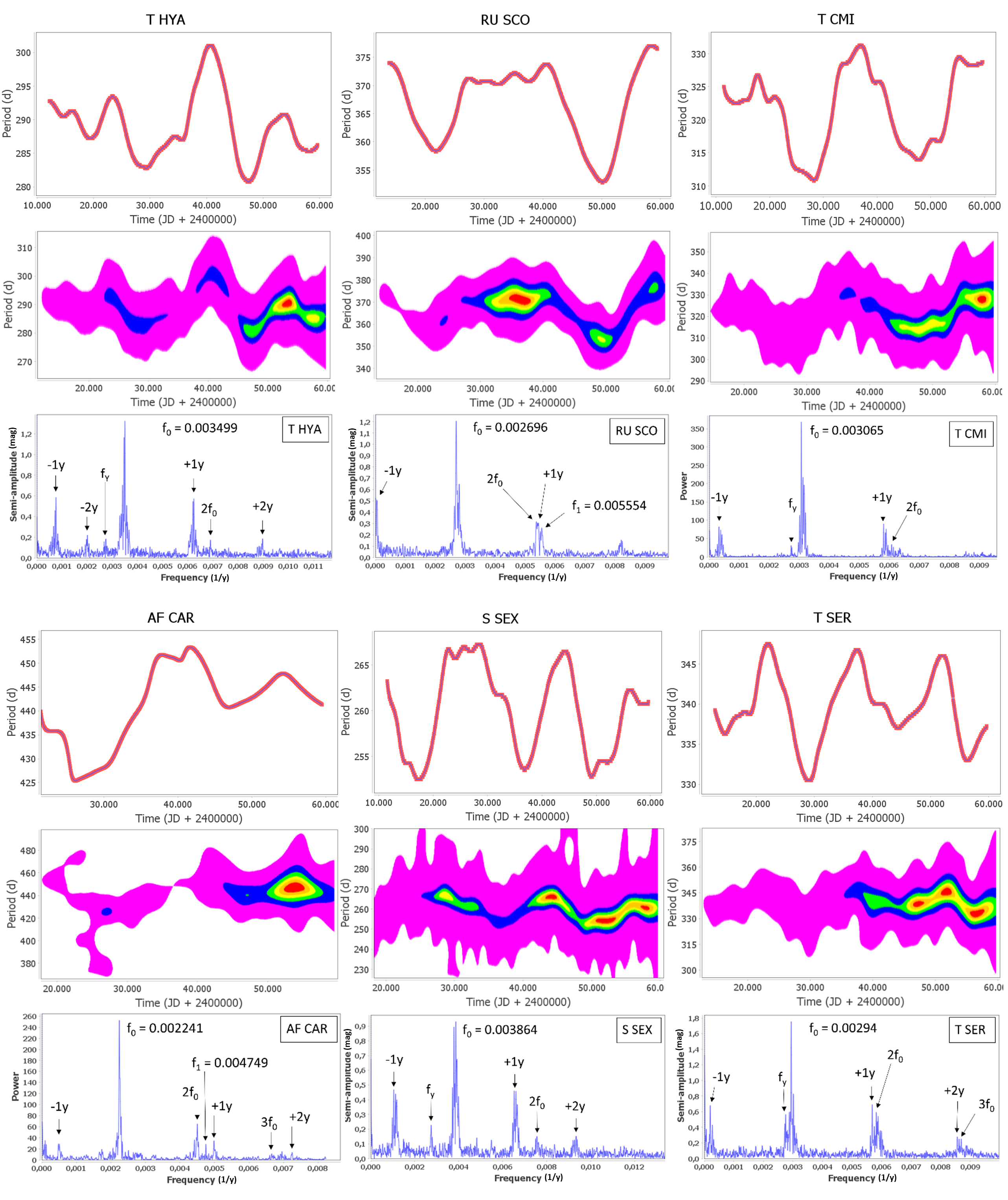}\\
\includegraphics[width=17cm, height=22cm]{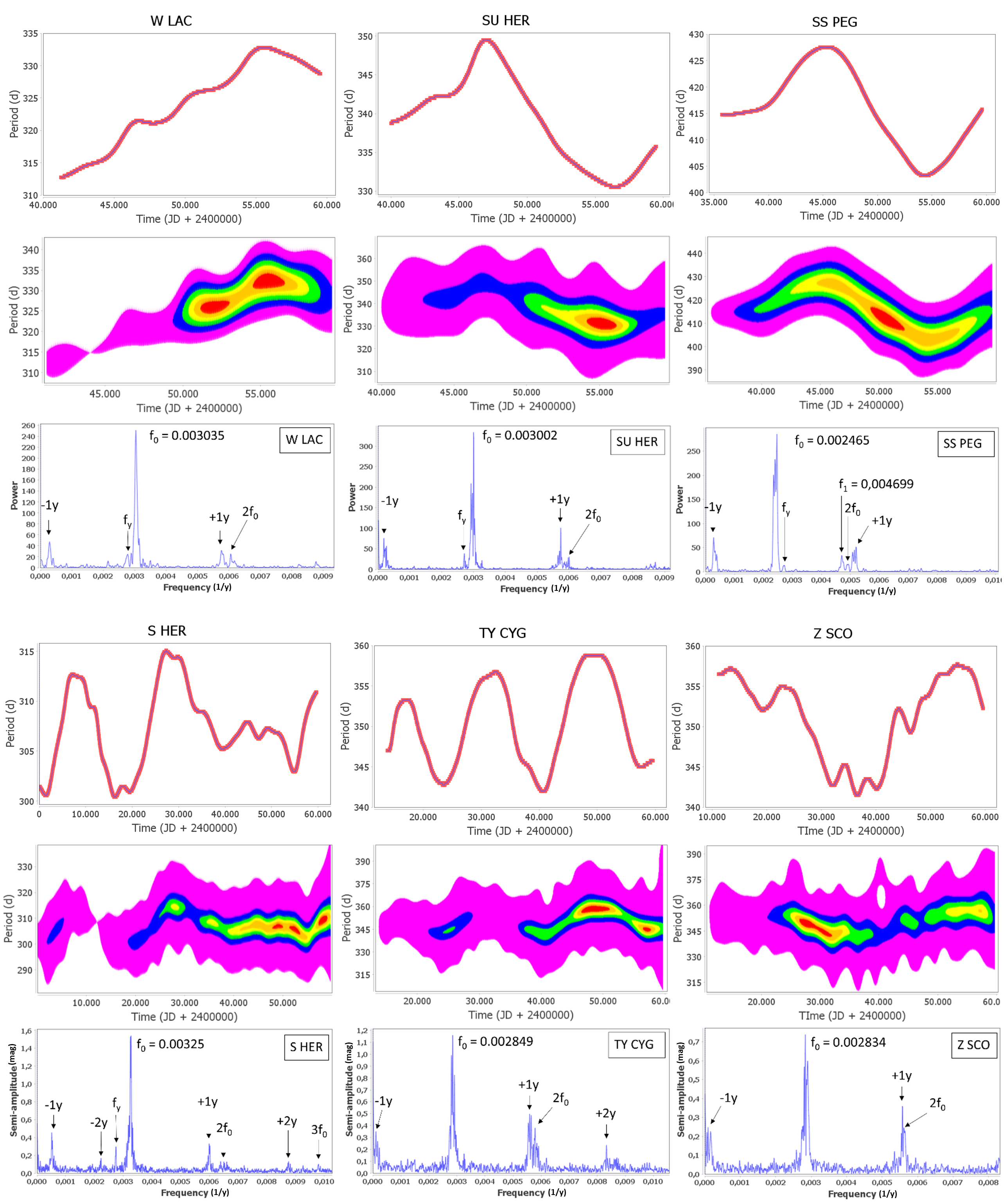}\\
\caption{Period as a function of time and WWZ, Fourier spectra and identification of the relevant peaks for the 18 stars with significant MPCs. From left to right and from top to bottom they are arranged in decreasing order of $\Delta P/\left<P\right>$}
\end{longfigure}

\onecolumn

\section{Appendix Tables}

\begin{longtable}[t]{c c c c c c c c c}
\label{Tab:All data}\\
\caption{Data of the 548 sample Mira stars.}\\
\hline\\
GCVS & S.Type & Period & $\Delta P/P$ & Bumps & Tc & $^{12}$C/$^{13}$C & $K-[22]$ &  [3.4]--[22]\\\\
\hline\\
\endfirsthead
\hline\\
GCVS & S.Type & Period & $\Delta P/P$ & Bumps & Tc & $^{12}$C/$^{13}$C & $K-[22]$ &  [3.4]--[22]\\\\
\hline\\
\endhead
\hline\\
\endfoot
\hline\\
\endlastfoot

AK And & S: & 321.3 & 3.5 & 0 &  &  & 1.388 & 1.371\\
AZ And & M4 & 195.2 & 3.1 & 0 &  &  & 2.206 & 1.549\\
BG And & S6.5.5e & 291.0 & 3.1 & 1 &  &  & 0.636 & 0.727\\
BU And & M7e & 380.4 & 4.5 & 1 &  &  & 2.359 & 2.173\\
R And & S3.5e-S8.8e(M7e) & 410.0 & 3.2 & 1 & yes$^{a}$ & $40\pm15^{1}$ & 2.905 & 2.259\\
RR And & S6.5.2e & 327.9 & 2.1 & 1 &  &  & 1.300 & 1.162\\
RW And & M5e-M10e(S6.2e) & 429.9 & 2.3 & 1 &  &  & 2.602 & 2.166\\
RY And & M8 & 392.5 & 1.8 & 1 &  &  & 2.807 & 2.867\\
SV And & M5e-M7e & 315.3 & 3.5 & 0 &  &  & 1.926 & 2.014\\
SX And & M6.5e & 332.7 & 2.4 & 1 &  &  & 1.612 & 1.325\\
SZ And & M2e & 342.6 & 2.6 & 0 &  &  & 2.522 & 2.360\\
T And & M4e-M7.5e & 280.9 & 2.8 & 0 &  &  & 2.027 & 1.581\\
TU And & M5e & 318.2 & 3.2 & 1 &  &  & 1.674 & 1.572\\
U And & M6e & 346.1 & 3.2 & 1 &  &  & 2.197 & 2.026\\
UW And & M5 & 239.7 & 2.1 & 0 &  &  & 1.461 & 1.451\\
UZ And & M7e & 314.6 & 1.9 & 0 &  &  & 2.346 & 2.108\\
V And & M2e-M3e & 257.9 & 2.7 & 0 & no$^{b}$ &  & 1.247 & 1.073\\
W And & S6.1e-S9.2e & 395.8 & 1.8 & 1 & yes$^{a}$ &  & 1.802 & 1.820\\
X And & S2.9e-S5.5e & 345.7 & 2.3 & 0 &  &  & 2.206 & 2.067\\
Y And & M3e-M4.5e & 221.0 & 1.8 & 0 & no$^{a}$ &  & 1.633 & 1.663\\
YZ And & M5e & 207.7 & 1.4 & 0 &  &  & 2.393 & 2.047\\
AC Ant & Me & 213.8 & 2.3 & 0 &  &  & 1.553 & 1.216\\
V Ant & M7IIIe & 303.2 & 1.0 & 0 &  &  & 2.826 & 2.690\\
X Ant & M2-M6e & 162.2 & 1.2 & 0 &  &  & 1.464 & 1.468\\
T Aps & M3e & 261.3 & 1.9 & 0 &  &  & 2.245 & 2.119\\
CY Aql & M8e & 348.6 & 2.9 & 0 &  &  & 2.886 & 2.335\\
DX Aql & Me & 314.5 & 2.6 & 0 &  &  & 2.772 & 2.141\\
EU Aql & M9 & 324.4 & 2.4 & 1 &  &  & 2.015 & 2.033\\
LO Aql & M0 & 156.4 & 3.2 & 0 &  &  & 2.504 & 2.263\\
R Aql & M5e-M9IIIe & 280.7 & 16.7 & 1 & no$^{a}$ & 8$^{1}$ & 2.270 & 2.233\\
RR Aql & M6e-M9 & 395.7 & 3.8 & 0 & no$^{a}$ &  & 3.707 & 3.044\\
RS Aql & M5e-M8 & 409.2 & 7.2 & 1 &  &  & 2.531 & 2.082\\
RT Aql & M6e-M9p(S) & 327.8 & 3.1 & 0 & no$^{a}$ &  & 2.726 & 2.271\\
RU Aql & M5e & 274.9 & 2.5 & 0 &  &  & 2.051 & 1.962\\
RV Aql & M2e-M7:e & 218.5 & 1.8 & 0 &  &  & 2.170 & 1.527\\
RZ Aql & M5e & 334.1 & 2.7 & 0 &  &  & 2.206 & 1.893\\
SV Aql & M8 & 251.8 & 3.1 & 0 &  &  & 4.337 & 3.589\\
SY Aql & M5e-M7e & 355.8 & 2.8 & 0 & no$^{a}$ &  & 3.511 & 2.795\\
TU Aql & M2e-M9 & 270.8 & 3.4 & 0 &  &  & 1.875 & 1.815\\
TV Aql & M4e & 242.5 & 2.4 & 0 &  &  & 2.065 & 1.628\\
W Aql & S3.9e-S6.9e & 487.3 & 3.2 & 1 & yes$^{a}$ & 26$^{2}$ & 3.016 & 3.444\\
WW Aql & M3IIIe & 175.8 & 4.0 & 0 &  &  & 2.253 & 1.864\\
X Aql & M6e & 345.9 & 2.9 & 1 &  &  & 1.983 & 1.790\\
Z Aql & M3e & 129.5 & 2.3 & 0 &  &  & 2.490 & 2.465\\
RR Aqr & M2e-M4e & 182.4 & 2.2 & 0 &  &  & 1.064 & 1.171\\
RT Aqr & M5e-M6e & 245.5 & 3.2 & 1 &  &  & 1.483 & 1.403\\
RW Aqr & M2e-M4e & 139.8 & 1.4 & 0 &  &  & 1.846 & 1.888\\
S Aqr & M4e-M6e & 279.4 & 2.5 & 1 & dbfl$^{b}$ &  & 1.522 & 1.958\\
T Aqr & M2e-M5.5e & 201.8 & 2.0 & 0 & no$^{a}$ &  & 1.324 & 1.476\\
W Aqr & M6-M8e & 380.9 & 3.5 & 1 & poss$^{b}$ &  & 2.355 & 1.985\\
X Aqr & S6.3e:(M4e-M6.5e) & 311.8 & 1.9 & 1 & prob$^{b}$ &  & 1.993 & 1.455\\
Y Aqr & M6.5e-M9 & 381.9 & 3.1 & 0 &  &  & 3.136 & 2.714\\
U Ara & M3IIep & 225.2 & 2.2 & 0 &  &  & 1.658 & 1.776\\
R Ari & M3e-M6e & 186.9 & 1.6 & 0 &  &  & 1.555 & 1.202\\
S Ari & M4e-M5e & 292.6 & 2.1 & 0 &  &  & 1.895 & 1.654\\
T Ari & M6e-M8e & 320.4 & 4.1 & 1 & no$^{a}$ & $15\pm5^{1}$ & 1.677 & 1.606\\
U Ari & M4e-M7.5e & 372.0 & 2.2 & 1 & yes$^{a}$ &  & 2.056 & 2.064\\
AA Aur & M3e & 268.9 & 2.2 & 0 &  &  & 1.611 & 1.304\\
AC Aur & M5e & 310.3 & 1.3 & 0 &  &  & 2.182 & 2.288\\
AQ Aur & M7 & 340.8 & 2.3 & 1 &  &  & 1.758 & 1.888\\
AU Aur & C6-7.3e(N0e) & 400.5 & 3.2 & 1 & yes$^{a}$ &  & 2.162 & 1.289\\
AW Aur & M5-M9 & 447.5 & 2.3 & 1 &  &  & 2.456 & 2.575\\
AY Aur & M6.5 & 388.8 & 4.2 & 1 &  &  & 1.717 & 1.745\\
AZ Aur & C7.1e-C8.2-3(N0e) & 415.4 & 3.2 & 0 & yes$^{a}$ &  & 2.963 & 2.008\\
ET Aur & M2e & 204.5 & 1.9 & 0 &  &  & 1.535 & 1.638\\
GQ Aur & M3 & 305.5 & 2.3 & 0 &  &  & 1.994 & 1.390\\
HT Aur & M & 301.9 & 2.0 & 0 &  &  & 1.718 & 1.802\\
R Aur & M6.5e-M9.5e & 457.3 & 4.4 & 1 & yes$^{c}$ & $33\pm13^{1}$ & 2.355 & 1.769\\
RR Aur & M3e-M7e & 309.9 & 2.6 & 1 &  &  & 1.700 & 1.769\\
RU Aur & M7e-M9e & 469.6 & 3.7 & 1 &  &  & 4.000 & 3.155\\
SZ Aur & M8e & 455.5 & 1.8 & 1 & yes$^{a}$ &  & 1.749 & 1.911\\
U Aur & M7e-M9e & 408.4 & 3.8 & 0 & no$^{a}$ &  & 3.026 & 2.385\\
UV Aur & C6.2-C8.2Jep(Ne) & 394.7 & 1.3 & 0 & yes$^{a}$ &  & 3.012 & 2.529\\
V Aur & C6.2e(N3e) & 352.6 & 3.2 & 0 &  &  & 1.719 & 1.473\\
VX Aur & M5e & 327.4 & 4.6 & 1 & no$^{a}$ &  & 1.167 & 1.517\\
VY Aur & M8e & 398.8 & 3.2 & 1 &  &  & 2.616 & 2.163\\
W Aur & M3e-M8e & 273.8 & 1.8 & 0 &  &  & 2.922 & 2.290\\
X Aur & M3e-M7e & 164.2 & 3.0 & 0 &  &  & 1.432 & 1.514\\
R Boo & M3e-M8e & 223.6 & 1.8 & 0 & no$^{a}$ &  & 1.582 & 1.793\\
RR Boo & M2e-M6e & 194.7 & 1.5 & 0 &  &  & 1.844 & 1.629\\
RT Boo & M6.5e-M8e & 270.2 & 3.6 & 1 &  &  & 1.887 & 1.810\\
S Boo & M3e-M6e & 271.2 & 3.0 & 0 & dbfl$^{b}$ &  & 1.401 & 1.433\\
Z Boo & M5e-M6e & 280.9 & 1.8 & 0 &  &  & 2.502 & 2.062\\
R Cae & M6e-M9e & 391.9 & 3.1 & 1 &  &  & 3.420 & 2.668\\
AI Cam & M5 & 187.4 & 1.6 & 0 &  &  & 3.830 & 3.241\\
R Cam & S2.8e-S8.7e & 270.0 & 4.4 & 1 &  &  & 0.205 & 0.623\\
RT Cam & M6e & 364.7 & 2.2 & 1 &  &  & 1.552 & 1.661\\
S Cam & C7.3e(R8e) & 329.9 & 2.7 & 1 &  & 14$^{3}$ & 0.929 & 1.217\\
SU Cam & M5 & 285.1 & 2.4 & 0 &  &  & 2.319 & 1.790\\
SW Cam & M5e & 255.1 & 2.3 & 0 &  &  & 1.677 & 1.371\\
T Cam & S4.7e-S8.5.8e & 373.7 & 3.8 & 1 & yes$^{a}$ & $31\pm10^{1}$ & 0.776 & 0.508\\
TT Cam & M0-M7 & 253.5 & 2.0 & 0 &  &  & 2.471 & 2.005\\
TX Cam & M8-M10 & 549.4 & 3.6 & 1 &  & $21\pm6^{1}$ & 3.826 & 2.698\\
V Cam & M7e & 523.2 & 2.9 & 1 &  &  & 3.668 & 3.414\\
W Cam & M7 & 281.6 & 2.8 & 0 &  &  & 2.272 & 1.728\\
WY Cam & S2e & 416.2 & 1.0 & 1 &  &  & 1.443 & 1.267\\
R Cap & C(Ne) & 344.9 & 2.6 & 0 &  &  & 3.359 & 2.506\\
RR Cap & M5e-M6e & 278.6 & 1.4 & 0 &  &  & 3.338 & 2.269\\
RU Cap & M9e & 348.1 & 1.7 & 0 &  &  & 4.014 & 3.716\\
ST Cap & M6IIIe & 268.1 & 3.0 & 0 &  &  & 1.758 & 1.726\\
T Cap & M2e-M6e & 269.8 & 2.6 & 0 & no$^{a}$ &  & 1.974 & 1.873\\
U Cap & Me & 203.1 & 3.5 & 0 &  &  & 1.832 & 1.737\\
V Cap & M5e-M6e & 275.3 & 2.2 & 0 &  &  & 3.133 & 2.881\\
W Cap & M5e & 208.0 & 4.3 & 0 &  &  & 1.661 & 1.700\\
Z Cap & M1e-M4e & 181.4 & 1.6 & 0 &  &  & 1.735 & 1.471\\
AF Car & M8e & 446.5 & 6.3 & 1 &  &  & 1.889 & 1.983\\
KK Car & M5ep & 422.5 & 4.1 & 1 &  &  & 2.230 & 1.959\\
KL Car & S2.5e & 348.1 & 2.0 & 1 &  &  & 1.207 & 0.961\\
R Car & M4e-M8e & 308.0 & 1.3 & 1 &  &  & 1.438 & 1.401\\
RV Car & M6e & 367.6 & 3.2 & 0 &  &  & 2.729 & 2.420\\
RW Car & M4e-M7e & 319.0 & 2.5 & 0 &  &  & 2.559 & 1.773\\
RY Car & S7.8e & 420.8 & 3.5 & 1 &  &  & 2.766 & 1.955\\
RZ Car & M4e-M8e & 272.4 & 1.5 & 0 &  &  & 2.796 & 2.529\\
SU Car & M5-M8e & 573.1 & 3.5 & 1 &  &  & 3.743 & 3.382\\
Z Car & M6e & 384.2 & 3.4 & 1 &  &  & 2.494 & 2.041\\
IW Cas & S4.5.9e & 397.1 & 4.4 & 0 &  &  & 3.706 & 3.481\\
R Cas & M6e-M10e & 430.6 & 2.1 & 1 & no$^{a}$ & $12\pm4^{1}$ & 2.393 & 1.836\\
RR Cas & M5e & 298.9 & 4.0 & 1 &  &  & 1.498 & 1.067\\
RV Cas & M5.5e-M6.5Se & 332.5 & 1.2 & 0 &  &  & 2.790 & 2.466\\
S Cas & S3.4e-S5.8e & 612.7 & 3.3 & 1 &  & 32$^{2}$ & 4.126 & 3.140\\
SS Cas & M3e-M7e & 139.7 & 2.8 & 0 &  &  & 2.297 & 1.795\\
T Cas & M6e-M9e & 444.1 & 2.5 & 1 & yes$^{c}$ & $33\pm5^{1}$ & 2.111 & 1.717\\
U Cas & S3.5e-S8.6e & 276.7 & 1.8 & 0 &  &  & 1.321 & 0.954\\
V Cas & M5e-M8.5e & 229.2 & 3.0 & 0 &  &  & 1.614 & 1.619\\
VZ Cas & M0e-M3e & 169.2 & 1.3 & 0 &  &  & 1.607 & 1.328\\
W Cas & C7.1e(SC5/9e) & 407.9 & 2.2 & 1 & yes$^{a}$ & 25$^{3}$ & 0.743 & 1.186\\
WY Cas & S6.5pe+G & 479.3 & 1.2 & 0 & yes$^{a}$ &  & 3.604 & 2.819\\
X Cas & C5.4e(N1e) & 423.8 & 3.6 & 1 &  & 19$^{3}$ & 1.578 & 1.279\\
Y Cas & M6e-M8e & 413.1 & 3.6 & 0 &  &  & 3.241 & 2.656\\
Z Cas & M7e & 497.1 & 2.8 & 1 &  &  & 2.821 & 2.626\\
AQ Cen & Me & 388.9 & 1.3 & 0 &  &  & 4.096 & 3.439\\
BE Cen & Me & 199.8 & 3.0 & 0 &  &  & 1.733 & 1.652\\
R Cen & M4e-M9.5 & 571.5 & 13.9 & 1 & no$^{a}$ &  & 2.163 & 2.063\\
RS Cen & M2e-M6.5e & 164.3 & 1.8 & 0 &  &  & 2.636 & 2.398\\
RT Cen & M6e-M9e & 255.6 & 3.9 & 0 &  &  & 1.566 & 1.372\\
RV Cen & C(N3e) & 447.0 & 4.0 & 1 &  &  & 1.663 & 1.478\\
RX Cen & M5e-M8e & 328.6 & 2.1 & 0 &  &  & 2.774 & 2.431\\
TT Cen & SC5:/8-Ce & 454.1 & 1.5 & 1 &  & 10$^{2}$ & 3.026 & 2.678\\
U Cen & M3II:e-M6.5e & 220.0 & 2.7 & 0 &  &  & 1.642 & 1.371\\
W Cen & M3e-M8(III)e & 201.5 & 2.0 & 0 &  &  & 1.907 & 1.689\\
X Cen & M5e-M8e & 316.0 & 1.6 & 0 &  &  & 2.351 & 2.334\\
YY Cen & M5e & 371.2 & 2.1 & 1 &  &  & 2.451 & 2.175\\
AX Cep & C(N) & 397.6 & 1.8 & 1 &  &  & 2.742 & 2.573\\
BF Cep & M7 & 428.3 & 3.2 & 1 &  &  & 2.255 & 1.982\\
PQ Cep & C6.3e(N) & 441.9 & 0.9 & 1 &  &  & 2.616 & 1.846\\
RR Cep & M6e-M8e & 384.2 & 3.1 & 1 &  &  & 2.327 & 2.053\\
S Cep & C7.4e(N8e) & 484.4 & 3.1 & 1 &  & $224\pm130^{1}$ & 2.720 & 1.667\\
SZ Cep & S3.5.8e-S4.8:e & 328.8 & 2.4 & 1 &  &  & 1.999 & 1.472\\
T Cep & M5-M9IIIe & 390.2 & 7.0 & 1 & yes$^{a}$ & $33\pm10^{1}$ & 1.246 & 1.200\\
X Cep & M4e-M7e & 537.0 & 2.1 & 1 &  &  & 3.477 & 2.762\\
Y Cep & M5e-M6e & 332.7 & 1.5 & 0 &  &  & 2.041 & 2.181\\
Z Cep & M2e-M7e & 279.1 & 3.2 & 0 &  &  & 2.801 & 2.119\\
R Cet & M4e-M9 & 166.1 & 1.8 & 0 & no$^{a}$ &  & 3.385 & 3.366\\
S Cet & M3e-M6.5e: & 320.0 & 2.8 & 1 &  &  & 2.129 & 1.715\\
U Cet & M2e-M6e & 234.6 & 2.1 & 0 & no$^{a}$ &  & 2.027 & 1.576\\
V Cet & M3e & 260.6 & 2.3 & 0 &  &  & 1.714 & 1.495\\
W Cet & S6.3e-S9.2e & 351.0 & 3.7 & 1 & ye$^{a}$s &  & 1.145 & 0.971\\
X Cet & M2e(S)-M6e & 177.7 & 2.3 & 0 & dbfl$^{b}$ &  & 1.348 & 1.530\\
Z Cet & M1e-M6.5e & 184.8 & 2.2 & 0 &  &  & 1.779 & 1.341\\
R Cha & M4e-M7e & 333.7 & 2.1 & 1 &  &  & 1.802 & 1.455\\
SY CMa & M4-6 & 215.0 & 1.9 & 0 &  &  & 2.229 & 1.877\\
R CMi & C0ev & 340.0 & 2.4 & 1 & yes$^{a}$ & 11$^{3}$ & 1.084 & 1.428\\
S CMi & M6e-M8e & 333.0 & 2.4 & 1 & yes$^{a}$ & $18\pm5^{1}$ & 1.998 & 1.691\\
T CMi & M4Se-M8 & 316.2 & 6.5 & 1 &  &  & 1.867 & 1.475\\
U CMi & M4e & 411.6 & 7.0 & 1 & no$^{b}$ &  & 2.018 & 2.097\\
V CMi & M4e-M10 & 366.3 & 1.9 & 0 & yes$^{a}$ &  & 2.501 & 2.037\\
VX CMi & Me & 279.0 & 3.3 & 0 &  &  & 1.550 & 1.465\\
WX CMi & Se & 420.4 & 1.9 & 0 &  &  & 2.968 & 2.117\\
R Cnc & M6e-M9e & 362.0 & 4.1 & 1 & no$^{a}$ & $17\pm4^{1}$ & 1.964 & 1.810\\
RR Cnc & M3e & 297.8 & 2.4 & 0 & no$^{b}$ &  & 2.116 & 1.648\\
SU Cnc & M6 & 190.0 & 2.1 & 0 &  &  & 1.512 & 1.267\\
SZ Cnc & M2 & 319.6 & 1.6 & 0 &  &  & 3.447 & 2.743\\
U Cnc & M2e & 305.1 & 2.3 & 0 &  &  & 1.656 & 1.527\\
UY Cnc & M6.5 & 229.2 & 2.2 & 0 &  &  & 1.951 & 2.067\\
V Cnc & S0e-S7.9e & 272.2 & 1.9 & 1 & yes$^{d}$ &  & 1.750 & 1.001\\
W Cnc & M6.5e-M9e  & 393.3 & 2.5 & 1 & no$^{a}$ &  & 2.702 & 2.643\\
R Col & M3e-M7 & 327.8 & 2.1 & 1 & yes$^{a}$ &  & 2.221 & 1.881\\
S Col & M6e-M8 & 325.8 & 1.8 & 0 &  &  & 2.720 & 2.376\\
T Col & M3e-M6e & 226.2 & 3.1 & 0 & no$^{a}$ &  & 1.869 & 1.502\\
R Com & M5e-M8ep & 362.5 & 1.9 & 0 & dbfl$^{b}$ &  & 2.245 & 2.371\\
U CrA & M2Ibe & 147.6 & 2.0 & 0 &  &  & 1.449 & 1.268\\
S CrB & M6e-M8e & 360.1 & 1.9 & 0 & poss$^{b}$ &  & 2.690 & 2.806\\
V CrB & C6.2e(N2e) & 357.5 & 3.6 & 0 & yes$^{a}$ &  & 2.302 & 1.860\\
W CrB & M2e-M5e & 238.7 & 2.9 & 0 & dbfl$^{b}$ &  & 2.084 & 1.623\\
X CrB & M5e-M7e & 241.0 & 2.9 & 0 & no$^{a}$ &  & 1.571 & 1.757\\
Z CrB & M4e-M5e & 251.0 & 1.6 & 0 &  &  & 2.057 & 2.289\\
U Crt & M0e & 168.0 & 1.8 & 0 &  &  & 1.650 & 1.237\\
BH Cru & SC4.5/8-e-SC7/8-e & 528.7 & 9.5 & 1 & yes$^{a}$ & 8$^{4}$ & 1.573 & 1.372\\
ST Cru & M6e & 438.3 & 1.6 & 1 &  &  & 2.261 & 2.185\\
U Cru & M4e-M6e & 345.4 & 1.8 & 0 &  &  & 3.248 & 2.487\\
V Cru & Ce(Ne) & 376.5 & 2.9 & 1 &  &  & 1.303 & 1.305\\
R Crv & M4.5e-M9:e & 317.3 & 2.2 & 1 & poss$^{b}$ &  & 1.852 & 1.535\\
U Crv & M5e & 283.4 & 1.4 & 0 &  &  & 1.772 & 1.799\\
R CVn & M6e-M9e & 328.9 & 3.3 & 1 & poss$^{b}$ &  & 2.111 & 1.974\\
RT CVn & M5e & 254.0 & 1.6 & 0 &  &  & 2.280 & 2.143\\
U CVn & M7e & 345.4 & 3.5 & 0 &  &  & 3.439 & 2.867\\
AU Cyg & M6e-M7e & 434.9 & 1.6 & 1 &  &  & 3.703 & 2.930\\
BG Cyg & M7e-M8e & 287.4 & 3.8 & 0 &  &  & 2.008 & 1.758\\
BU Cyg & M0 & 158.0 & 2.5 & 0 &  &  & 1.653 & 1.306\\
CHI Cyg & S6.8e(M4e) & 408.3 & 2.0 & 1 & yes$^{a}$ & 36$\pm4^{1}$ & 2.407 & 1.657\\
CM Cyg & SC2-S4e & 254.7 & 2.4 & 1 &  &  & 1.874 & 0.896\\
CN Cyg & M2-M7e(S) & 199.0 & 1.0 & 0 &  &  & 2.018 & 1.372\\
CU Cyg & M6e & 214.5 & 1.4 & 0 &  &  & 1.686 & 1.449\\
DD Cyg & M0e & 147.9 & 2.0 & 0 &  &  & 1.503 & 1.155\\
DR Cyg & M5e & 313.9 & 2.2 & 0 &  &  & 1.746 & 1.830\\
FF Cyg & S6.8e(M4e) & 325.3 & 2.2 & 1 &  &  & 0.887 & 1.011\\
LX Cyg & SC3e-S5.5e: & 477.0 & 22.6 & 1 & yes$^{a}$ & 32$\pm8^{7}$ & 1.602 & 1.690\\
R Cyg & S2.5.9e-S6.9e(Tc) & 427.2 & 2.3 & 0 & yes$^{a}$ & 29$\pm10^{5}$ & 2.874 & 2.887\\
RT Cyg & M2e-M8.8eIb & 190.2 & 1.1 & 0 & dbfl$^{b}$ &  & 1.624 & 1.267\\
S Cyg & S2.5.1e(M3.5-M7e) & 323.2 & 2.8 & 1 &  &  & 0.745 & 0.895\\
ST Cyg & M5.5e-M8.0e & 337.7 & 3.6 & 1 &  &  & 1.847 & 1.594\\
SX Cyg & M7e & 411.2 & 1.5 & 1 &  &  & 2.319 & 2.224\\
TU Cyg & M3e-M6e & 219.6 & 1.8 & 0 &  &  & 2.509 & 1.531\\
TW Cyg & M6.5-M10ep & 342.9 & 3.8 & 0 & poss$^{b}$ &  & 2.340 & 1.998\\
TY Cyg & M6e-M8e & 357.0 & 5.1 & 1 &  &  & 1.868 & 1.889\\
U Cyg & C7.2e-C9.2(Npe) & 463.2 & 4.5 & 1 & yes$^{b}$ & 14$\pm6^{1}$ & 1.923 & 2.200\\
UX Cyg & M4e-M6.5e & 568.1 & 1.9 & 1 &  &  & 4.198 & 4.484\\
V Cyg & C5.3e-C7.4e(Npe) & 421.2 & 2.1 & 0 &  & 20$^{6}$ & 3.243 & 2.264\\
WX Cyg & C8.2JLi(N3e) & 410.5 & 1.9 & 1 & yes$^{a}$ & 5$^{4}$ & 1.511 & 1.512\\
WY Cyg & M5e-M6e & 304.6 & 2.6 & 0 &  &  & 2.273 & 1.680\\
Z Cyg & M5e-M9e & 263.9 & 3.4 & 0 &  &  & 3.897 & 3.327\\
BR Del & M8e & 336.1 & 2.7 & 0 &  &  & 2.544 & 2.910\\
ES Del & M10 & 361.8 & 4.3 & 1 &  &  & 2.434 & 2.085\\
R Del & M5e-M6e & 285.6 & 1.4 & 0 & no$^{a}$ &  & 2.239 & 1.754\\
RU Del & M3 & 260.0 & 1.9 & 0 &  &  & 1.815 & 1.463\\
RX Del & M2e & 185.6 & 2.7 & 0 &  &  & 1.884 & 1.425\\
S Del & M5e-M8 & 279.1 & 2.5 & 1 &  &  & 1.553 & 1.412\\
T Del & M3e-M6e & 331.7 & 3.0 & 1 & poss$^{b}$ &  & 1.290 & 1.785\\
V Del & M4e-M6e & 533.5 & 3.7 & 1 &  &  & 1.213 & 2.864\\
X Del & M4e-M6e & 281.7 & 2.1 & 0 &  &  & 2.062 & 2.008\\
Y Del & M8e & 469.2 & 3.0 & 1 &  &  & 3.095 & 2.586\\
Z Del & S5.2.5e-S7.2e: & 304.9 & 1.6 & 1 & yes$^{a}$ &  & 1.584 & 1.249\\
T Dor & M5IIe & 169.9 & 3.6 & 0 &  &  & 2.074 & 1.588\\
U Dor & M8IIIe & 395.1 & 2.0 & 0 &  &  & 3.219 & 2.896\\
R Dra & M5e-M9eIII & 246.0 & 2.0 & 0 & dbfl$^{b}$ &  & 2.304 & 1.879\\
RT Dra & M5 & 277.7 & 3.2 & 0 &  &  & 1.771 & 1.922\\
RV Dra & M1e-M3e & 208.8 & 1.9 & 0 &  &  & 1.739 & 1.610\\
SV Dra & M7e & 256.3 & 2.0 & 0 &  &  & 1.899 & 1.703\\
T Dra & C6.2e-C8.3e(N0e) & 421.6 & 4.0 & 1 & yes$^{a}$ & 24$^{4}$ & 3.377 & 2.621\\
U Dra & M6e-M8 & 322.3 & 4.4 & 1 &  &  & 1.970 & 1.747\\
V Dra & M4e & 278.8 & 2.9 & 0 &  &  & 1.716 & 1.539\\
W Dra & M3e-M4e & 279.3 & 14.9 & 0 & no$^{a}$ &  & 3.033 & 2.413\\
X Dra & M5e: & 257.5 & 2.3 & 0 &  &  & 2.321 & 1.655\\
Y Dra & M5e & 326.0 & 2.8 & 0 &  &  & 2.544 & 2.278\\
ZZ Dra & M7e & 267.9 & 2.2 & 0 &  &  & 2.393 & 2.002\\
R Equ & M3e-M4e & 260.6 & 1.9 & 0 &  &  & 1.382 & 1.609\\
RS Eri & M7e & 299.4 & 2.4 & 0 &  &  & 1.870 & 1.842\\
RT Eri & M7e & 371.0 & 1.9 & 1 &  & 15$\pm7^{1}$ & 2.771 & 2.158\\
T Eri & M3e-M5e & 252.8 & 2.4 & 0 & no$^{a}$ &  & 2.104 & 1.719\\
U Eri & M4e & 274.3 & 1.5 & 0 & no$^{a}$ &  & 1.652 & 1.541\\
W Eri & M7e-M9 & 376.1 & 3.2 & 0 & no$^{a}$ &  & 2.769 & 2.098\\
R For & C4.3e(Ne) & 388.7 & 2.1 & 1 &  &  & 3.206 & 2.188\\
BP Gem & M8e & 243.9 & 2.0 & 0 &  &  & 1.763 & 2.203\\
R Gem & S2.9e-S8.9e(Tc) & 370.1 & 1.6 & 0 & yes$^{a}$ & 22$^{2}$ & 1.934 & 1.985\\
S Gem & M4e-M8e & 292.8 & 1.7 & 0 &  &  & 2.765 & 2.239\\
T Gem & S1.5.5e-S9.5e & 287.5 & 3.8 & 1 & yes$^{a}$ &  & 0.115 & 0.761\\
V Gem & M4(S)e-M8 & 275.1 & 4.7 & 1 & yes$^{a}$ &  & 1.463 & 1.477\\
VX Gem & C7.2e-C9.1e(Nep) & 379.4 & 1.8 & 1 & yes$^{a}$ & 9$^{3}$ & 1.581 & 1.625\\
WZ Gem & M3e & 333.1 & 2.7 & 1 &  &  & 1.790 & 1.547\\
X Gem & M5e-M8e(Tc:) & 264.1 & 2.6 & 0 & poss$^{b}$ &  & 1.661 & 1.599\\
R Gru & M5e-M7II-IIIe & 332.6 & 2.4 & 0 &  &  & 2.171 & 1.943\\
S Gru & M5e-M8IIIe & 400.9 & 2.7 & 1 & yes$^{a}$ &  & 1.956 & 1.858\\
T Gru & M1Iae-M2Ibe & 136.8 & 1.5 & 0 &  &  & 1.434 & 1.005\\
AE Her & M4e & 249.0 & 1.2 & 0 &  &  & 2.082 & 2.299\\
AS Her & M2e & 268.2 & 1.9 & 0 &  &  & 1.583 & 1.811\\
AZ Her & M4 & 270.8 & 1.1 & 0 &  &  & 2.063 & 1.686\\
BG Her & M3e & 349.2 & 2.9 & 1 &  &  & 2.115 & 2.195\\
CF Her & M0 & 306.7 & 2.9 & 0 & dbfl$^{b}$ &  & 1.927 & 1.925\\
R Her & M6e & 317.9 & 1.6 & 0 &  &  & 3.058 & 2.503\\
RS Her & M4e-M8: & 219.4 & 2.3 & 0 &  &  & 2.093 & 1.638\\
RT Her & M4e & 297.9 & 2.3 & 0 &  &  & 3.001 & 2.648\\
RU Her & M6e-M9 & 486.4 & 4.5 & 1 & yes$^{a}$ & 25$\pm5^{1}$ & 2.217 & 2.440\\
RV Her & M2e & 205.6 & 1.5 & 0 &  &  & 2.834 & 2.235\\
RY Her & M4e-M6e & 221.5 & 2.3 & 0 &  &  & 1.839 & 1.523\\
RZ Her & M5e-M6e & 328.5 & 2.1 & 0 & prob$^{b}$ &  & 2.619 & 1.951\\
S Her & M4.Se-M7.5.Se & 305.5 & 5.2 & 1 & yes$^{a}$ &  & 1.032 & 1.219\\
SS Her & M0e-M5e & 106.7 & 2.8 & 0 &  &  & 2.158 & 1.829\\
SU Her & M6e & 341.6 & 5.7 & 1 &  &  & 1.963 & 2.018\\
SV Her & M5e & 238.3 & 2.1 & 0 & no$^{a}$ &  & 1.588 & 1.507\\
SY Her & M1e-M6e & 116.7 & 1.7 & 0 & dbfl$^{b}$ &  & 1.802 & 1.503\\
T Her & M2.5e-M8e & 164.9 & 1.8 & 0 & no$^{a}$ &  & 1.286 & 1.572\\
TV Her & M4e & 304.7 & 2.3 & 0 &  &  & 2.896 & 2.062\\
U Her & M6.5e-M9.5e & 406.0 & 2.2 & 1 & no$^{c}$ & 19$\pm8^{1}$ & 2.442 & 2.478\\
UV Her & M6e-M6.5e & 345.4 & 2.6 & 1 & dbfl$^{b}$ &  & 1.950 & 2.002\\
UZ Her & M5e & 264.6 & 1.5 & 0 &  &  & 2.203 & 2.154\\
VY Her & M5e: & 300.1 & 2.0 & 0 &  &  & 2.453 & 2.547\\
W Her & M3e-M5e & 279.9 & 3.2 & 0 &  &  & 1.744 & 1.492\\
XZ Her & M0 & 171.0 & 2.9 & 0 &  &  & 1.299 & 1.449\\
R Hor & M5e-M8eII-III & 404.7 & 2.9 & 1 & yes$^{a}$ &  & 2.429 & 1.826\\
S Hor & M7II:e & 337.4 & 1.8 & 0 &  &  & 3.532 & 2.824\\
T Hor & M5IIe & 217.5 & 2.8 & 0 & no$^{a}$ &  & 1.899 & 1.527\\
U Hor & M6IIIe & 351.2 & 2.0 & 0 &  &  & 3.203 & 2.633\\
R Hya & M6e-M9eS & 388.0 & 15.4 & 1 & yes$^{a}$ & 26$\pm4^{1}$ & 0.839 & 0.897\\
RR Hya & M3.0e-M8e & 341.8 & 4.1 & 1 & yes$^{a}$ &  & 1.624 & 1.504\\
RS Hya & M6e & 335.9 & 3.5 & 1 &  &  & 1.817 & 1.681\\
RU Hya & M6e-M8.8e & 331.6 & 2.1 & 0 & no$^{a}$ &  & 3.521 & 3.038\\
S Hya & M4e-M8.0e & 257.0 & 3.9 & 1 &  &  & 1.510 & 1.468\\
ST Hya & Me & 304.1 & 1.6 & 1 &  &  & 1.489 & 1.648\\
T Hya & M3e-M9:e & 285.7 & 6.7 & 1 & prob$^{b}$ &  & 1.351 & 1.463\\
TU Hya & M5e & 279.4 & 3.6 & 0 &  &  & 1.929 & 1.365\\
X Hya & M7e-M8.5e & 300.2 & 3.7 & 1 & no$^{a}$ &  & 2.565 & 1.945\\
R Ind & M2e-M4(II)e & 216.5 & 1.4 & 0 &  &  & 1.569 & 1.346\\
RZ Ind & Me & 255.4 & 1.6 & 0 &  &  & 2.264 & 1.782\\
S Ind & M6e-M8eII-Ib: & 400.2 & 2.0 & 1 &  &  & 1.873 & 1.854\\
X Ind & M4e-M5IIe & 225.8 & 2.7 & 0 &  &  & 1.995 & 1.653\\
Y Ind & M6(II))e-M7e & 303.6 & 2.0 & 1 &  &  & 1.782 & 1.836\\
R Lac & M5e-M8.5e & 300.8 & 3.0 & 0 &  &  & 2.587 & 1.785\\
S Lac & M4e-M8.2e & 240.0 & 2.9 & 0 & no$^{a}$ &  & 2.107 & 1.916\\
W Lac & M7e-M8e & 320.2 & 5.8 & 1 &  &  & 1.955 & 1.748\\
R Leo & M6e-M8IIIe-M9.5e & 311.9 & 3.2 & 1 & no$^{a}$ & 10$\pm3^{1}$ & 1.347 & 0.951\\
RS Leo & M5e & 207.7 & 1.9 & 0 &  &  & 1.828 & 2.095\\
S Leo & M3e-M6e: & 189.6 & 2.6 & 0 &  &  & 1.290 & 1.214\\
V Leo & M5e & 273.5 & 1.5 & 0 &  &  & 2.706 & 2.097\\
W Leo & M5.5e-M7e & 387.6 & 3.1 & 1 &  &  & 2.668 & 2.826\\
R Lep & C7.6e(N6e) & 427.1 & 4.7 & 1 &  & 34$\pm5^{1}$ & 3.014 & 2.007\\
T Lep & M6e-M9e & 369.1 & 2.7 & 1 & dbfl$^{b}$ & 19$\pm4^{1}$ & 1.994 & 2.143\\
R Lib & M5e & 242.1 & 1.7 & 0 &  &  & 2.607 & 1.988\\
RR Lib & M4e-M8e & 277.6 & 2.2 & 0 & no$^{a}$ &  & 2.122 & 1.711\\
RS Lib & M7e-M8.5e & 218.1 & 2.8 & 0 & dbfl$^{b}$ & 16$\pm4^{1}$ & 1.625 & 1.918\\
RT Lib & M2.5pe-M8.2e & 251.6 & 2.0 & 0 & dbfl$^{b}$ &  & 2.113 & 1.835\\
RU Lib & M5e-M6e & 316.4 & 2.5 & 1 &  &  & 2.075 & 1.626\\
S Lib & M1.0e-M6.0e & 193.3 & 3.6 & 0 &  &  & 1.184 & 1.112\\
SX Lib & M6e: & 332.0 & 1.8 & 0 &  &  & 2.329 & 2.160\\
T Lib & M4e-M5.5e & 237.9 & 2.1 & 0 &  &  & 1.991 & 1.523\\
U Lib & M3e-M8.0e & 227.3 & 2.2 & 0 &  &  & 1.998 & 1.897\\
V Lib & M5e-M8.0e & 255.7 & 1.2 & 0 &  &  & 2.533 & 1.919\\
X Lib & M4e & 165.4 & 4.3 & 0 &  &  & 2.027 & 2.020\\
Y Lib & M5e-M8.2e & 276.2 & 2.2 & 0 & no$^{a}$ &  & 3.158 & 2.971\\
YY Lib & Me & 230.1 & 1.7 & 0 &  &  & 2.890 & 2.370\\
Z Lib & M3e & 301.8 & 3.0 & 0 &  &  & 2.142 & 1.870\\
R LMi & M6.5e-M9.0e(Tc:) & 372.9 & 2.1 & 0 & no$^{a}$ & 12$\pm2^{1}$ & 3.047 & 2.374\\
S LMi & M2.0e-M8.2e & 233.6 & 2.6 & 0 & dbfl$^{b}$ &  & 2.417 & 2.346\\
GI Lup & S7.8e & 325.0 & 3.2 & 1 &  &  & 1.528 & 1.618\\
R Lup & M5e & 234.8 & 2.5 & 0 &  &  & 1.618 & 1.646\\
RT Lup & Me & 365.2 & 2.2 & 0 &  &  & 2.647 & 2.375\\
S Lup & Se & 342.8 & 3.2 & 1 & yes$^{a}$ &  & 1.971 & 1.328\\
Y Lup & M7e & 402.0 & 2.8 & 0 &  &  & 3.144 & 3.020\\
R Lyn & S2.5.5e-S6.8e: & 378.4 & 2.4 & 1 & yes$^{a}$ &  & 1.602 & 1.132\\
RT Lyn & M6e & 395.3 & 3.0 & 1 & yes$^{a}$ &  & 1.810 & 1.710\\
S Lyn & M6e-M8.2e & 297.2 & 2.7 & 0 &  &  & 2.205 & 1.761\\
T Lyn & C5.2e-C7.1e(NOe) & 409.0 & 6.4 & 1 &  &  & 2.148 & 1.786\\
U Lyn & M7e-M9.5:e & 435.6 & 3.0 & 1 & no$^{a}$ &  & 2.919 & 2.648\\
W Lyn & M6 & 294.6 & 1.7 & 0 &  &  & 2.794 & 2.707\\
X Lyn & M5e & 322.4 & 2.8 & 0 &  &  & 2.017 & 2.090\\
RS Lyr & M5e & 304.6 & 3.3 & 1 &  &  & 1.253 & 1.410\\
RT Lyr & M5e & 252.7 & 2.8 & 0 &  &  & 2.454 & 1.764\\
RU Lyr & M6e:-M8e & 368.6 & 3.8 & 1 &  &  & 2.150 & 1.932\\
RW Lyr & M7e & 502.7 & 3.2 & 1 &  &  & 3.744 & 3.574\\
RY Lyr & M5e-M6e & 325.9 & 2.8 & 1 &  &  & 1.655 & 1.807\\
SS Lyr & M5IIIe & 351.7 & 3.8 & 0 &  &  & 2.417 & 2.241\\
ST Lyr & M4IIIe & 300.0 & 2.0 & 0 &  &  & 1.770 & 1.339\\
TW Lyr & M6 & 378.7 & 2.1 & 1 &  &  & 2.665 & 2.280\\
TX Lyr & M2e & 222.6 & 2.2 & 0 &  &  & 1.491 & 1.540\\
TY Lyr & M8e & 336.2 & 1.8 & 1 &  &  & 2.026 & 1.898\\
U Lyr & C4.5e(N0e) & 451.7 & 3.5 & 1 &  & 23$^{3}$ & 2.402 & 1.752\\
V Lyr & M7e & 374.2 & 1.3 & 0 & poss$^{b}$ &  & 2.824 & 2.756\\
W Lyr & M2e-M8e & 196.9 & 3.0 & 0 &  &  & 1.986 & 1.512\\
WZ Lyr & M9e & 375.9 & 2.1 & 1 &  &  & 2.303 & 2.035\\
Z Lyr & M4e-M5.5:e & 288.3 & 3.4 & 0 &  &  & 1.974 & 1.729\\
R Mic & M4e & 139.0 & 2.2 & 0 &  &  & 1.278 & 1.743\\
S Mic & M3e-M5.5 & 208.8 & 1.4 & 0 &  &  & 2.654 & 2.035\\
U Mic & M5e-M7e & 334.8 & 1.8 & 0 & no$^{a}$ &  & 3.254 & 3.223\\
V Mic & M3e-M6e & 384.8 & 1.8 & 0 &  &  & 4.047 & 3.146\\
BC Mon & M3e & 273.1 & 2.9 & 0 &  &  & 2.056 & 1.293\\
RR Mon & S7.2e-S8.2e/M6-10 & 394.3 & 2.5 & 1 &  &  & 2.881 & 2.310\\
RS Mon & M3e-M6e: & 263.8 & 2.7 & 0 &  &  & 2.558 & 1.947\\
RX Mon & M6e-M9 & 340.7 & 2.9 & 0 &  &  & 2.021 & 1.674\\
SY Mon & M6e-M9 & 422.6 & 2.4 & 1 & no$^{a}$ &  & 2.717 & 2.508\\
TT Mon & M5e-M8 & 320.7 & 2.2 & 0 &  &  & 3.018 & 2.690\\
V Mon & M5e-M8e & 334.3 & 2.6 & 1 & prob$^{b}$ &  & 1.841 & 1.811\\
Y Mon & M4e-M8.2e & 229.8 & 2.6 & 0 &  &  & 1.683 & 1.540\\
R Nor & M3e-M6II & 497.8 & 4.0 & 1 & no$^{a}$ &  & 2.212 & 2.440\\
T Nor & M3e-M6e & 242.7 & 1.7 & 0 &  &  & 2.508 & 3.156\\
R Oct & M5.5e & 405.6 & 2.7 & 1 &  &  & 1.944 & 1.969\\
RT Oct & Me & 179.0 & 2.8 & 0 &  &  & 1.835 & 1.533\\
S Oct & M4(II)e-M5e & 258.9 & 1.9 & 0 &  &  & 1.919 & 2.077\\
T Oct & M2e-M4(II:)e & 219.3 & 2.7 & 0 &  &  & 1.855 & 2.282\\
U Oct & M4e-M6(II-III)e & 302.9 & 2.6 & 1 & no$^{a}$ &  & 1.855 & 1.826\\
BC Oph & M6e & 308.3 & 1.3 & 0 &  &  & 2.403 & 2.080\\
R Oph & M4e-M6e & 303.2 & 1.6 & 0 & no$^{a}$ &  & 2.279 & 1.853\\
RR Oph & M3e-M7 & 293.6 & 2.7 & 0 & dbfl$^{b}$ &  & 1.871 & 1.578\\
RT Oph & M7e(C) & 427.3 & 2.6 & 1 &  &  & 2.211 & 2.294\\
RU Oph & M3e-M5e & 202.1 & 2.5 & 0 &  &  & 1.991 & 1.444\\
RX Oph & M5 & 322.4 & 1.9 & 0 &  &  & 2.830 & 2.495\\
RY Oph & M3e-M6e & 150.4 & 1.3 & 0 & no$^{a}$ &  & 1.678 & 1.790\\
SS Oph & M5e & 180.0 & 1.7 & 0 &  &  & 1.894 & 1.714\\
SV Oph & M2e & 215.5 & 1.4 & 0 &  &  & 2.465 & 1.738\\
T Oph & M6.5e & 366.7 & 3.3 & 1 &  &  & 2.414 & 1.810\\
UX Oph & M4e & 116.6 & 1.7 & 0 &  &  & 1.575 & 1.408\\
V Oph & C5.2-C7.4e(N3e) & 297.2 & 4.4 & 1 & yes$^{a}$ & 11$^{3}$ & 1.639 & 1.412\\
VW Oph & M5e & 285.6 & 2.5 & 0 &  &  & 2.139 & 1.923\\
W Oph & M8e & 331.3 & 3.3 & 1 &  &  & 2.261 & 2.026\\
X Oph & M5e-M9e & 333.6 & 3.6 & 1 & no$^{a}$ & 12$\pm4^{1}$ & 2.109 & 1.793\\
BK Ori & M7e & 336.9 & 4.0 & 1 &  &  & 1.374 & 1.692\\
EP Ori & M10e & 345.9 & 4.2 & 0 &  &  & 3.205 & 2.351\\
EU Ori & M4 & 328.3 & 1.8 & 1 &  &  & 1.696 & 1.469\\
R Ori & C8.2e(Ne) & 379.2 & 2.9 & 0 &  &  & 2.007 & 1.491\\
RR Ori & M6.5e-M8 & 252.5 & 2.0 & 0 &  &  & 1.824 & 1.737\\
S Ori & M6.5e-M9.5e & 424.3 & 8.5 & 1 & yes$^{a}$ & 45$\pm13^{1}$ & 1.733 & 1.860\\
U Ori & M6e-M9.5e & 371.5 & 2.4 & 1 & poss$^{b}$ & 25$\pm10^{1}$ & 2.861 & 2.264\\
V Ori & M3e-M8.0e & 267.8 & 4.2 & 0 &  &  & 1.939 & 1.421\\
R Pav & M3e-M5(II)e & 230.4 & 2.6 & 0 &  &  & 2.177 & 1.846\\
SU Pav & M4(II)e-M6II-IIIe & 246.1 & 2.0 & 0 &  &  & 1.549 & 1.850\\
T Pav & M4e & 243.9 & 1.2 & 0 &  &  & 2.768 & 2.066\\
W Pav & M4e-M7e & 283.4 & 2.1 & 0 &  &  & 3.108 & 2.813\\
AN Peg & M5 & 272.2 & 2.9 & 0 &  &  & 2.051 & 1.781\\
DG Peg & M4e & 146.6 & 2.7 & 0 &  &  & 1.820 & 1.632\\
DL Peg & M0 & 181.6 & 2.2 & 0 &  &  & 2.215 & 1.739\\
FF Peg & M5 & 251.7 & 1.6 & 0 &  &  & 1.658 & 1.639\\
R Peg & M6e-M9e & 377.4 & 2.6 & 1 & no$^{a}$ & 10$\pm4^{1}$ & 2.443 & 2.233\\
RR Peg & M4e-M8e & 264.5 & 1.5 & 0 &  &  & 2.476 & 2.417\\
RS Peg & M6e-M9e & 414.2 & 2.2 & 1 &  &  & 2.356 & 2.318\\
RT Peg & M3e-M6e & 215.0 & 3.7 & 0 &  &  & 1.568 & 0.880\\
RV Peg & M6e & 390.4 & 2.5 & 0 & poss$^{b}$ &  & 4.334 & 3.969\\
RZ Peg & C9.1e(Ne)(Tc)/CSe & 438.7 & 1.8 & 1 & yes$^{a}$ & 9$\pm5^{1}$ & 2.154 & 2.337\\
S Peg & M5e-M8.5e & 319.4 & 1.6 & 1 & no$^{a}$ &  & 1.394 & 1.563\\
SS Peg & M6e-M7e & 416.3 & 5.7 & 1 &  &  & 1.962 & 1.724\\
SX Peg & S3.9e-S4.5.9e & 306.7 & 2.6 & 1 &  &  & 0.777 & 0.818\\
T Peg & M6e-M8e & 374.0 & 2.4 & 1 & prob$^{b}$ &  & 2.247 & 1.502\\
TU Peg & M7e-M8e & 322.4 & 1.9 & 1 & dbfl$^{b}$ &  & 1.923 & 1.715\\
TV Peg & M0e & 246.5 & 2.8 & 0 &  &  & 1.936 & 1.584\\
TZ Peg & M3e & 216.2 & 1.9 & 0 &  &  & 1.676 & 1.444\\
V Peg & M3e-M7e & 303.1 & 1.7 & 0 &  &  & 2.321 & 1.846\\
W Peg & M6e-M8e & 345.4 & 3.5 & 1 & no$^{a}$ & 17$\pm7^{1}$ & 2.117 & 1.942\\
X Peg & M2e-M5e & 200.9 & 2.5 & 0 & no$^{a}$ &  & 1.401 & 1.243\\
Y Peg & M3e-M5e & 207.0 & 1.9 & 0 &  &  & 2.420 & 1.474\\
Z Peg & M6e-M8.5e(Tc) & 327.5 & 4.2 & 1 & yes$^{a}$ &  & 1.627 & 1.724\\
R Per & M2e-M5e & 210.1 & 2.4 & 0 &  &  & 1.140 & 1.364\\
RR Per & M6e-M7e & 390.6 & 1.8 & 0 & poss$^{b}$ &  & 3.027 & 2.329\\
RX Per & M3e & 419.4 & 3.6 & 1 &  &  & 2.926 & 2.508\\
RZ Per & S4.9e & 354.8 & 1.7 & 1 &  &  & 0.695 & 0.723\\
TW Per & M2e & 337.9 & 1.2 & 0 &  &  & 3.190 & 3.113\\
U Per & M5e-M7e & 318.9 & 4.4 & 1 & prob$^{b}$ &  & 1.602 & 1.580\\
Y Per & C4.3e(R4e) & 248.6 & 4.0 & 1 &  &  & 1.616 & 1.060\\
R Phe & M2(II)e-M4(III)e & 268.4 & 3.0 & 0 &  &  & 1.770 & 1.506\\
T Phe & M5e & 282.2 & 2.1 & 0 &  &  & 2.432 & 1.897\\
V Phe & M4e & 257.7 & 2.7 & 0 &  &  & 2.043 & 1.768\\
W Phe & M5e-M6e & 332.4 & 2.4 & 1 &  &  & 2.006 & 2.199\\
Z Phe & Me & 261.1 & 2.7 & 0 &  &  & 1.832 & 1.573\\
S Pic & M6.5e-M8III-IIe & 424.6 & 2.6 & 1 &  &  & 2.917 & 2.961\\
T Pic & M6IIIe & 201.0 & 1.5 & 0 &  &  & 1.780 & 1.336\\
R PsA & M3(II)e-M5IIe & 292.8 & 1.3 & 0 &  &  & 1.839 & 1.746\\
RY PsA & Me & 226.7 & 2.2 & 0 &  &  & 1.268 & 1.517\\
S PsA & M3e-M5IIe & 272.3 & 2.2 & 0 &  &  & 2.672 & 2.470\\
ST PsA & Me & 180.5 & 2.8 & 0 &  &  & 1.625 & 1.402\\
R Psc & M3e-M6e & 344.5 & 2.0 & 0 & prob$^{b}$ &  & 2.448 & 2.259\\
S Psc & M5e-M7e & 406.4 & 1.7 & 1 &  &  & 2.399 & 2.496\\
U Psc & M4e & 173.4 & 1.7 & 0 &  &  & 1.860 & 1.245\\
X Psc & M5e-M6e & 352.3 & 1.7 & 0 &  &  & 3.092 & 1.916\\
AS Pup & M7e-M9 & 327.4 & 1.8 & 1 &  &  & 1.761 & 1.962\\
CH Pup & Me & 499.8 & 2.8 & 1 &  &  & 3.361 & 3.315\\
RV Pup & M1e-M9 & 188.4 & 2.1 & 0 &  &  & 1.653 & 1.475\\
RW Pup & M3e-M6e & 336.4 & 3.8 & 1 &  &  & 2.043 & 1.916\\
SU Pup & M(S4.2)e & 341.1 & 2.1 & 0 &  &  & 2.534 & 1.938\\
SV Pup & M5e & 168.2 & 3.0 & 0 &  &  & 4.014 & 3.412\\
TU Pup & M8 & 240.5 & 2.1 & 0 &  &  & 2.207 & 1.756\\
U Pup & M5e-M8e & 316.6 & 2.5 & 0 &  &  & 3.139 & 3.285\\
W Pup & M1e-M6e & 120.3 & 0.8 & 0 &  &  & 1.823 & 1.403\\
Z Pup & M4e-M9e & 511.6 & 2.6 & 1 & dbfl$^{b}$ &  & 3.247 & 3.213\\
R Pyx & C(R)e & 364.7 & 3.8 & 1 &  &  & 2.555 & 1.738\\
S Pyx & M3e-M5e & 206.2 & 1.5 & 0 &  &  & 1.878 & 1.441\\
R Ret & M4e-M7.5e & 278.3 & 2.2 & 0 &  &  & 2.227 & 1.895\\
S Scl & M3e-M9e(Tc) & 365.5 & 3.6 & 1 & prob$^{b}$ &  & 1.854 & 1.832\\
T Scl & M3-M6e & 203.6 & 2.5 & 0 &  &  & 1.605 & 1.327\\
U Scl & M5e & 333.8 & 2.4 & 0 &  &  & 2.313 & 2.448\\
V Scl & M4e-M6e & 296.6 & 2.0 & 1 &  &  & 1.540 & 1.866\\
BK Sco & M7e & 197.5 & 2.0 & 0 &  &  & 1.495 & 1.439\\
R Sco & M(3)e & 223.5 & 1.8 & 0 &  &  & 2.377 & 1.648\\
RS Sco & M5e-M9 & 319.4 & 2.2 & 0 &  &  & 2.503 & 2.495\\
RT Sco & S7.2(M6e-M7e) & 448.1 & 2.9 & 1 &  &  & 2.551 & 2.705\\
RU Sco & M4/6e-M7II-IIIe & 370.5 & 6.7 & 1 &  &  & 2.142 & 1.906\\
RW Sco & M5e & 389.6 & 2.3 & 0 &  &  & 3.257 & 2.359\\
S Sco & M(3)e & 177.6 & 2.2 & 0 &  &  & 1.727 & 1.600\\
SV Sco & M3e & 256.5 & 2.7 & 0 &  &  & 1.809 & 1.595\\
SY Sco & M6e & 236.0 & 1.3 & 0 &  &  & 2.693 & 1.914\\
X Sco & M2e & 199.3 & 2.0 & 0 &  &  & 1.361 & 1.162\\
Z Sco & M5.5e:-M7e & 345.3 & 5.0 & 1 &  &  & 1.633 & 1.756\\
AH Ser & M2 & 283.0 & 2.5 & 0 &  &  & 2.952 & 2.114\\
BC Ser & M3e-M5e & 245.3 & 2.0 & 0 &  &  & 2.103 & 2.112\\
R Ser & M5IIIe-M9e & 355.6 & 1.7 & 1 & yes$^{a}$ & 14$\pm4^{1}$ & 1.936 & 2.021\\
RX Ser & Me & 215.5 & 2.5 & 0 &  &  & 2.147 & 1.690\\
S Ser & M5e-M6e & 369.0 & 1.9 & 1 &  &  & 2.334 & 2.146\\
T Ser & M7e & 340.0 & 5.9 & 1 &  &  & 1.567 & 1.745\\
U Ser & M3e-M6e & 237.5 & 1.7 & 0 & poss$^{b}$ &  & 2.138 & 1.722\\
S Sex & M2e-M5e & 259.0 & 6.0 & 1 &  &  & 1.704 & 1.313\\
ST Sge & M4e & 191.7 & 1.6 & 0 &  &  & 1.185 & 1.022\\
W Sge & M4e-M6.5 & 278.5 & 1.4 & 0 &  &  & 2.926 & 2.829\\
R Sgr & M4e-M6e & 268.6 & 1.5 & 0 & no$^{a}$ &  & 1.529 & 1.448\\
RR Sgr & M4e-M9e & 334.3 & 2.1 & 1 & prob$^{b}$ &  & 2.168 & 2.034\\
RT Sgr & M5e-M7e & 306.5 & 1.6 & 0 & dbfl$^{b}$ &  & 2.309 & 1.905\\
RU Sgr & M3e-M6e & 240.3 & 2.1 & 0 &  &  & 1.609 & 1.876\\
RX Sgr & M5e & 333.3 & 2.1 & 1 & yes$^{a}$ &  & 1.643 & 1.459\\
S Sgr & M3e-M4.5e & 230.8 & 2.2 & 0 &  &  & 2.232 & 1.561\\
ST Sgr & C4.3e-S9.5e & 394.0 & 1.8 & 0 & yes$^{a}$ &  & 2.293 & 1.573\\
SW Sgr & M5e-M8 & 290.7 & 3.1 & 0 &  &  & 1.869 & 1.877\\
T Sgr & S4.5.8e-S5.5.8e & 392.8 & 3.5 & 1 & yes$^{a}$ &  & 1.654 & 1.178\\
TW Sgr & M2e-M3e & 220.6 & 2.3 & 0 &  &  & 2.092 & 1.496\\
TY Sgr & M3e & 325.9 & 1.8 & 0 & prob$^{b}$ &  & 2.249 & 1.827\\
R Tau & M5e-M9e & 323.2 & 2.2 & 0 & no$^{a}$ &  & 2.516 & 2.359\\
RU Tau & M3.5e-M6.5 & 588.6 & 9.5 & 1 &  &  & 1.685 & 2.121\\
RX Tau & M6e-M7e & 334.5 & 3.3 & 0 &  &  & 2.803 & 2.256\\
S Tau & M6.5e-M9e & 374.1 & 1.9 & 0 &  &  & 2.637 & 1.957\\
TZ Tau & M9 & 266.6 & 4.5 & 0 &  &  & 1.923 & 1.590\\
V Tau & M0e-M4.5e & 170.0 & 1.8 & 0 &  &  & 2.002 & 1.650\\
VX Tau & M8e & 301.3 & 2.7 & 1 &  &  & 1.310 & 1.312\\
Z Tau & S7.5.1e(M7e) & 458.5 & 11.8 & 1 &  &  & 2.633 & 1.824\\
NT Tel & S1.9e-S5.9e & 255.4 & 3.1 & 1 &  &  & 1.256 & 0.582\\
R Tel & M5IIe-M7e & 463.6 & 2.6 & 1 &  &  & 2.633 & 2.432\\
W Tel & M5e-M8(III:)e & 307.3 & 1.9 & 0 &  &  & 2.579 & 2.301\\
X Tel & M5e-M8e & 306.4 & 2.9 & 0 &  &  & 1.976 & 2.020\\
W Tra & Me & 250.3 & 1.2 & 0 &  &  & 2.280 & 1.705\\
R Tri & M4IIIe-M8e & 266.7 & 1.1 & 0 & no$^{a}$ &  & 1.795 & 1.473\\
R Tuc & M5e & 286.2 & 1.7 & 0 & no$^{a}$ &  & 2.461 & 2.563\\
S Tuc & M3e-M5II-Ibe & 240.8 & 1.2 & 0 &  &  & 2.403 & 2.050\\
T Tuc & M3IIe-M6IIe & 250.9 & 3.6 & 0 &  &  & 1.331 & 1.665\\
U Tuc & M3e-M7e & 260.4 & 2.6 & 0 &  &  & 2.036 & 1.820\\
UU Tuc & M4e & 323.4 & 4.5 & 0 &  &  & 4.188 & 3.451\\
R UMa & M3e-M9e & 301.5 & 1.7 & 0 &  &  & 2.935 & 2.828\\
RR UMa & M4e & 231.1 & 2.6 & 0 & no$^{a}$ &  & 1.782 & 1.506\\
RS UMa & M4e-M6e & 259.6 & 1.9 & 0 &  &  & 1.689 & 1.831\\
RU UMa & M3e-M5e & 251.1 & 1.2 & 0 &  &  & 2.756 & 2.029\\
S UMa & S0.9e-S5.9e & 226.0 & 3.5 & 1 & yes$^{a}$ &  & 1.188 & 1.060\\
T UMa & M4IIIe-M7e & 256.3 & 2.3 & 0 & no$^{a}$ &  & 2.402 & 2.285\\
X UMa & M0III & 249.3 & 1.6 & 0 &  &  & 1.601 & 1.780\\
T Umi & M4e-M6e & 317.3 & 38.2 & 1 & no$^{a}$ &  & 1.212 & 1.492\\
U UMi & M6e-M8e & 325.6 & 3.6 & 1 &  &  & 1.477 & 1.674\\
W Vel & M5-M8IIIe & 394.2 & 3.5 & 1 &  &  & 2.597 & 2.298\\
Y Vel & M8e-M9.5 & 444.9 & 3.8 & 1 &  &  & 2.551 & 2.221\\
Z Vel & M9e & 411.4 & 7.0 & 1 &  &  & 2.327 & 1.892\\
R Vir & M3.5IIIe-M8.5e & 146.0 & 2.7 & 0 & no$^{a}$ &  & 1.989 & 1.628\\
RS Vir & M6IIIe-M8e & 353.9 & 1.7 & 0 & no$^{a}$ &  & 3.538 & 2.954\\
RU Vir & C8.1e(R3ep) & 433.2 & 3.2 & 1 & yes$^{a}$ &  & 4.507 & 2.790\\
RV Vir & M5e & 267.5 & 1.9 & 0 &  &  & 2.799 & 1.776\\
S Vir & M6IIIe-M9.5e & 377.5 & 2.9 & 1 & yes$^{a}$ &  & 1.977 & 1.935\\
SU Vir & M2e-M5.5e & 207.2 & 1.9 & 0 &  &  & 1.884 & 1.428\\
SV Vir & M4e & 295.6 & 2.4 & 0 &  &  & 2.681 & 2.333\\
SY Vir & M6: & 237.0 & 1.3 & 0 &  &  & 1.904 & 1.822\\
T Vir & M6e & 339.4 & 2.1 & 0 &  &  & 2.567 & 2.468\\
U Vir & M2e-M8e: & 206.7 & 2.4 & 0 & dbfl$^{b}$ &  & 1.007 & 1.266\\
V Vir & M3e-M6e & 249.4 & 2.0 & 0 &  &  & 2.687 & 2.404\\
Y Vir & M2e-M5e & 218.1 & 1.8 & 0 &  &  & 1.761 & 1.369\\
Z Vir & M5e & 304.0 & 2.3 & 0 &  &  & 2.865 & 2.443\\
R Vol & C(N)e & 453.6 & 2.2 & 1 &  &  & 4.069 & 2.559\\
S Vol & M4e & 394.8 & 4.3 & 1 &  &  & 2.142 & 1.930\\
T Vol & M5II-IIIe & 176.8 & 3.4 & 0 &  &  & 1.903 & 1.589\\
BD Vul & C6-7.3e(Ne) & 430.0 & 3.0 & 1 &  &  & 1.818 & 1.556\\
R Vul & M3e-M7e & 136.7 & 2.2 & 0 & no$^{a}$ &  & 2.037 & 1.587\\
YZ Vul & M5e-M9 & 369.8 & 4.0 & 1 &  &  & 2.347 & 2.162\\
\hline
\end{longtable}
\tablefoot{Meaning of Column 5: 0: Stars with symmetric light curves (Group~S); 1: Stars with asymmetric light curves (Group~A). Literature sources of Tc content in Column 6: (a) \citet{Uttenthaler2019}; (b) \citet{Little1987}; (c) Uttenthaler et al., in preparation; (d) \citet{Vanture2007}. Literature sources of $^{12}$C/$^{13}$C content in Column 7: (1) \citet{Hinkle2016}; (2) \citet{Ramstedt2014}; (3) \citet{Ohnaka1996}; (4) \citet{Abia1997}; (5) \citet{Lebzelter2019}; (6) \citet{Schoier2000}; (7) \citet{Dominy1987}.}

\twocolumn

\end{appendix}

\end{document}